# An Efficient Integrator Scheme for Sampling the (Quantum) Isobaric-Isothermal Ensemble in (Path Integral) Molecular Dynamics Simulations

*Weihao Liang[1,†], Sihan Wang[1,†], Cong Wang[1,2,†], Weizhou Wang[1], Xinchen She[1], Chongbin Wang[1], Jiushu Shao[2], Jian Liu[1,]\**

1. Beijing National Laboratory for Molecular Sciences, Institute of Theoretixcal and Computational Chemistry, College of Chemistry and Molecular Engineering, Peking University, Beijing 100871, China

2. College of Chemistry and Center for Advanced Quantum Studies, Key Laboratory of Theoretical and Computational Photochemistry, Ministry of Education, Beijing Normal University, Beijing, 100875, China

AUTHOR INFORMATION

**Corresponding Author**

\* Electronic mail: jianliupku@pku.edu.cn

**Author Contributions**

† W.Liang, S. Wang, and C. Wang contributed equally.

**ABSTRACT.**

Because most chemical or biological experiments are performed under conditions of controlled pressure and temperature, it is important to simulate the isobaric-isothermal ensemble at the atomic level to reveal the microscopic mechanism. By extending our configuration sampling protocol for the canonical ensemble, we propose a unified "middle" scheme to sample the coordinate (configuration) and volume distribution and thereby are able to accurately simulate either classical or quantum isobaric-isothermal processes. Various barostats and thermostats can be employed in the unified "middle" scheme for simulating real molecular systems with or without holonomic constraints. In particular, we demonstrate the recommended "middle" scheme by employing the Martyna-Tuckerman-Tobias-Klein barostat and stochastic cell-rescaling barostat, with the Langevin thermostat, in molecular simulation packages (DL_POLY, Amber, Gromacs, *etc.*). Benchmark numerical tests show that, without additional numerical effort, the "middle" scheme is competent in increasing the time interval by a factor of 5~10 to achieve the same accuracy of converged results for most thermodynamic properties in (path integral) molecular dynamics simulations.

**TOC GRAPHICS**

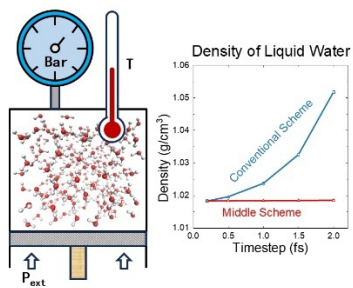

1. **Introduction**

Most real experiments or processes occur under conditions of constant pressure and temperature. Recent progress on pressure-driven phenomena suggests that exotic chemical, electronic, magnetic, and structural properties can be unlocked in molecular solids and condensed matter materials[1-5], which include superconductivity, quantum Hall effects, abnormal phase transition, to name but a few. The isobaric-isothermal ensemble, where the number of particles (N), external pressure (P), and temperature (T) are fixed, has been extensively used for studying real complex molecular systems *via* (path integral) molecular dynamics (MD) in physics, chemistry, biology, astronomy, and environmental science[6-8]. The time interval $\Delta t$ is the most fundamental factor that determines both the accuracy and efficiency of the MD simulation of the isobaric-isothermal ensemble. A larger time interval produces more efficient sampling of the ensemble, but often decreases the accuracy of or even breaks down the simulation. It is important yet challenging to design robust integrators/algorithms for the isobaric-isothermal ensemble, in which larger time intervals are allowed for considerably improving the sampling efficiency while maintaining the same accuracy.

The purpose of the paper is to present a simple, efficient, and robust integrator scheme for sampling the isobaric-isothermal (constant-NPT) ensemble for real molecular systems by MD or path integral MD simulations. The outline of the paper is as follows. Section 2 begins by briefly reviewing the theory on the isobaric-isothermal ensemble in classical mechanics as well as in quantum mechanics. It then introduces the strategy for rationally constructing the MD integrator by employing the barostatting methods *via* MD or Monte Carlo (MC)[7, 9-30]. In Section 3, the optimized "middle" scheme is numerically implemented and compared to conventional NPT algorithms used in molecular simulation packages. Benchmark test examples include the Lennard-Jones liquid, liquid *para*-Hydrogen, and liquid water. Finally, conclusion remarks are presented in Section 4.

## 2. Theory and Methodologies

We consider a general real molecular system that includes $N_{atom}$ atoms in the three-dimensional space, which is described by the (time-independent) Hamiltonian of standard Cartesian form

$$H(\mathbf{x},\mathbf{p}) = \mathbf{p}^T \mathbf{M}^{-1} \mathbf{p}/2 + U(\mathbf{x}) = K + U(\mathbf{x}) \quad, \tag{1}$$

where $\mathbf{M}$ is the diagonal 'mass matrix' with elements $\{m_j; 1 \leq j \leq 3N_{atom}\}$, $\mathbf{x}$ and $\mathbf{p}$ are the $3N_{atom}$-dimensional coordinate and momentum vectors, respectively, $U(\mathbf{x})$ is the potential energy, and $K$ is the kinetic energy.

### 2.1 NPT Ensemble in Classical Statistical Mechanics

In classical statistical mechanics, $\tilde{H}_{ins} = PV + H$ defines the instantaneous enthalpy and the average of any physical property of the isobaric-isothermal ensemble reads

$$\langle B \rangle_{NPT}^{CM} = \frac{I_N}{V_0 Z_{NPT}} \int_0^\infty dV \int_{D(V)} d\mathbf{x} \int d\mathbf{p} \exp\left[-\beta(PV + H(\mathbf{x},\mathbf{p}))\right] B(\mathbf{x},\mathbf{p};V), \tag{2}$$

where the partition function $Z_{NPT}$ of the isobaric-isothermal ensemble is

$$Z_{NPT} = \frac{I_N}{V_0} \int_0^\infty dV \int_{D(V)} d\mathbf{x} \int d\mathbf{p} \exp\left[-\beta(PV + H(\mathbf{x},\mathbf{p}))\right] \quad. \tag{3}$$

Here, $D(V)$ is the coordinate space of the volume $V$ of the system, $I_N = (2\pi\hbar)^{-3N_{atom}}$ is the normalization factor for the (coordinate-momentum) phase space integral, $\beta = 1/k_B T$ with the Boltzmann constant $k_B$, and $V_0$ is the unit volume. In eq (2), $B(\mathbf{x},\mathbf{p};V)$ is the estimator for the corresponding property $B$. For instance, the virial expression of the internal pressure in classical statistical mechanics is

$$P_{\text{int}} = \frac{1}{dV}\left(\mathbf{p}^T\mathbf{M}^{-1}\mathbf{p} - \mathbf{x}^T\frac{\partial U}{\partial \mathbf{x}}\right) - \frac{\partial U}{\partial V}, \tag{4}$$

where $d$ is the dimensionality of the space, in which the molecular system lies. (In the paper, $d = 3$ as we study the molecular system in the three-dimensional space.) The last term $\frac{\partial U}{\partial V}$ on the RHS of eq (4), which represents the explicit volume dependence of the potential energy function, is usually zero for molecular systems. (The last term of eq (4) should be ignored with caution when more general cases are studied.) As will be discussed in Sub-Section 2.4, the evaluation of the virial expression of the internal pressure is often requested in MD barostats. More discussion on the estimator of other physical properties is presented in Section S1 of the Supporting Information.

As the partition function of the canonical (constant-NVT) ensemble is $Z_{\text{NVT}} = I_N \int_{D(V)} \mathrm{d}\mathbf{x} \int \mathrm{d}\mathbf{p} \exp[-\beta H(\mathbf{x},\mathbf{p})]$, we have the relation

$$Z_{\text{NPT}} = \frac{1}{V_0}\int_0^\infty \mathrm{d}V \exp[-\beta PV] Z_{\text{NVT}}(N,V,T) \ . \tag{5}$$

Equation (5) can also be expressed as

$$e^{-\beta G(N,P,T)} = \int_0^\infty \mathrm{d}V e^{-\beta(PV+F(N,V,T))} , \tag{6}$$

where the Gibbs free energy is $G(N,P,T) = -\ln Z_{\text{NPT}}/\beta$ and Helmholtz free energy is $F(N,V,T) = -\ln Z_{\text{NVT}}/\beta$. Note that physical properties such as density, isothermal compressibility, isobaric heat capacity, thermal expansion coefficient, and so forth can be estimated only in the isobaric-isothermal ensemble rather than in the canonical ensemble. The accurate sampling of the integrand of the right-hand side (RHS) of eq (5), i.e., the partial probability distribution function of the volume, is the essential task of the barostat for controlled pressure. Equation (2) indicates that in classical statistical mechanics the accurate sampling of the joint distribution of the volume ($V$) and coordinate ($\mathbf{x}$) variables is the key for sampling the isobaric-isothermal ensemble, because most physical properties of interest depend only on the coordinate ($\mathbf{x}$) and volume ($V$). Note that the Maxwell momentum distribution is simply a Gaussian function that is independent of the coordinate and volume. It is straightforward to first use MD to obtain only the accurate joint distribution of the volume ($V$) and coordinate ($\mathbf{x}$) and then employ MC to faithfully obtain the independent momentum distribution, which produces accurate evaluation of any thermodynamic properties that involve both coordinate/volume and momentum variables in classical statistical mechanics.

## 2.2 NPT Ensemble in Quantum Statistical Mechanics

Equation (5) can also be justified in quantum statistical mechanics in the thermodynamic limit, where the canonical partition function is defined as

$$Z_{\mathrm{NVT}} = \mathrm{Tr}\left[\exp\left(-\beta \hat{H}\right)\right] = \int_{D(V)} d\mathbf{x} \langle \mathbf{x} | \exp\left(-\beta \hat{H}\right) | \mathbf{x} \rangle \tag{7}$$

and the corresponding average physical property of eq (2) becomes

$$\langle \hat{B} \rangle_{\mathrm{NPT}}^{\mathrm{QM}} = \frac{1}{V_0 Z_{\mathrm{NPT}}} \int_0^\infty dV \exp[-\beta PV] \int_{D(V)} d\mathbf{x} \langle \mathbf{x} | \exp\left(-\beta \hat{H}\right) \hat{B} | \mathbf{x} \rangle \ . \tag{8}$$

After decomposing $\exp(-\beta\hat{H})$ into $L$ equal parts $\left[\exp(-\beta\hat{H}/L)\right]^L$, substituting the resolution of identity $\hat{1} \equiv \int_{D(V)} d\mathbf{x}_j |\mathbf{x}_j\rangle\langle\mathbf{x}_j|$ into each neighbor parts, and applying the symmetric trotter factorization $\exp(-\beta\hat{H}/L) = \exp(-\beta\hat{U}/2L)\exp(-\beta\hat{K}/L)\exp(-\beta\hat{U}/2L)$, the sampling of the probability distribution function of eq (8) can often be accomplished by imaginary time path integral[31-34], namely,

$$\langle\hat{B}\rangle_{NPT}^{QM} = \frac{I_{PI}/V_0 \lim_{L\to\infty}\int_0^\infty dV e^{-\beta PV} \int_{D(V)} d\mathbf{x}_1 \int_{D(V)} d\mathbf{x}_2 \cdots \int_{D(V)} d\mathbf{x}_L \exp\{-\beta U_{eff}(\mathbf{x}_1,\cdots,\mathbf{x}_L)\}\tilde{B}(\mathbf{x}_1,\cdots,\mathbf{x}_L;V)}{I_{PI}/V_0 \lim_{L\to\infty}\int_0^\infty dV e^{-\beta PV} \int_{D(V)} d\mathbf{x}_1 \int_{D(V)} d\mathbf{x}_2 \cdots \int_{D(V)} d\mathbf{x}_L \exp\{-\beta U_{eff}(\mathbf{x}_1,\cdots,\mathbf{x}_L)\}} \quad (9)$$

where $L$ is the number of path integral beads (replica),

$$U_{eff}(\mathbf{x}_1,\cdots,\mathbf{x}_L) = \frac{L}{2\beta^2\hbar^2}\sum_{i=1}^{L}\left[(\mathbf{x}_{i+1}-\mathbf{x}_i)^T \mathbf{M}(\mathbf{x}_{i+1}-\mathbf{x}_i)\right] + \frac{1}{L}\sum_{i=1}^{L}U(\mathbf{x}_i) \quad (10)$$

is the effective potential of path integral beads, the estimator $\tilde{B}(\mathbf{x}_1,\cdots,\mathbf{x}_L;V)$ for any operator $\hat{B}$ is a function of only the coordinate vector of path integral beads and the volume, $I_{PI} = (L/2\pi\beta\hbar^2)^{3N_{atom}L/2}|\mathbf{M}|^{L/2}$ is a normalization factor, and the denominator of the RHS of eq (9) is the path integral expression of the partition function $Z_{NPT}$ of the isobaric-isothermal ensemble.

When the staging transformation[35, 36] is used, the transformation of coordinates and that of forces read

$$\begin{aligned}\xi_1 &= \mathbf{x}_1, \\ \xi_i &= \mathbf{x}_i - \frac{(i-1)\mathbf{x}_{i+1}+\mathbf{x}_1}{i} \quad (i=2,\cdots,L),\end{aligned} \quad (11)$$

and

$$\frac{\partial \phi}{\partial \xi_1} = \frac{1}{L}\sum_{i=1}^{L}U'(\mathbf{x}_i),$$
$$\frac{\partial \phi}{\partial \xi_i} = \frac{\partial \phi}{\partial \mathbf{x}_i} + \frac{i-2}{i-1}\frac{\partial \phi}{\partial \xi_{i-1}} \quad (i=2,\cdots,L), \tag{12}$$

where $\phi$ is defined as $\phi(\xi_1,\cdots,\xi_L) = \frac{1}{L}\sum_{i=1}^{L}U\left[\mathbf{x}_i(\xi_1,\cdots,\xi_L)\right]$. The effective potential (of eq (10)) after the staging transformation becomes

$$U_{\text{eff}}(\xi_1,\cdots,\xi_L) = \sum_{i=1}^{L}\frac{1}{2}\omega_L^2 \xi_i^T \overline{\mathbf{M}}_i \xi_i + \frac{1}{L}\sum_{i=1}^{L}U\left[\mathbf{x}_i(\xi_1,\cdots,\xi_L)\right], \tag{13}$$

The corresponding mass matrix of the staging transformation is

$$\overline{\mathbf{M}}_1 = 0,$$
$$\overline{\mathbf{M}}_i = \frac{i}{i-1}\mathbf{M} \quad (i=2,\cdots,L). \tag{14}$$

The integral of eq (9) can be performed by MD[36-39] after introducing the artificial momentum and mass of the path integral bead and substituting the identity of the multi-dimensional Gaussian integral $(2\pi)^{-3N_{\text{atom}}/2}\left|\widetilde{\mathbf{M}}_j\right|^{-1/2}\int d\mathbf{p}_j \exp\left(-\frac{1}{2}\mathbf{p}_j^T \widetilde{\mathbf{M}}_j^{-1}\mathbf{p}_j\right) = 1$, that is,

$$\langle \hat{B}\rangle_{\text{NPT}}^{\text{QM}} = \frac{\lim_{L\to\infty}\int_0^\infty dV \prod_{k=1}^{L}\left(\int d\mathbf{p}_k\right)\prod_{j=1}^{L}\left(\int_{D(V)} d\mathbf{x}_j\right) e^{-\beta\left[PV + H_{\text{eff}}(\mathbf{p}_1,\cdots,\mathbf{p}_L;\mathbf{x}_1,\cdots,\mathbf{x}_L)\right]}\tilde{B}(\mathbf{x}_1,\cdots,\mathbf{x}_L;V)}{\lim_{L\to\infty}\int_0^\infty dV \prod_{k=1}^{L}\left(\int d\mathbf{p}_k\right)\prod_{j=1}^{L}\left(\int_{D(V)} d\mathbf{x}_j\right) e^{-\beta\left[PV + H_{\text{eff}}(\mathbf{p}_1,\cdots,\mathbf{p}_L;\mathbf{x}_1,\cdots,\mathbf{x}_L)\right]}}, \tag{15}$$

where the effective Hamiltonian reads

$$H_{\text{eff}}(\mathbf{x}_1,\cdots,\mathbf{x}_L;\mathbf{p}_1,\cdots,\mathbf{p}_L) = \sum_{i=1}^{L}\frac{1}{2}\mathbf{p}_i^T \widetilde{\mathbf{M}}_i^{-1}\mathbf{p}_i + U_{\text{eff}}(\mathbf{x}_1,\cdots,\mathbf{x}_L). \tag{16}$$

In eq (16), the fictitious mass $\widetilde{\mathbf{M}}$ can be chosen as

$$\widetilde{\mathbf{M}}_1 = \mathbf{M},$$
$$\widetilde{\mathbf{M}}_i = \frac{i}{i-1}\mathbf{M} \quad (i = 2,\cdots,L). \tag{17}$$

Please see Section S1 of the Supporting Information for more discussion. In quantum statistical mechanics, the accurate sampling of the joint distribution of the volume and the coordinate vector of path integral beads of the isobaric-isothermal ensemble (of eq (9) or eq (15)) is then the only key for evaluating all thermodynamic properties.

In Sub-Section S1-C, we explicitly describe the estimator $B(\mathbf{x},\mathbf{p};V)$ of eq (2) in classical statistical mechanics as well as the estimator $\tilde{B}(\mathbf{x}_1,\cdots,\mathbf{x}_L;V)$ of either eq (9) or eq (15) in quantum statistical mechanics, when the density, enthalpy, isobaric heat capacity, isothermal compressibility, and thermal expansion coefficient are simulated.

## 2.3    Barostat for the Volume Distribution

Either MC or MD methods[7, 9-30] can be used for the barostat for sampling the volume distribution of eq (2) or eq (15). Note that the volume distribution of eq (2) or eq (15) includes not only $e^{-\beta PV}$ but also the potential energy in the confined volume (e.g., the box with periodic boundary conditions when the bulk system is investigated).

When MD is employed for the barostat, one approach is to substitute a classical resolution of identity, i.e., replacing a unit by one-dimensional Gaussian integral $[\beta/(2\pi W)]^{1/2}\int dp_V \exp[-\beta p_V^2/(2W)] = 1$ on the RHS of eq(2), which yields

$$\langle B \rangle_{\mathrm{NPT}}^{\mathrm{CM}} = \frac{\int dp_V \int_0^\infty dV \int_{D(V)} d\mathbf{x} \int d\mathbf{p}\, \exp\left[-\beta\left(\frac{p_V^2}{2W} + PV + H(\mathbf{x},\mathbf{p})\right)\right] B(\mathbf{x},\mathbf{p};V)}{\int dp_V \int_0^\infty dV \int_{D(V)} d\mathbf{x} \int d\mathbf{p}\, \exp\left[-\beta\left(\frac{p_V^2}{2W} + PV + H(\mathbf{x},\mathbf{p})\right)\right]}, \tag{18}$$

where $p_V$ and $W$ are the "momentum" variable and "mass" of the fictitious "piston" introduced for controlling the volume fluctuation and then sampling the volume distribution by MD trajectories. Similarly, eq (15) becomes

$$\langle \hat{B} \rangle_{\text{NPT}}^{\text{QM}} = \frac{\lim_{L \to \infty} \int \mathrm{d}p_V \int_0^\infty \mathrm{d}V \prod_{k=1}^{L} \left( \int \mathrm{d}\mathbf{p}_k \right) \prod_{j=1}^{L} \left( \int_{D(V)} \mathrm{d}\mathbf{x}_j \right) e^{-\beta \left[ \frac{p_V^2}{2W} + PV + H_{\text{eff}}(\mathbf{p}_1,\cdots,\mathbf{p}_L;\mathbf{x}_1,\cdots,\mathbf{x}_L) \right]} \tilde{B}(\mathbf{x}_1,\cdots,\mathbf{x}_L;V)}{\lim_{L \to \infty} \int \mathrm{d}p_V \int_0^\infty \mathrm{d}V \prod_{k=1}^{L} \left( \int \mathrm{d}\mathbf{p}_k \right) \prod_{j=1}^{L} \left( \int_{D(V)} \mathrm{d}\mathbf{x}_j \right) e^{-\beta \left[ \frac{p_V^2}{2W} + PV + H_{\text{eff}}(\mathbf{p}_1,\cdots,\mathbf{p}_L;\mathbf{x}_1,\cdots,\mathbf{x}_L) \right]}} . \quad (19)$$

In addition to the extended-system approach discussed above, another approach is to weakly couple the system to a pressure bath without the fictitious "piston" momentum, which includes the Berendsen barostat[40] and stochastic cell-rescaling (SCR) barostat[41]. During each pressure control step, the volume is scaled by a factor determined by the pressure difference between the bath and the system. While the Berendsen barostat fails to yield the correct volume distribution[42], the SCR barostat is capable of faithfully producing the isobaric-isothermal ensemble[41, 43] and will be used in the paper.

Because PIMD essentially employs MD of the effective Hamiltonian system of eq (16) while using different estimators for physical properties in quantum statistical mechanics from those in classical statistical mechanics, both MD and PIMD can in principle share the same integrator for the isobaric-isothermal ensemble. In the following we only use MD integrators to demonstrate the procedure for simplicity. It is straightforward to implement the same strategy for PIMD integrators.

## 2.4   Molecular Dynamics Integrators

Numerical MD integrators for a finite time interval $\Delta t$ often consist of a step for updating the coordinate   $\mathbf{x}(t+\Delta t) \leftarrow \mathbf{x}(t) + \mathbf{M}^{-1}\mathbf{p}(t)\Delta t$  ,   that   for   updating   the   momentum $\mathbf{p}(t+\Delta t) \leftarrow \mathbf{p}(t) - U'(\mathbf{x})\Delta t$, that for the thermostat for controlled temperature, and that for the barostat for controlled external pressure. Use $e^{\mathcal{L}_{\mathbf{x}}\Delta t}$, $e^{\mathcal{L}_{\mathbf{p}}\Delta t}$, $e^{\mathcal{L}_{T}\Delta t}$, and $e^{\mathcal{L}_{Bar}\Delta t}$ to represent the phase space propagators for the four steps, respectively. Here, $\mathcal{L}_{Bar}$ is the Kolmogorov operator for the barostat, which is defined in the volume space *via* MC or on "extended" phase space $(V, p_V)$ *via* MD, $\mathcal{L}_T$ is the Kolmogorov operator for the thermostat, and $\mathcal{L}_{\mathbf{x}}$ and $\mathcal{L}_{\mathbf{p}}$ are the relevant Kolmogorov operators on phase space $(\mathbf{x}, \mathbf{p})$ related to the system Hamiltonian,

$$\mathcal{L}_{\mathbf{x}}\rho = -\mathbf{p}^T \mathbf{M}^{-1} \frac{\partial \rho}{\partial \mathbf{x}} \quad \text{and} \quad \mathcal{L}_{\mathbf{p}}\rho = \left(\frac{\partial U}{\partial \mathbf{x}}\right)^T \frac{\partial \rho}{\partial \mathbf{p}}$$

, with $\rho$ being the corresponding probability distribution function of the ensemble.

Note that the characteristic time scale of the evolution of coordinate-momentum variables $(\mathbf{x}, \mathbf{p})$ of the molecular system is usually several orders of magnitude smaller than that of the fluctuation of the volume variable, $V$. It is expected that an efficient algorithm for the isobaric-isothermal (constant-NPT) ensemble should be rationally constructed from that for the canonical (constant-NVT) ensemble. Recently, we have shown that the unified middle thermostat scheme leads to efficient and robust (path integral) molecular dynamics algorithms for sampling (the coordinate space of) the canonical ensemble, regardless of whether the thermostat for controlled temperature is stochastic, deterministic, or mixed[36, 44, 45]. For instance, it is straightforward in the unified middle thermostat scheme to derive the relation[46] between the algorithm on Langevin dynamics by Leimkuhler and Matthews[47] and that by Grønbech-Jensen and Farago independently[48]. More importantly, the unified thermostat scheme with phase space evolution operators is capable of describing thermostatting algorithms that lead to *no* corresponding differential equations in the limit of $\Delta t \to 0$ .[46, 49] The middle thermostat scheme suggests $e^{\mathcal{L}_{\mathrm{NVT}}^{\mathrm{Middle}}\Delta t} = e^{\mathcal{L}_\mathbf{p}\Delta t/2}e^{\mathcal{L}_\mathbf{x}\Delta t/2}e^{\mathcal{L}_T\Delta t}e^{\mathcal{L}_\mathbf{x}\Delta t/2}e^{\mathcal{L}_\mathbf{p}\Delta t/2}$ based on the velocity-Verlet algorithm or $e^{\mathcal{L}_{\mathrm{NVT}}^{\mathrm{Middle}}\Delta t} = e^{\mathcal{L}_\mathbf{x}\Delta t/2}e^{\mathcal{L}_T\Delta t}e^{\mathcal{L}_\mathbf{x}\Delta t/2}e^{\mathcal{L}_\mathbf{p}\Delta t}$ based on the leap-frog algorithm. The operations of the phase space propagators are in a *right-to-left* sequence throughout the paper. Conventional MD algorithms often employ the "side" scheme $e^{\mathcal{L}_{\mathrm{NVT}}^{\mathrm{Side}}\Delta t} = e^{\mathcal{L}_T\Delta t/2}e^{\mathcal{L}_\mathbf{p}\Delta t/2}e^{\mathcal{L}_\mathbf{x}\Delta t}e^{\mathcal{L}_\mathbf{p}\Delta t/2}e^{\mathcal{L}_T\Delta t/2}$ or the "end"/ "beginning" schemes[44-46] $e^{\mathcal{L}_{\mathrm{NVT}}^{\mathrm{End}}\Delta t} = e^{\mathcal{L}_T\Delta t}e^{\mathcal{L}_\mathbf{p}\Delta t/2}e^{\mathcal{L}_\mathbf{x}\Delta t}e^{\mathcal{L}_\mathbf{p}\Delta t/2}$, which produces accurate sampling of the (trivial) momentum distribution but significant errors for sampling the coordinate distribution that is often much more important. The most prominent feature of the unified middle thermostat is that both the accuracy and efficiency of sampling are insensitive to the value of the thermostat parameter(s) in a broad range for a relatively large time interval[45, 46, 49]. In comparison to the unified "middle" thermostat scheme, when conventional MD algorithms are used, the numerical error of sampling the coordinate distribution is significantly greater for a finite time interval, and the characteristic correlation time of sampling is rather sensitive to the choice of the thermostat parameter(s)[45, 46, 49].

As demonstrated in refs [44, 50], when the integrator $e^{\mathcal{L}_{\mathrm{NVT}}^{\mathrm{Middle}}\Delta t} = e^{\mathcal{L}_\mathbf{p}\Delta t/2}e^{\mathcal{L}_\mathbf{x}\Delta t/2}e^{\mathcal{L}_T\Delta t}e^{\mathcal{L}_\mathbf{x}\Delta t/2}e^{\mathcal{L}_\mathbf{p}\Delta t/2}$ based on the velocity-Verlet algorithm is used, we obtain the accurate coordinate marginal distribution as well as the accurate momentum marginal distribution in two positions: One is immediately after the first momentum-updating step for a half time interval $e^{\mathcal{L}_\mathbf{p}\Delta t/2}$ and before $e^{\mathcal{L}_\mathbf{x}\Delta t/2}e^{\mathcal{L}_T\Delta t}e^{\mathcal{L}_\mathbf{x}\Delta t/2}$ is applied, the other is immediately after $e^{\mathcal{L}_\mathbf{x}\Delta t/2}e^{\mathcal{L}_T\Delta t}e^{\mathcal{L}_\mathbf{x}\Delta t/2}$ and before the second momentum-updating step for a half time interval $e^{\mathcal{L}_\mathbf{p}\Delta t/2}$. The two positions are demonstrated in Figure 1. Specifically, in the harmonic limit, both the coordinate marginal distribution and the momentum marginal distribution are *exact* in either of the two positions in Figure 1. Similarly, when the integrator $e^{\mathcal{L}_{\mathrm{NVT}}^{\mathrm{Middle}}\Delta t} = e^{\mathcal{L}_\mathbf{x}\Delta t/2}e^{\mathcal{L}_T\Delta t}e^{\mathcal{L}_\mathbf{x}\Delta t/2}e^{\mathcal{L}_\mathbf{p}\Delta t}$ based on the leap-frog algorithm is employed, we achieve the accurate coordinate marginal distribution as well as the accurate momentum marginal distribution in two positions: One is immediately after the momentum-updating step $e^{\mathcal{L}_p \Delta t}$ and before $e^{\mathcal{L}_x \Delta t/2} e^{\mathcal{L}_T \Delta t} e^{\mathcal{L}_x \Delta t/2}$ is applied, and the other is immediately after $e^{\mathcal{L}_x \Delta t/2} e^{\mathcal{L}_T \Delta t} e^{\mathcal{L}_x \Delta t/2}$.

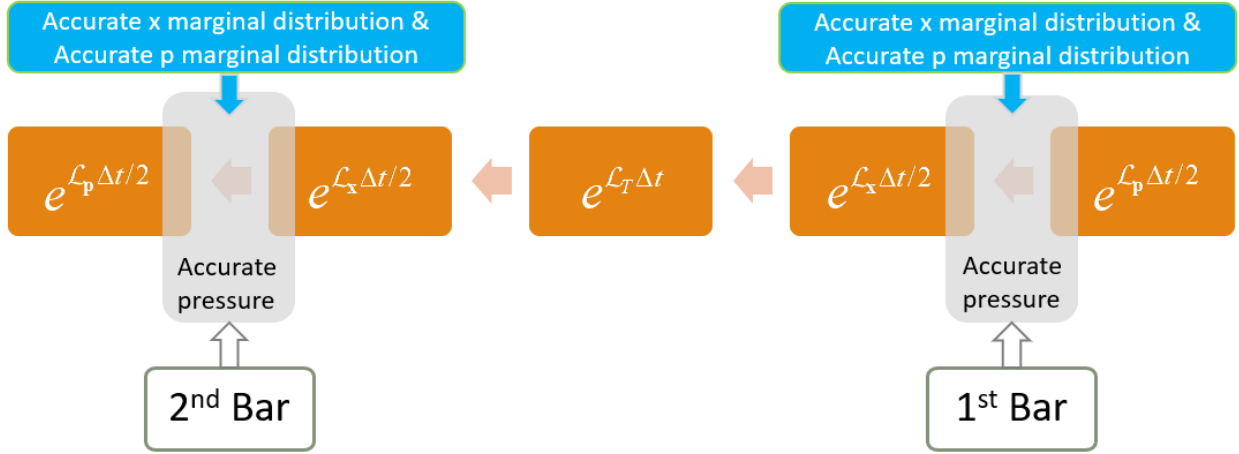

**Figure 1.** Schematic representation of the "middle" scheme for the isobaric-isothermal ensemble. When both the instantaneous kinetic energy and the virial term in the evaluation of the internal pressure are required, the barostat step should be applied in either of the two positions listed in the right-to-left sequence. When the evaluation of the force or potential is NOT required in the barostat step in the first position, the barostat step can also be applied in both positions to achieve a symmetric decomposition $e^{\mathcal{L}_{\text{Bar}(2)} \Delta t/2} e^{\mathcal{L}_x \Delta t/2} e^{\mathcal{L}_T \Delta t} e^{\mathcal{L}_x \Delta t/2} e^{\mathcal{L}_{\text{Bar}(1)} \Delta t/2}$ (for higher accuracy) before and after the momentum-updating step(s).

It is anticipated that the efficient MD integrator for a finite time interval $\Delta t$ for the isobaric-isothermal ensemble should approach the "middle" thermostat scheme for the canonical ensemble, when the barostat effectively vanishes. Actually, various integrators for the isobaric-isothermal ensemble can be designed to reach the "middle" thermostat scheme in the canonical-ensemble limit. Because the barostat step updates the volume and then the potential or force, it should be applied when the coordinate (configuration) marginal distribution is accurate. That is, the phase space evolution operator for the barostat step should be applied before or after $e^{\mathcal{L}_x \Delta t/2} e^{\mathcal{L}_T \Delta t} e^{\mathcal{L}_x \Delta t/2}$.

When the instantaneous kinetic energy is involved in the evaluation of the internal pressure in eq (4), the barostat step ought to be applied when the momentum marginal distribution is also accurate. This suggests that the barostat step should be designed in the first or second position of the right-to-left sequence as listed in Figure 1. The update of the force or/and the potential is only required in the second position for the barostat step before the momentum-updating step. If the evaluation of the force or potential is not requested in the barostat step in the first position, then the barostat step can also be applied in the first position such that a symmetric decomposition $e^{\mathcal{L}_{\text{Bar}(2)}^{\text{MD}} \Delta t/2} e^{\mathcal{L}_x \Delta t/2} e^{\mathcal{L}_T \Delta t} e^{\mathcal{L}_x \Delta t/2} e^{\mathcal{L}_{\text{Bar}(1)}^{\text{MD}} \Delta t/2}$ is achieved for higher accuracy before and after the momentum-updating step(s).

## 2.4A  Integrators with the Monte Carlo Barostat

When MC is used for the barostat[6, 9, 10], the evaluation of the potential is always necessary in each move trial of the Metropolis algorithm for sampling the volume distribution. We then recommend that, either the "middle" scheme of

$$e^{\mathcal{L}_{\text{NPT}}^{\text{Middle}} \Delta t} = e^{\mathcal{L}_p \Delta t/2} e^{\mathcal{L}_{\text{Bar}}^{\text{MC}} \Delta t} e^{\mathcal{L}_x \Delta t/2} e^{\mathcal{L}_T \Delta t} e^{\mathcal{L}_x \Delta t/2} e^{\mathcal{L}_p \Delta t/2} \qquad (20)$$

based on the velocity-Verlet algorithm or of

$$e^{\mathcal{L}_{\text{NPT}}^{\text{Middle}} \Delta t} = e^{\mathcal{L}_{\text{Bar}}^{\text{MC}} \Delta t} e^{\mathcal{L}_x \Delta t/2} e^{\mathcal{L}_T \Delta t} e^{\mathcal{L}_x \Delta t/2} e^{\mathcal{L}_p \Delta t} \qquad (21)$$

based on the leap-frog algorithm, should be used to sample the isobaric-isothermal ensemble when the MC barostat is used. Because the MC barostat does not employ the virial expression of the internal pressure (e.g., eq (4)), it is convenient to employ the MC barostat for general cases, e.g., even for *ab initio* dynamics simulations of condensed phase molecular systems, where periodic boundary conditions are applied.

When it is straightforward to evaluate the virial expression for the internal pressure, MD barostats are often used as they readily approach the NPT equilibrium, even when the initial density is far from the density of the isobaric-isothermal ensemble. In the following we focus on MD barostats.

### 2.4B  Integrators with Molecular Dynamics Barostats

#### 2.4B-1  Martyna-Tuckerman-Tobias-Klein Barostat

When the MD barostat is used for the volume sampling under a constant external pressure condition, "extended" phase space variables $(V, p_V)$ for the barostat and coordinate-momentum phase space variables $(\mathbf{x}, \mathbf{p})$ of the molecular system often propagate simultaneously. Equation (18) (or eq (19)) indicates that the "force" for the update of $\mathbf{p}$ depends on both the volume $V$ and the coordinate $\mathbf{x}$ during the propagation of the MD trajectory. As discussed in Section S2 of the Supporting Information, the "middle" scheme of

$$e^{\mathcal{L}_{\text{NPT}}^{\text{Middle}} \Delta t} = e^{\mathcal{L}_{\mathbf{p}} \Delta t/2} e^{\mathcal{L}_{\text{Bar}(2)}^{\text{MD}} \Delta t/2} e^{\mathcal{L}_{\mathbf{x}} \Delta t/2} e^{\mathcal{L}_T \Delta t} e^{\mathcal{L}_{\mathbf{x}} \Delta t/2} e^{\mathcal{L}_{\text{Bar}(1)}^{\text{MD}} \Delta t/2} e^{\mathcal{L}_{\mathbf{p}} \Delta t/2} \tag{22}$$

based on the velocity-Verlet algorithm or of

$$e^{\mathcal{L}_{\text{NPT}}^{\text{Middle}} \Delta t} = e^{\mathcal{L}_{\text{Bar}(2)}^{\text{MD}} \Delta t/2} e^{\mathcal{L}_{\mathbf{x}} \Delta t/2} e^{\mathcal{L}_T \Delta t} e^{\mathcal{L}_{\mathbf{x}} \Delta t/2} e^{\mathcal{L}_{\text{Bar}(1)}^{\text{MD}} \Delta t/2} e^{\mathcal{L}_{\mathbf{p}} \Delta t} \tag{23}$$

based on the leap-frog algorithm is recommended for the isobaric-isothermal ensemble. The force is updated only in the step $e^{\mathcal{L}_{\text{Bar}(2)}^{\text{MD}} \Delta t/2}$ during each whole integration for the finite time interval $\Delta t$.

We use the Martyna-Tuckerman-Tobias-Klein (MTTK) barostat[7, 22-26, 51] as an example. In this case, the first and second barostat steps of the right-to-left sequence in eq (22) or eq (23) are

$$e^{\mathcal{L}_{\text{Bar}(1)}^{\text{MD}} \Delta t} = e^{\mathcal{L}_B \Delta t} e^{\mathcal{L}_{\mathbf{x}_r} \Delta t} e^{\mathcal{L}_V \Delta t} e^{\mathcal{L}_{p_\varepsilon} \Delta t} e^{\mathcal{L}_{\mathbf{p}_r} \Delta t} \quad \text{and} \quad e^{\mathcal{L}_{\text{Bar}(2)}^{\text{MD}} \Delta t} = e^{\mathcal{L}_{\mathbf{p}_r} \Delta t} e^{\mathcal{L}_{p_\varepsilon} \Delta t} e^{\mathcal{L}_V \Delta t} e^{\mathcal{L}_{\mathbf{x}_r} \Delta t} e^{\mathcal{L}_B \Delta t} . \quad \text{Apparently,} \quad e^{\mathcal{L}_{\text{Bar}(1)}^{\text{MD}} \Delta t} \quad \text{and}$$

$e^{\mathcal{L}_{\text{Bar}(2)}^{\text{MD}} \Delta t}$ are symmetric to achieve accuracy of the full integrator for the finite time interval $\Delta t$. The Kolmogorov operator for the MTTK barostat consists of five terms,

$$\mathcal{L}_{\mathbf{x}_r} \rho = -\frac{p_\varepsilon}{W} \nabla_\mathbf{x} \cdot (\mathbf{x}\rho), \tag{24}$$

$$\mathcal{L}_{\mathbf{p}_r} \rho = \left(1 + \frac{d}{N_f}\right) \frac{p_\varepsilon}{W} \nabla_\mathbf{p} \cdot (\mathbf{p}\rho), \tag{25}$$

$$\mathcal{L}_V \rho = -d \frac{p_\varepsilon}{W} \frac{\partial}{\partial V}(V\rho), \tag{26}$$

$$\mathcal{L}_{p_\varepsilon} \rho = -dV\left(P_{\text{int}} - P_{\text{ext}}\right) \frac{\partial \rho}{\partial p_\varepsilon} - \frac{d}{N_f} \mathbf{p}^T \mathbf{M}^{-1} \mathbf{p} \frac{\partial \rho}{\partial p_\varepsilon}, \tag{27}$$

$$\mathcal{L}_B \rho = \gamma_{\text{Lang}}^V \frac{\partial}{\partial p_\varepsilon}(p_\varepsilon \rho) + \frac{\gamma_{\text{Lang}}^V W}{\beta} \frac{\partial^2 \rho}{\partial p_\varepsilon^2}. \tag{28}$$

Here, $N_f$ is the total degrees of freedom of the system ($N_f \equiv d \cdot N_{\text{atom}} = 3N_{\text{atom}}$ in this paper). Moreover, the mass $W$ of the fictitious "piston" is sufficiently large to ensure the convergence of estimation of physical properties of interest. Typically, this value is set to

$$W = (N_f + 1)k_B T \tau_b^2, \tag{29}$$

where $\tau_b$ is the relaxation time of the barostat, which is often set to 1 ps as suggested in ref [7]. In eq (27) the internal pressure ($P_{\text{int}}$) estimator of the system is presented in eq (4).

In the MTTK barostat, the volume-updating step $e^{\mathcal{L}_V \Delta t}$ related to eq (26) does not request the update of the virial expression of the internal pressure (or the update of the force). As shown in Figure 2, it is required to update the force and internal pressure only *once* after the operation $e^{\mathcal{L}_V \Delta t} e^{\mathcal{L}_{\mathbf{x}_r} \Delta t} e^{\mathcal{L}_B \Delta t}$ and before the operation $e^{\mathcal{L}_{\mathbf{p}_r} \Delta t} e^{\mathcal{L}_{p_\varepsilon} \Delta t}$ in the second MTTK barostat step in the whole integration for the finite time interval $\Delta t$.

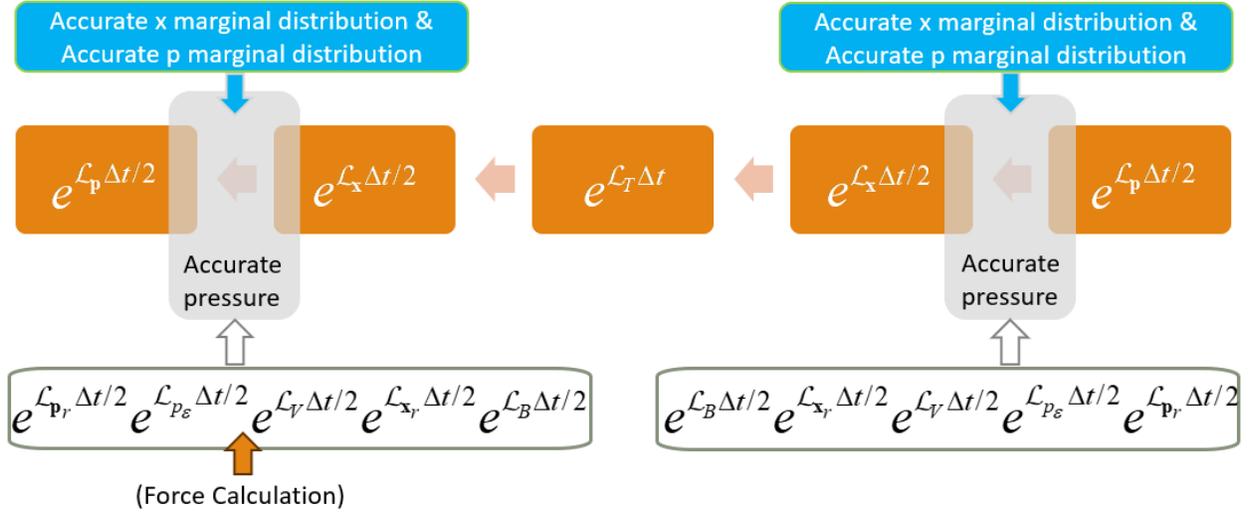

**Figure 2.** Schematic representation of the "middle" scheme with the MTTK barostat for the isobaric-isothermal ensemble. When the MTTK barostat step (for a half time interval $\Delta t/2$) is applied in the first position, the evaluation of the force or potential is NOT required. The MTTK barostat step should then be applied in the two positions listed in the right-to-left sequence. The update of the force or/and the potential is explicitly performed, before the piston-momentum-updating step of $e^{\mathcal{L}_{\text{Bar}(2)}^{\text{MD}}\Delta t/2}$, in the second position for the barostat step.

It is easy to implement the MTTK barostat to PIMD.[7] The internal pressure is an analogy of the classical form using the effective potential $U_{\text{eff}}(\mathbf{x}_1,\cdots,\mathbf{x}_L)$, and the method employing the corresponding equations of motion is called the "all-mode scaled" method. Martyna *et al.* proposed another set of equations of motion leading to the "reduced dynamics" method. Details of these two approaches are presented in Sub-Section S1-B of the Supporting Information.

In Section S2 of the Supporting Information (which is based on our earlier investigations[52,53]), we explicitly show that the "middle" scheme of eq (22) or eq (23) with the MTTK barostat is the optimally designed strategy by the analytical analysis and by numerical investigations of the one-dimensional nanowire model and liquid water. In addition, although the recommended scheme of eq (22) or eq (23) has equivalent variants as discussed in Section S2 of the Supporting Information, eq (22) or eq (23) presents a clean-cut unified form for implementing other extended-system barostatting methods that have the same appropriate properties of the MTTK barostat.

**2.4B-2   Stochastic Cell Rescaling Barostat**

In addition to extended-system barostatting methods, weak coupling methods without the fictitious "piston" momentum for the barostat can also be used in the "middle" scheme. Here, we use the SCR barostat as an example for illustration.

The update of the volume in the SCR barostat always requires the new evaluation of the virial expression of the internal pressure (or of the force), which is often the most time-consuming factor in the MD simulation. As suggested in Figure 3, the "middle" scheme with the SCR barostat can be realized with either

$$e^{\mathcal{L}_{\text{NPT}}^{\text{Middle}} \Delta t} = e^{\mathcal{L}_{\text{p}} \Delta t/2} e^{\mathcal{L}_{\text{Bar}}^{\text{MD}} \Delta t} e^{\mathcal{L}_{\text{x}} \Delta t/2} e^{\mathcal{L}_{T} \Delta t} e^{\mathcal{L}_{\text{x}} \Delta t/2} e^{\mathcal{L}_{\text{p}} \Delta t/2} \tag{30}$$

based on the velocity-Verlet algorithm or

$$e^{\mathcal{L}_{\text{NPT}}^{\text{Middle}} \Delta t} = e^{\mathcal{L}_{\text{Bar}}^{\text{MD}} \Delta t} e^{\mathcal{L}_{\text{x}} \Delta t/2} e^{\mathcal{L}_{T} \Delta t} e^{\mathcal{L}_{\text{x}} \Delta t/2} e^{\mathcal{L}_{\text{p}} \Delta t} \tag{31}$$

based on the leap-frog algorithm. The phase space evolution operator of the SCR barostat (for a finite time interval $\Delta t$) reads

$$e^{\mathcal{L}_{\text{Bar}}^{\text{MD}} \Delta t} = e^{\mathcal{L}_{\text{p}_r} \Delta t} e^{\mathcal{L}_{\text{x}_r} \Delta t} e^{\mathcal{L}_{\varepsilon} \Delta t} \quad , \tag{32}$$

where $\varepsilon$ is usually defined as $\varepsilon = \ln V / V_0$ with $V_0$ as the unit volume. In eq (32) the Kolmogorov operator, $\mathcal{L}_{\varepsilon}$, is

$$\mathcal{L}_{\varepsilon} \rho = -\partial_{\varepsilon} \left[ \frac{\kappa_T}{\tau_P} \left( P_{\text{int}} - P_{\text{ext}} \right) \right] \rho + \partial_{\varepsilon}^2 \left( \frac{\kappa_T}{\tau_P \beta V} \rho \right) \quad , \tag{33}$$

where $\tau_P$ the characteristic time of the barostat, and $\kappa_T$ the isothermal compressibility of the system. The value of $\kappa_T$ in eq (33) is typically obtained from experimental data. When the experimental value of $\kappa_T$ is not available, it can be estimated from primitive simulations or from the known value of a similar system. For instance, the value for ambient water ($4.5 \times 10^{-5}$ bar$^{-1}$) can be used for the simulation of a molecular system in solution. The corresponding phase space evolution operator for eq (33) reads

$$e^{\mathcal{L}_\varepsilon \Delta t} \rho(\mathbf{x},\mathbf{p};\varepsilon) = \rho(\mathbf{x},\mathbf{p};\varepsilon + \Delta\varepsilon), \tag{34}$$

where

$$\Delta\varepsilon \approx \frac{\kappa_T}{\tau_P}(P_{\text{int}} - P_{\text{ext}})\Delta t + \sqrt{\frac{2\kappa_T \Delta t}{\beta \tau_P V}}\eta(t) \quad, \tag{35}$$

with $\eta(t)$ being the standard Gaussian random number at a fixed time $t$ (with zero mean $\langle \eta(t) \rangle = 0$ and unit deviation $\langle (\eta(t))^2 \rangle = 1$). After the operation of $e^{\mathcal{L}_\varepsilon \Delta t}$, the other two phase space evolution operators are

$$e^{\mathcal{L}_{\mathbf{x}_r} \Delta t} \rho(\mathbf{x},\mathbf{p};\varepsilon) = \rho(\exp(\Delta\varepsilon/d)\mathbf{x},\mathbf{p};\varepsilon) \tag{36}$$

$$e^{\mathcal{L}_{\mathbf{p}_r} \Delta t} \rho(\mathbf{x},\mathbf{p};\varepsilon) = \rho(\mathbf{x},\exp(-\Delta\varepsilon/d)\mathbf{p};\varepsilon) \quad, \tag{37}$$

respectively. The time scale of operation $e^{\mathcal{L}_{\mathbf{x}_r} \Delta t}$ is same as that of operation $e^{\mathcal{L}_\varepsilon \Delta t}$, although they are well-separable from the time scale of the atomic motion (where operations $e^{\mathcal{L}_{\mathbf{x}} \Delta t}$ and $e^{\mathcal{L}_{\mathbf{p}} \Delta t}$ are applied). The volume-updating step $e^{\mathcal{L}_\varepsilon \Delta t}$ of the SCR barostat (related to eq (33)) always directly involves the internal pressure ($P_{\text{int}}$) and the force. Under such a circumstance, the SCR barostat should be applied only once in the whole integration for the finite time interval $\Delta t$, as suggested by Figure 3. More discussion and numerical results are available in Section S3 of the Supporting Information.

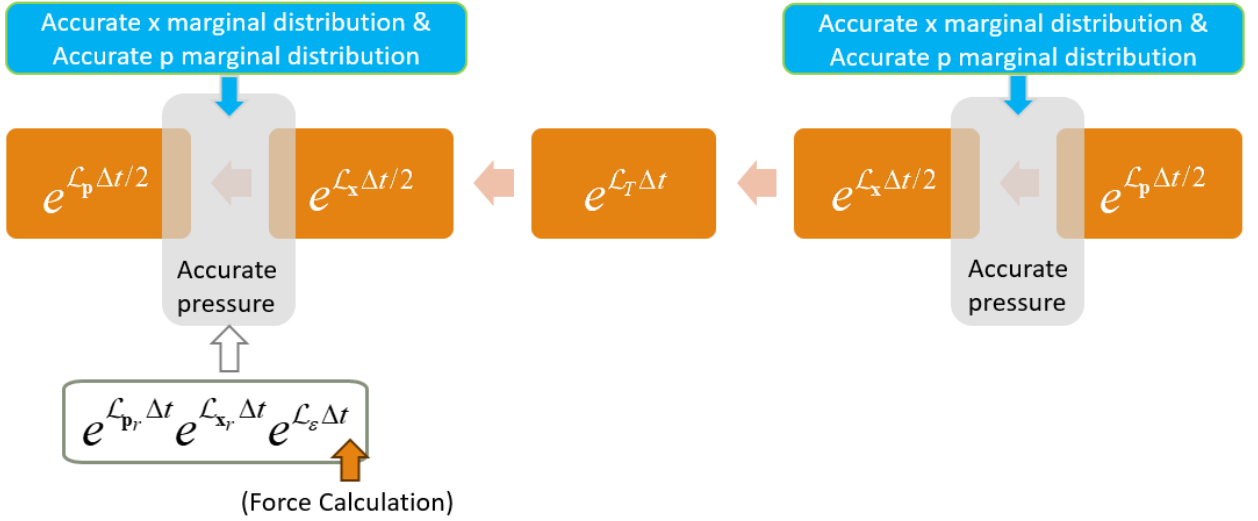

**Figure 3.** Schematic representation of the "middle" scheme with the SCR barostat for the isobaric-isothermal ensemble. The barostat step should be applied in only the second position in the right-to-left sequence. The update of the force or/and the potential is explicitly performed, before the volume-updating step $e^{\mathcal{L}_\varepsilon \Delta t}$ of $e^{\mathcal{L}_{\mathrm{Bar}}^{\mathrm{MD}} \Delta t}$, in the second position for the barostat step.

We apply the proposed MD/PIMD integrators of the "middle" scheme (for the isobaric-isothermal ensemble) to different benchmark molecular systems to test their accuracy and efficiency. Section S5 of the Supporting Information presents the explicit numerical algorithms of the "middle" scheme with the MTTK or SCR barostat for the MD/PIMD simulation of the isobaric-isothermal ensemble.

When the MTTK barostat is used, the first conventional scheme employed for comparison reads

$$e^{\mathcal{L}_{\mathrm{NPT}}^{\mathrm{Side-1}} \Delta t} = e^{\mathcal{L}_B \Delta t/2} e^{\mathcal{L}_T \Delta t/2} e^{\mathcal{L}_{p_\varepsilon} \Delta t/2} e^{\mathcal{L}_\mathbf{p} \Delta t/2} e^{\mathcal{L}_V \Delta t} e^{\mathcal{L}_\mathbf{x} \Delta t} e^{\mathcal{L}_\mathbf{p} \Delta t/2} e^{\mathcal{L}_{p_\varepsilon} \Delta t/2} e^{\mathcal{L}_T \Delta t/2} e^{\mathcal{L}_B \Delta t/2}. \qquad (38)$$

The second one is

$$e^{\mathcal{L}_{\mathrm{NPT}}^{\mathrm{Side-2}} \Delta t} = e^{\mathcal{L}_B \Delta t/2} e^{\mathcal{L}_{p_\varepsilon} \Delta t/2} e^{\mathcal{L}_\mathbf{p} \Delta t/2} e^{\mathcal{L}_T \Delta t/2} e^{\mathcal{L}_V \Delta t} e^{\mathcal{L}_\mathbf{x} \Delta t} e^{\mathcal{L}_T \Delta t/2} e^{\mathcal{L}_\mathbf{p} \Delta t/2} e^{\mathcal{L}_{p_\varepsilon} \Delta t/2} e^{\mathcal{L}_B \Delta t/2}. \qquad (39)$$

Both schemes are implemented in the open-source and commercial software, GROMACS[54], and are referred to as "vv" and "vv-avek". When the SCR barostat is employed, the Euler integrator

$$e^{\mathcal{L}_{\mathrm{NPT}}^{\mathrm{Euler}} \Delta t} = e^{\mathcal{L}_T \Delta t/2} e^{\mathcal{L}_{\mathrm{Bar}}^{\mathrm{MD}} \Delta t} e^{\mathcal{L}_\mathbf{p} \Delta t/2} e^{\mathcal{L}_\mathbf{x} \Delta t} e^{\mathcal{L}_\mathbf{p} \Delta t/2} e^{\mathcal{L}_T \Delta t/2} \qquad (40)$$

described in ref [41] is also of the conventional "side" scheme. It is compared to the "middle" scheme of eq (30) with the SCR barostat.

In Section S5 of the Supporting Information, we present the details of the MD/PIMD algorithms of the recommended "middle" scheme with the MTTK/SRC barostat for sampling the isobaric-isothermal ensemble.

### 3. Numerical Examples

We first test the Lennard-Jones (LJ) fluid and liquid para-hydrogen near its triple point with the Silvera-Goldman potential[55]. These two systems only include non-electrostatic intermolecular interactions. We then consider liquid water where all intramolecular, polar/nonpolar and electrostatic interactions are included. We test the q-SPC/fw flexible water model[56], SPC/E model where both intramolecular O-H bonds and H-O-H bond angles are rigidly fixed[57], and accurate flexible and polarizable MB-pol model[58-60].

In Section S1 and Section S2 of the Supporting Information, we show that, in classical statistical mechanics, when the evaluation of the physical property includes the components that are functions of only the momentum, it is useful to take advantage of the known Maxwell momentum distribution to yield more accurate estimators. In classical MD simulations, while we employ eqs (S36) and (S44) of the Supporting Information to compute the density and isothermal compressibility, we use eqs (S49), (S51), and (S52) of the Supporting Information for the evaluation of the enthalpy, isobaric heat capacity, and thermal expansion coefficient. When PIMD is used for quantum statistical mechanics, the evaluation of any physical property depends on only the path integral bead coordinate and the volume. In PIMD simulations, we utilize eqs (S34), (S36), (S42), (S44), and (S48) of the Supporting Information as the estimators for calculating the enthalpy, density, isobaric heat capacity, isothermal compressibility, and thermal expansion coefficient.

## 3.1 Lennard-Jones Molecular Liquid

The Lennard-Jones (LJ) potential reads $u(r) = 4\varepsilon\left[\left(\frac{\sigma}{r}\right)^{12} - \left(\frac{\sigma}{r}\right)^{6}\right]$, where $r$ is the distance between two particles. The LJ potential is commonly used to describe the two-body interaction between nonpolar atoms/molecules in various force fields. For instance, the LJ parameters for liquid Argon are $\varepsilon = k_B \times 119.8\,\text{K}$, $\sigma = 340.5\,\text{pm}$, and $m = 39.962\,A_r$ (relative atomic mass). Since the LJ liquid involves only the Van der Waals intermolecular interaction, it serves as a clean benchmark system in which intramolecular and electrostatic forces play no role. We study the state point at external pressure $P = 1.706\,\varepsilon/\sigma^3$ and inverse temperature $\beta = 0.4\,\varepsilon$. The reduced unit (r.u.), where $\varepsilon = \sigma = m = 1$, is used for convenience.

Classical MD simulations are performed with 256 particles in a cubic box with periodic boundary conditions applied using the minimum image convention. The time interval $\Delta t$ ranges from $10^{-3}$ r.u. to $10^{-2}$ r.u.. After the system is equilibrated for $2.5 \times 10^5$ r.u., 64 trajectories with each propagated up to $2.5 \times 10^5$ r.u. are employed for estimating thermodynamic properties. We test the three integrators with the MTTK barostat, namely, eq (23) of the recommended "middle" scheme, and eq (38) and eq (39) of the conventional scheme.

Figure 4 demonstrates that the three integrators approach the same converged results (within statistical error bars) as the time interval is decreased, which is consistent with the expectation that all numerical integrators are in principle equivalent as the time interval becomes zero. Fully converged results are obtained at $\Delta t = 10^{-3}$ r.u.. As shown in Figure 4(a)-(b), the absolute deviation of the density is only $2\times10^{-4}$ r.u. and that of the enthalpy is only $2\times10^{-3}$ r.u. when comparing the converged value to value at $\Delta t = 10^{-2}$ r.u. for the "middle" scheme. In comparison, the same quantities for the "side" schemes are an order of magnitude larger, $1.6\times10^{-3}$ r.u. for the density and $2.7\times10^{-2}$ r.u. for the enthalpy per atom, respectively. Panels (c)-(e) of Figure 4 display the results for isobaric heat capacity per atom, isothermal compressibility, and thermal expansion coefficient. Because the estimation of these properties involves second-order derivatives of the partition function, numerical simulations are more difficult to converge. Panels (c)-(e) of Figure 4 also show that the time interval for achieving the same accuracy is increased by a factor of 4-10 by the "middle" scheme over the conventional algorithms.

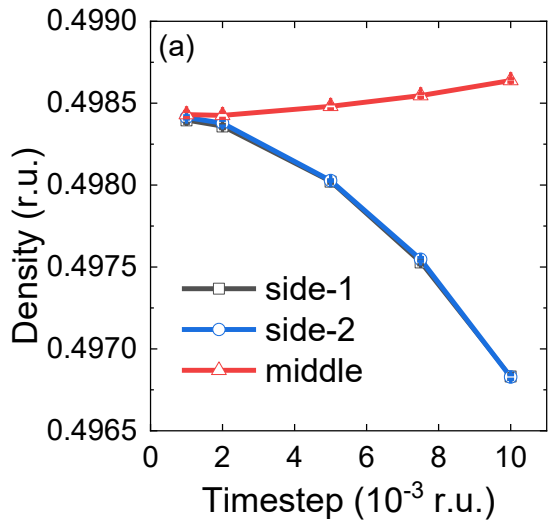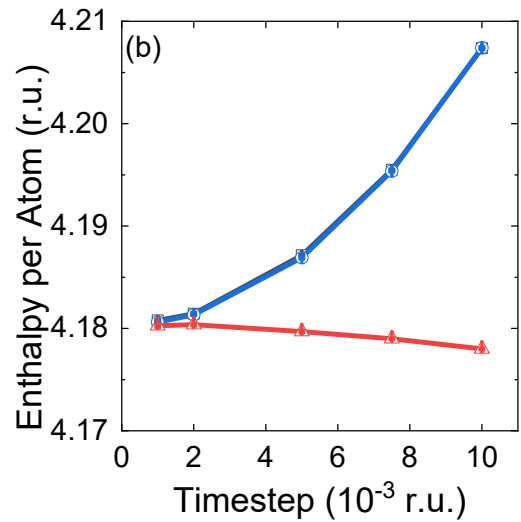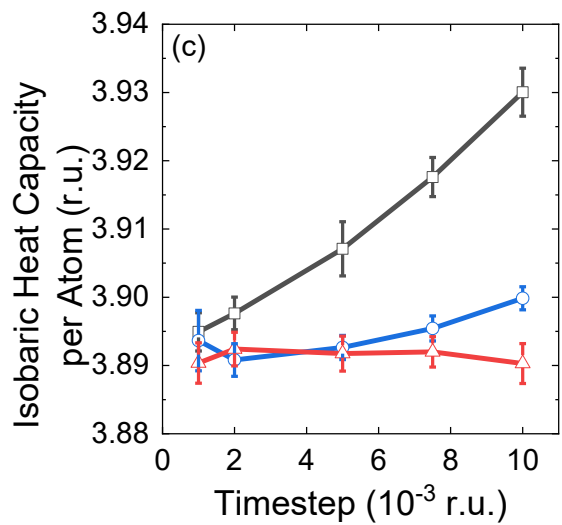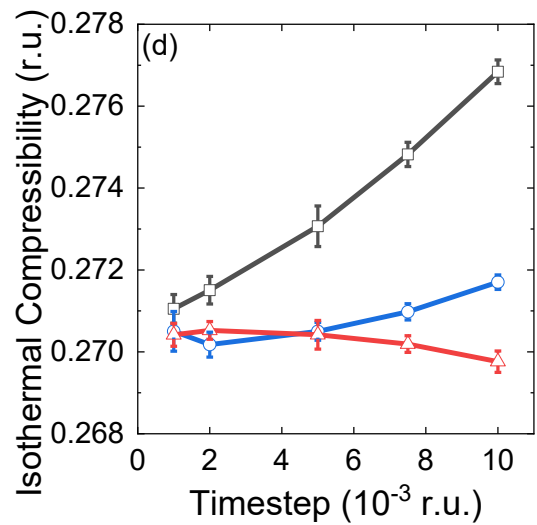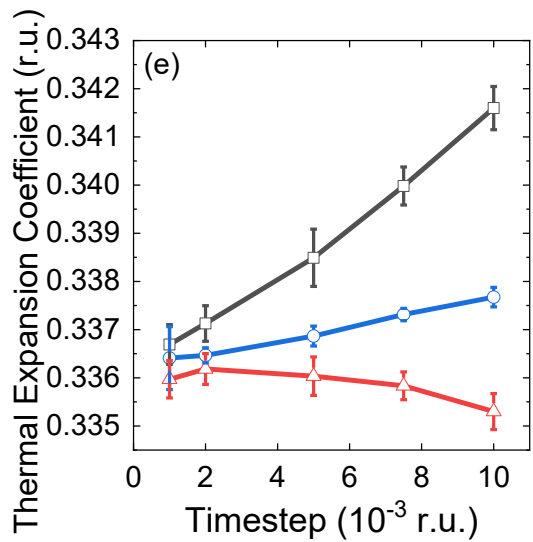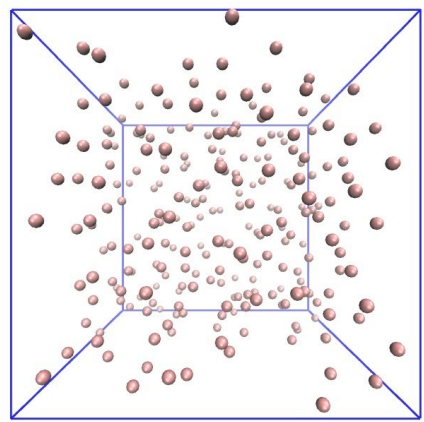

**Figure 4.** Results of NPT MD simulations for liquid argon using Lennard-Jones potential with increasing time intervals. The parameters of L-J potential are $\varepsilon = k_B \times 119.8$ K, $\sigma = 340.5$ pm and $m = 39.962\, A_r$. The state point is at $\beta = 0.4\,\varepsilon$ and $P_{ext} = 1.706\,\varepsilon/\sigma^3$. The MTTK barostat is used for the simulation. The calculated thermodynamic properties include (a) density, (b) enthalpy per atom, (c) isobaric heat capacity per atom, (d) isothermal compressibility and (e) thermal expansion coefficient. Red lines: the velocity-verlet-based "middle" scheme. Black and blue lines: two conventional "side" schemes. 64 trajectories are equilibrated for $2.5 \times 10^5$ r.u. and then propagated for $2.5 \times 10^5$ r.u., using a cubic box with periodic boundary conditions containing 256 particles. All data are divided into 16 groups to calculate the standard error. The friction parameter for the Langevin thermostat ($\gamma_{\text{Lang}}$) and that for the barostat ($\gamma_{\text{Lang}}^V$) are $5.0$ r.u. and $0.5$ r.u., respectively. The "piston" mass $W$ is set to $1000$ r.u. A smooth cutoff is chosen, i.e., the potential is truncated at $r_c = 3.0\,\sigma$ and is shifted to zero from $r_s = 2.5\,\sigma$. For liquid argon, the reduced unit of time is $1.000 \times 10^{-3}$ r.u. $= 1.000 \times 10^{-3}\,\sqrt{m\sigma^2/\varepsilon} = 2.1564$ fs. All MD simulations are performed by our independently developed MD/PIMD program. The data are verified by a modified DL_POLY_2 package[61].

## 3.2 Liquid *para*-Hydrogen

Consider liquid *para*-hydrogen at low temperature, a benchmark system that includes only nonpolar intermolecular interactions, but significant nuclear quantum effects. Because its rotational state is at the ground energy level $J = 0$, the *para*-hydrogen molecule can effectively be treated as a spherical particle. The three integrators (eq (23), eq (38), and eq (39)) are tested for the PIMD simulation with 125 *para*-hydrogen molecules in a cubic box with periodic boundary conditions applied using the minimum image convention. In the PIMD simulation, 72 beads are used for the state point at zero external pressure and constant temperature $T = 14$ K, which is nearly the critical point of the system. After the system approaches equilibrium, 64 PIMD trajectories with each propagated up to $\sim 40$ ns are employed for estimation of physical properties of the isobaric-isothermal ensemble. Both the "all-mode scaled" method and "reduced dynamics" method are used for the MTTK barostat for the PIMD simulation of this system.

Figure 5 shows that the "middle" scheme (eq (23)) is much more accurate and robust than the two conventional algorithms of the "side" scheme (eq (38) and eq (39)). The "side" scheme fails at the time interval $\Delta t = 9.6$ fs $(\sim 400$ a.u.$)$ and produces significant deviations from the converged results, while the "middle" scheme with $\Delta t = 9.6$ fs still leads to converged results. When the "all-mode scaled" method is used for estimation of the internal pressure of the MTTK barostat in PIMD, the absolute difference from the converged value of the molar volume is as large as $\sim 0.42$ cm$^3$ / mol for the result produced by the "side" scheme at $\Delta t = 9.6$ fs. In comparison, the absolute difference of the molar volume is only $\sim 7 \times 10^{-3}$ cm$^3$ / mol for $\Delta t = 9.6$ fs when the "middle" scheme is utilized. When the "reduced dynamics" method is employed to estimate the internal pressure of the MTTK barostat in PIMD, the two conventional algorithms of the "side" scheme also lead to considerable deviation—the absolute difference from the converged value of the molar volume is as large as $\sim 0.40$ cm$^3$ / mol at $\Delta t = 9.6$ fs, while the same quantity produced by the "middle" scheme is only $\sim 6 \times 10^{-3}$ cm$^3$ / mol for $\Delta t = 9.6$ fs. The "middle" scheme also significantly outperforms the conventional "side" schemes for estimation of other properties, e.g., enthalpy per molecule, isothermal compressibility, thermal expansion coefficient, and so forth. Since the state point is near the critical point of the system, it is rather demanding to converge the value for the molar isobaric heat capacity. It is expected that the performance trend is similar when fully converged results (for this property) are obtained for each time interval.

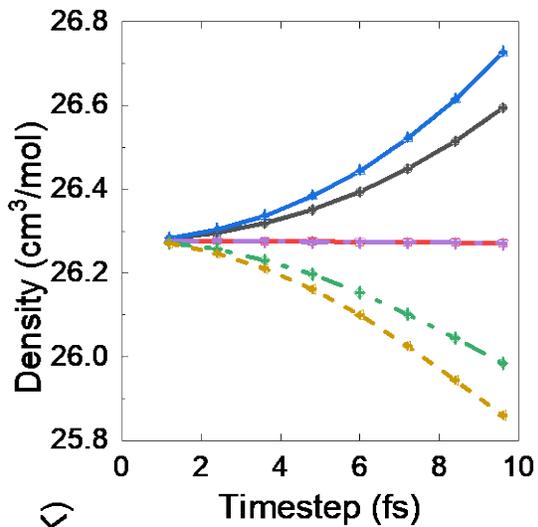
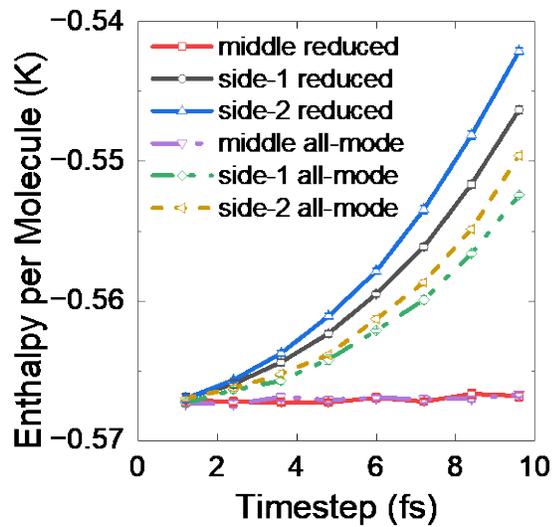
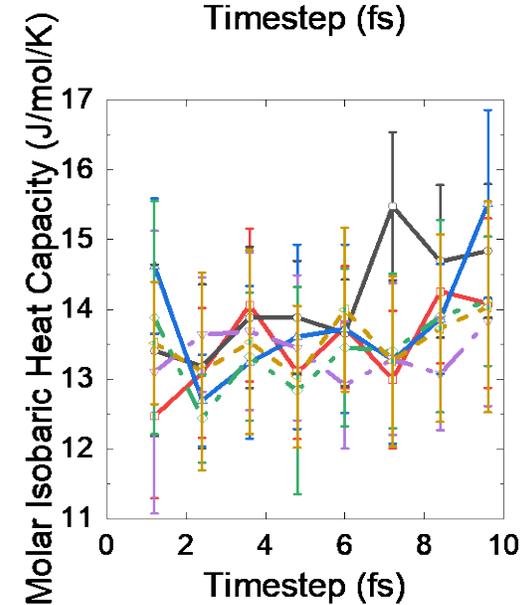
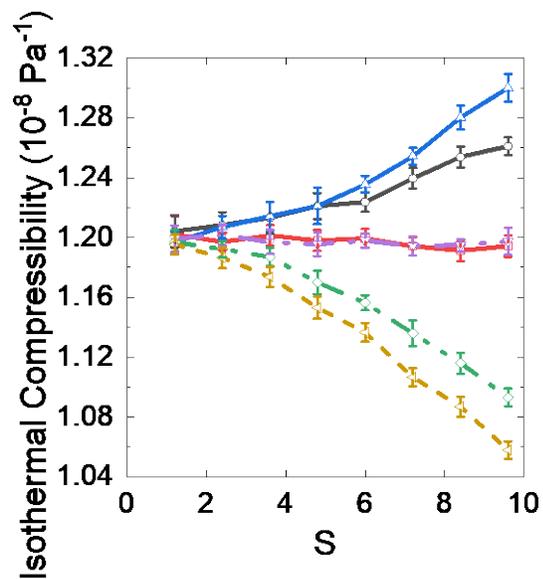
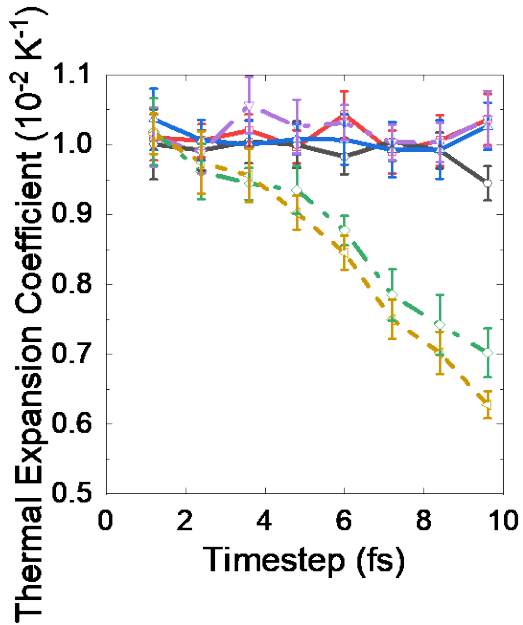
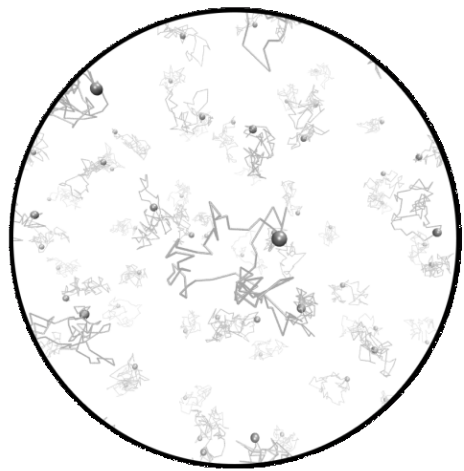

**Figure 5.** Results of NPT PIMD simulations for liquid para hydrogen at $T = 14 \text{ K}$ and $P_{\text{ext}} = 0$ using the Silvera-Goldman model[55] with increasing time intervals. The MTTK barostat is used for the simulation. The calculated thermodynamic properties include (a) density, (b) enthalpy per molecule using the virial form kinetic energy, (c) molar isobaric heat capacity, (d) isothermal compressibility and (e) thermal expansion coefficient. Red lines: the velocity-verlet-based "middle" scheme for the "reduced dynamics" method. Black and blue lines: conventional "side" schemes for the "reduced dynamics" method. Purple dashed lines: the velocity-verlet-based "middle" scheme for the "all-mode scaled" method. Green and yellow dashed lines: "side" schemes for the "all-mode scaled" method. 72 beads are employed in the PIMD simulation. 64 trajectories are equilibrated for ~ 40 ns and propagated for ~ 40 ns using a cubic box with periodic boundary conditions containing 125 particles. All data are divided into 16 groups to calculate the standard error. The friction parameter for the Langevin thermostat ($\gamma_{\text{Lang}}$) is $1 \times 10^{-3}$ a.u.$^{-1}$ for the first staging mode and $\sqrt{L}/\beta\hbar$ for other staging modes of PIMD as suggested in ref [36]. The friction parameter for the barostat ($\gamma_{\text{Lang}}^V$) is $1 \times 10^{-5}$ a.u.$^{-1}$. The "piston" mass $W$ is set to $1 \times 10^{10}$ a.u. The cutoff value for the Silvera-Goldman model is 8.466 Å. The atomic unit of time is $1 \text{ a.u.} = 0.024189 \text{ fs}$. All PIMD simulations are performed by our independently developed MD/PIMD program. The data are verified by a modified DL_POLY_2 package[61].

### 3.3 Liquid Water with the Flexible q-SPC/fw Model

We then consider liquid water at room temperature, where intramolecular, electrostatic, and Van der Waals inter-molecular interactions are involved. We use the flexible q-SPC/fw model for the demonstration. MD simulations are performed at constant temperature $T = 298.15 \text{ K}$ and constant external pressure $P = 1 \text{ atm}$ for a system of 216 molecules in a cubic box with periodic boundary conditions applied using the minimum image convention.

After the system reaches the equilibrium, 20 MD trajectories, each evolving up to 10 ns, are employed for evaluating thermodynamic properties. The time interval size $\Delta t$ ranges from 0.2 fs to 2.0 fs. All integrators lead to the same converged results (within the statistical error) at $\Delta t = 0.2 \text{ fs}$ for all tested properties. As shown in Figure 6(a), with a time interval of $\Delta t = 2.0 \text{ fs}$, two "side" schemes produce significant absolute deviation from the converged value of the density, which is $\sim 3.4\times 10^{-2}$ g·cm$^{-3}$. In contrast, the same quantity produced by the proposed "middle" scheme is less than $1\times 10^{-3}$ g·cm$^{-3}$. As shown in Panels (b)-(e) in Figure 6, the "middle" scheme significantly outperforms the "side" schemes for all other properties as well.

In addition to MD, we also perform PIMD for the same system. The "reduced dynamics" method (described in Sub-Section S1-B) is used for the barostat for the PIMD simulation of liquid water. Here, 24 path integral beads are employed in the simulation. Five independent PIMD trajectories are propagated for 10 ns after equilibration for 1 ns. Figure 7 demonstrates that the advantage of the "middle" scheme over the conventional "side" scheme is even more pronounced for PIMD in comparison to MD. The absolute deviation at $\Delta t = 1.5$ fs from the converged value of the density is less than $1\times 10^{-3}$ g·cm$^{-3}$ for the "middle" scheme of PIMD. In contrast, the corresponding deviation for the "side" scheme of PIMD is as large as $\sim 6.3\times 10^{-2}$ g·cm$^{-3}$, which is more than 60 times larger than the same quantity yield by the "middle" scheme of PIMD. When the enthalpy or isobaric heat capacity is considered, the absolute deviation for the "side" scheme increases significantly as the time interval increases. When the time interval $\Delta t = 1.5$ fs is used, the deviation of the molar isobaric heat capacity for the "side" scheme can reach 130-180 kcal/mol, while that for the "middle" scheme is only 0.6 kcal/mol. Figure 7 shows that the "middle" scheme always outperforms the "side" scheme for the PIMD simulation of any physicochemical properties. When we compare the PIMD results of Figure 7 to the corresponding MD results of Figure 6, it is evident that the superiority of the "middle" scheme to the conventional algorithms of the "side" scheme is even more prominent in the PIMD simulation than in the MD simulation.

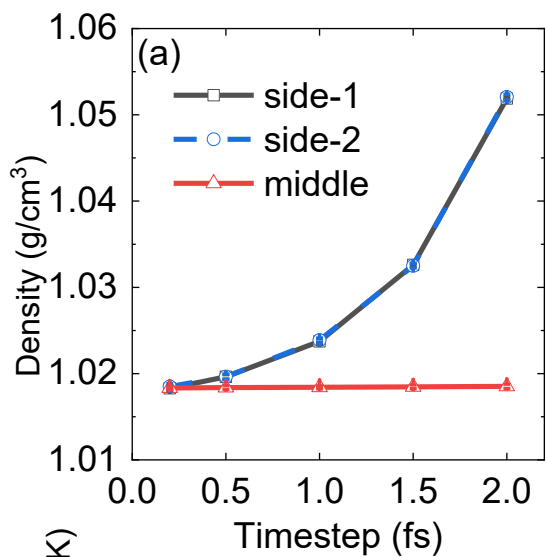
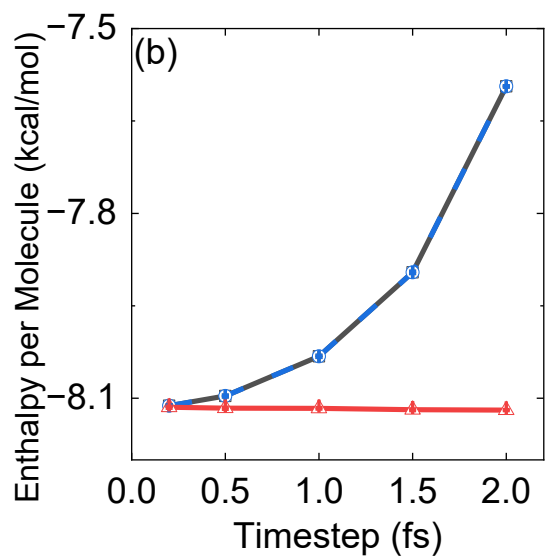
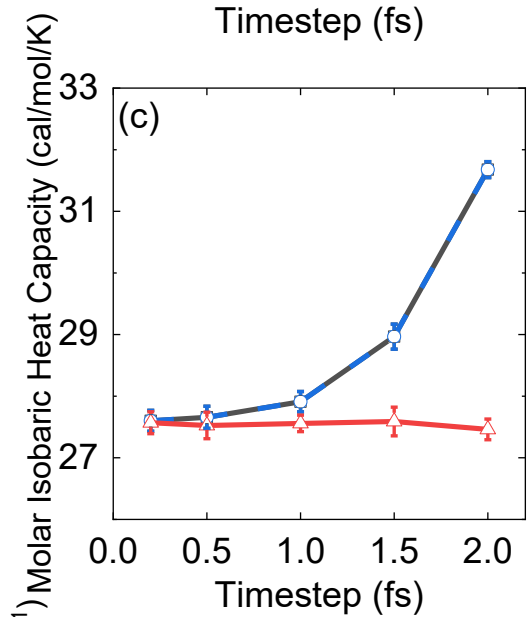
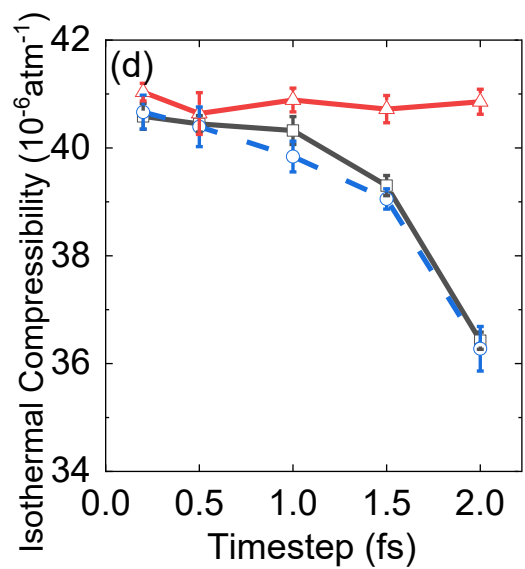
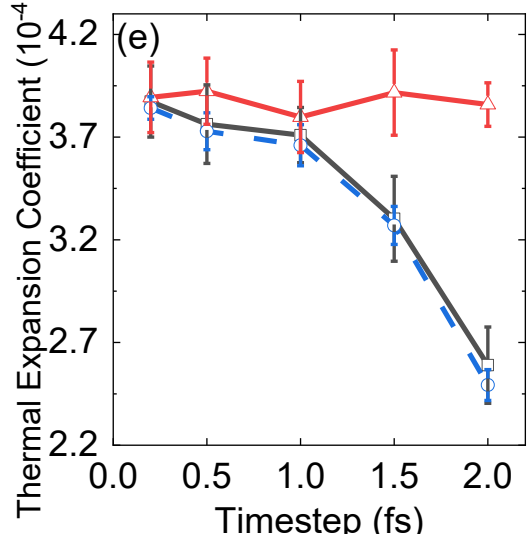
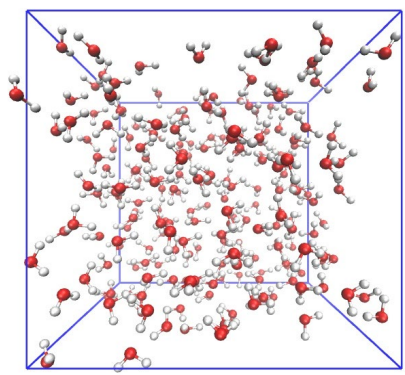

**Figure 6.** Results of NPT MD simulations for liquid water at $T = 298.15$ K and $P_{ext} = 1$ bar using q-SPC/fw model with increasing time intervals. The MTTK barostat is used for the MD simulation. The calculated thermodynamic properties include (a) density, (b) enthalpy per molecule, (c) molar isobaric heat capacity, (d) isothermal compressibility and (e) thermal expansion coefficient. Red lines: the velocity-verlet-based "middle" scheme. Black and blue lines: two conventional algorithms of the "side" scheme. 20 trajectories are equilibrated for 1 ns and propagated for 10 ns, using a cubic box with periodic boundary conditions containing 216 molecules. All data are divided into 20 groups to calculate the standard error. The friction parameter for the Langevin thermostat ($\gamma_{Lang}$) and that for the barostat ($\gamma_{Lang}^V$) are all $5.0$ ps$^{-1}$. The "piston" mass $W$ is set to the recommended value as shown in eq (29). The cutoff for the short-range interactions is 9 Å. All MD simulations are performed by our modified DL_POLY_2 package[61].

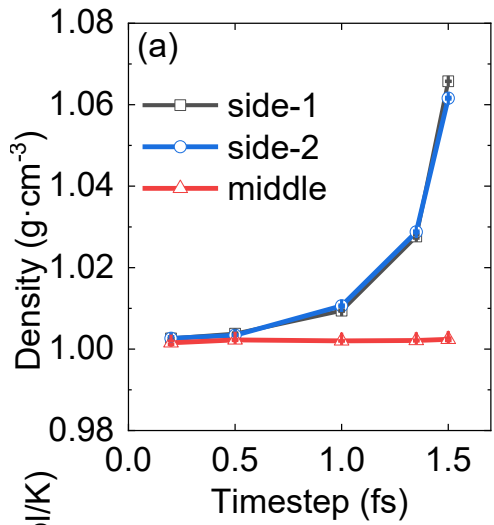
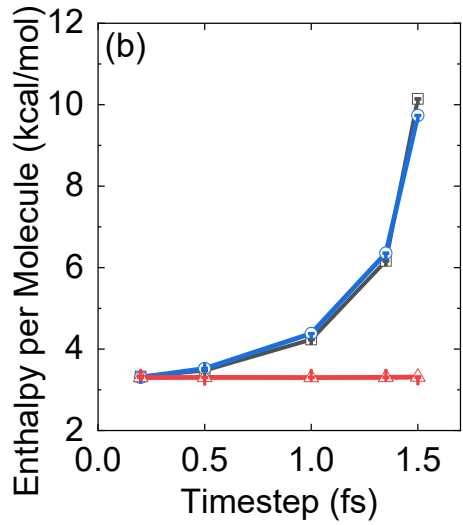
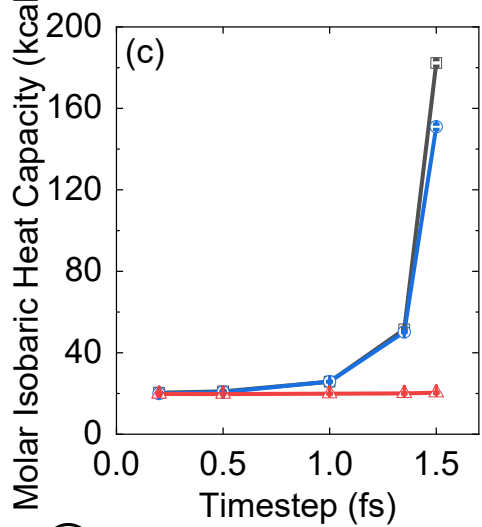
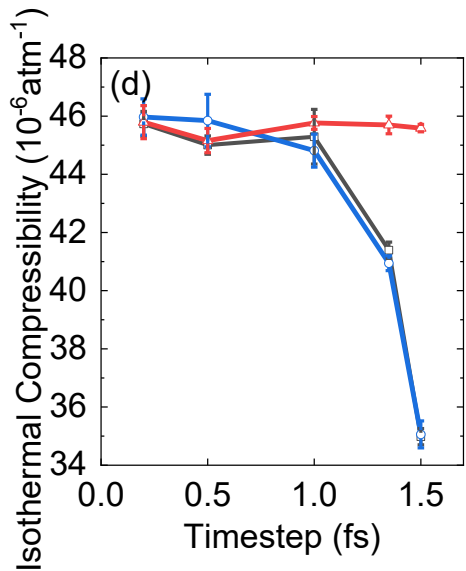
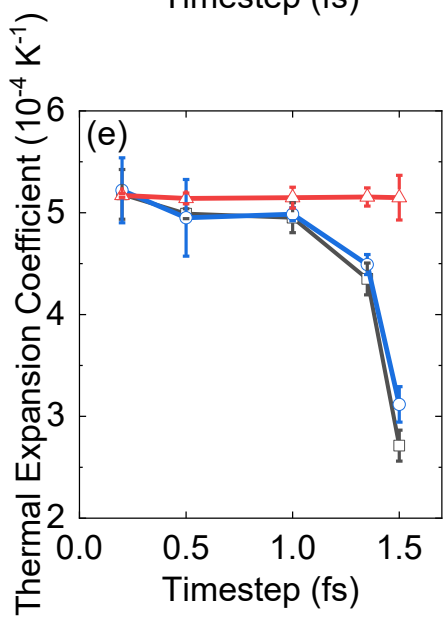
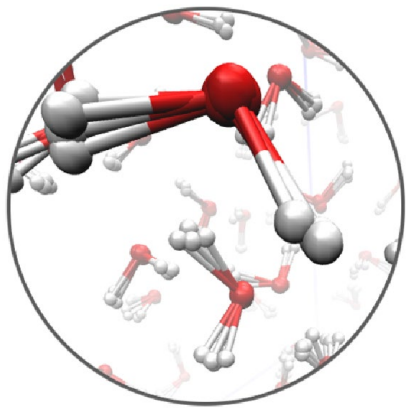

**Figure 7.** Results of NPT PIMD simulations for liquid water at $T = 298.15 \text{ K}$ and $P_{ext} = 1$ bar using q-SPC/fw model with increasing time intervals. The MTTK barostat is used for the PIMD simulation. The calculated thermodynamic properties include (a) density, (b) enthalpy per molecule using the virial form kinetic energy, (c) molar isobaric heat capacity, (d) isothermal compressibility and (e) thermal expansion coefficient. Red lines: the velocity-verlet-based "middle" scheme. Black and blue lines: conventional "side" schemes. 24 beads are employed and 5 trajectories are equilibrated for 1 ns and propagated for 10 ns using a cubic box with periodic boundary conditions containing 216 molecules. The friction parameter for the Langevin thermostat ($\gamma_{Lang}$) is $5.0 \text{ ps}^{-1}$ for the first staging mode and $\sqrt{L}/\beta\hbar$ for other staging modes of PIMD. The friction parameter for the barostat ($\gamma_{Lang}$) is $5.0 \text{ ps}^{-1}$. The "piston" mass $W$ is set to the recommended value as shown in eq (29). The cutoff for the short-range interactions is 9 Å. All PIMD simulations are performed by our modified DL_POLY_2 package[61].

### 3.4 Liquid Water Model with Holonomic Constraints

Constraints are widely used in the molecular simulation of biological systems to tackle multi-time-scale problems. By applying holonomic constraints to high-frequency modes, it is possible to significantly increase the time interval in the MD simulation. In light of our investigation on the canonical ensemble in ref [45], we propose the "middle" scheme with holonomic constraints for the isobaric-isothermal ensemble. The details of the decomposition and of the algorithm are shown in Section S4 of the Supporting Information. Here, we use the SCR barostat as an example for a liquid water model with fixed O-H bond lengths and H-O-H bond angles, where all intramolecular modes are frozen.

MD simulations with **the SPC/E water model** are performed at the state point with constant temperature $T = 298.15 \text{ K}$ and constant external pressure $P = 1 \text{ atm}$. We implement the "middle" scheme for the isobaric-isothermal ensemble in the AMBER software package[62, 63]. While a system of 324 water molecules in a cubic box with periodic boundary conditions is simulated with pmemd.MPI for CPU-based parallel computing, a system of 1,296 water molecules in a cubic box with periodic boundary conditions is investigated with pmemd.cuda_SPFP for GPU acceleration. In the MD simulations, 16 trajectories are equilibrated for 1 ns and propagated for 10 ns for estimating thermodynamic properties. All data are divided into 16 groups to calculate the standard error. The friction parameter for the Langevin thermostat ($\gamma_{\text{Lang}}$) is $5.0 \text{ ps}^{-1}$. The characteristic time $\tau_P$ for the SCR barostat is set to 2.0 ps and the isothermal compressibility $\kappa_T$ is $4.5 \times 10^{-5} \text{ bar}^{-1}$. The cutoff for the short-range interactions is 9 Å.

The results are shown in Figure 8. When the time interval $\Delta t = 6.0$ fs is used, the absolute deviation from the converged value of the density for the "middle" scheme is only $5 \times 10^{-4} \text{ g} \cdot \text{cm}^{-3}$, while the same quantity for the "side" scheme is as large as $6 \times 10^{-3} \text{ g} \cdot \text{cm}^{-3}$. Similarly, the "middle" scheme significantly outperforms the conventional "side" scheme in AMBER for calculating the enthalpy per molecule, molar isobaric heat capacity, isothermal compressibility, and thermal expansion coefficient. Figure 8 indicates that the "middle" scheme is superior to the conventional "side" scheme for simulating the isobaric-isothermal ensemble even for molecular systems with holonomic constraints.

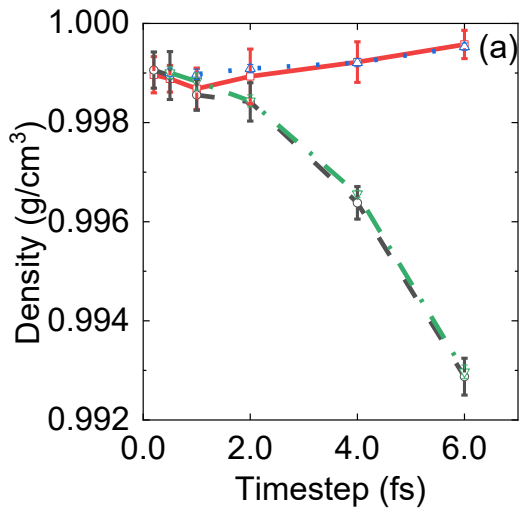
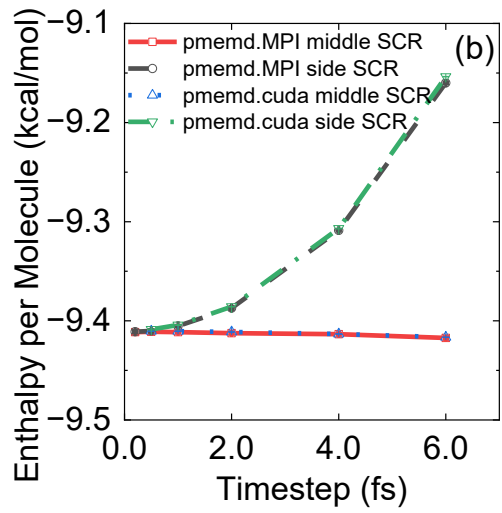
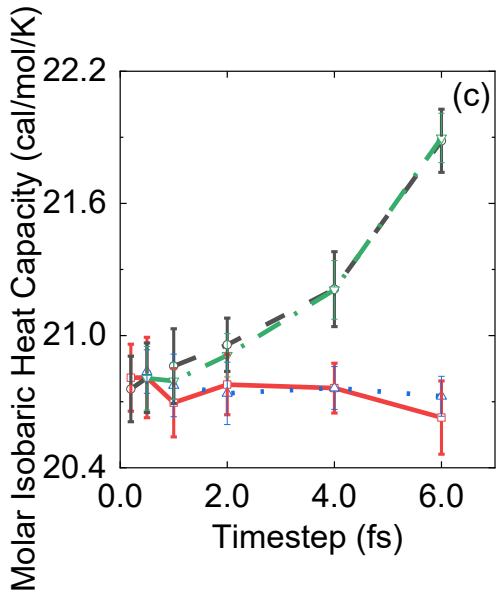
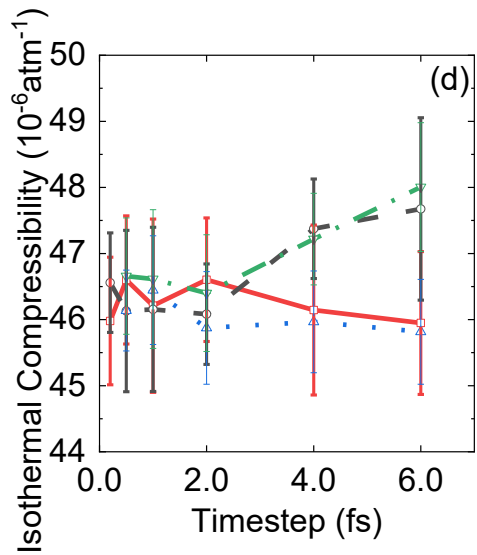
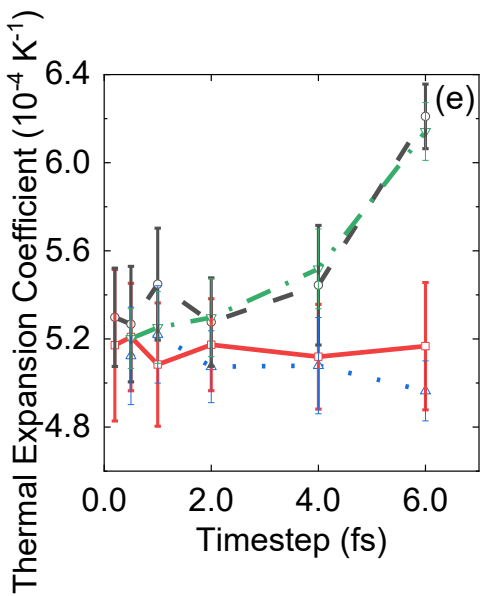
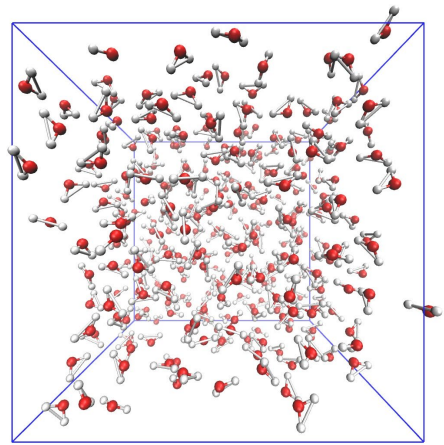

**Figure 8.** Results of liquid water with **the SPC/E model** at $T = 298.15 \text{ K}$ and $P_{\text{ext}} = 1 \text{ bar}$ with increasing time intervals. Intramolecular O-H bonds length and H-H bonds length corresponding to the fixed bond angle are constrained. The MD simulations are performed with AMBER. The SETTLE algorithm for constraining triatomic molecules is employed in the AMBER simulation[64]. The SCR barostat is implemented and used in AMBER. The calculated thermodynamic properties include (a) density, (b) enthalpy per molecule, (c) molar isobaric heat capacity, (d) isothermal compressibility and (e) thermal expansion coefficient. Red and green lines: the leap-frog-based "middle" scheme with CPU-based parallelization and with GPU-accelerated computing, respectively. Black and blue lines: the conventional "side" scheme with CPU-based parallelization and with GPU-accelerated computing, respectively. All MD simulations are performed using the modified AMBER software package[62, 63].

### 3.5    Liquid Water with the Flexible and Polarizable MB-pol Force Field

MB-pol is an accurate, flexible and polarizable water model developed by Paesani and coworkers[58-60]. Ref [65] shows that the density anomaly of water is indicated by the simulation results of liquid water at various temperatures computed by MB-pol. The MD simulations with the conventional "side" scheme reported in ref [65] was performed by the authors' in-house DL_POLY_2 software package, where a cubic box of 256 molecules with periodic boundary conditions was used and the total time length of the MD trajectory ranged from 1.9 ns to 8.0 ns with a time interval of $\Delta t = 0.2 \text{ fs}$. The Nose-Hoover chain thermostat and the MTTK barostat were applied. To obtain converged values, longer trajectories were required for simulations at lower temperatures.

Since ref [65] reported MD simulation results for the density and isobaric heat capacity produced by the conventional "side" scheme, we focus on the two thermodynamic properties for the comparison between the "middle" and "side" schemes. We perform MD simulations with MB-pol for liquid water under the condition of constant external pressure $P = 1 \text{ atm}$ for a system of 256 molecules in a cubic box with periodic boundary conditions. The Langevin thermostat and the MTTK barostat are employed. After each trajectory is propagated to 0.2 ns for approaching the equilibrium, it is further propagated for 1 ns for evaluating thermodynamic properties. Ten MD trajectories are employed for $T = 278$ K or lower temperatures, while five MD trajectories are used for temperatures above $T = 278$ K. The friction parameter for the Langevin thermostat ($\gamma_{\text{Lang}}$) and that for the barostat ($\gamma_{\text{Lang}}^V$) are $5.0 \, \text{ps}^{-1}$. The "piston" mass $W$ is set to the recommended value as described in eq (29). The cutoff for the short-range interactions is 9 Å.

Figure 9 shows that, when the "middle" scheme is used, a time interval of $\Delta t = 1.5 \, \text{fs}$ is sufficient to achieve the same converged results yielded by the conventional "side" scheme with $\Delta t = 0.2 \, \text{fs}$. This is consistent with the conclusion in Figure 6 with a simpler water force field. When $\Delta t = 1.5 \, \text{fs}$ is used with the "side" scheme instead, the results significantly deviate from the converged results. The "middle" scheme enhances the sampling efficiency for the isobaric-isothermal ensemble by a factor of 5-10 even for the accurate, but more computationally expensive MB-pol force field. It is expected that the "middle" scheme for the isobaric-isothermal ensemble will be generally useful to accelerate developing accurate (machine learning) force fields and to improve the efficiency of *ab initio* simulations for water/ice and other important chemical, materials, and biological molecular systems in solution or at interfaces/surfaces[58-60, 66-78].

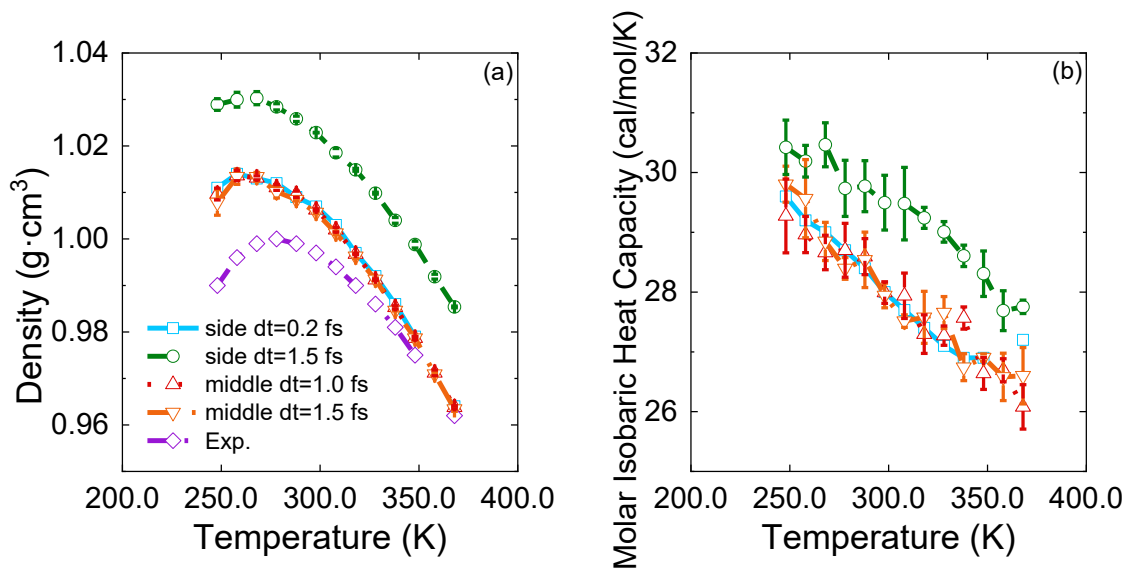

**Figure 9.** Simulation results for the temperature dependence of the density and that of the isobaric heat capacity for liquid water at $P_{\text{ext}} = 1 \text{ bar}$ using the MB-pol model. The MTTK barostat is used for the MD simulation. Red lines and orange lines: the velocity-verlet-based middle" scheme with two different time intervals. Green line: the conventional "side" scheme with the time interval of $\Delta t = 1.5 \text{ fs}$. Cyan line: data from ref [65] produced by the conventional "side" scheme with $\Delta t = 0.2 \text{ fs}$. Purple line: experiment data. All MD simulations are performed by our modified DL_POLY_2 software package[61].

## 4. Conclusion Remarks

In this paper we propose the unified "middle" scheme for MD and PIMD to sample the isobaric-isothermal (constant-NPT) ensemble, regardless of which barostatting or thermostatting method is used. As described in ref [44], a reasonable thermostat always produces the correct Maxwell momentum distribution after the operation of $e^{\mathcal{L}_T \Delta t}$. Similarly, a reasonable barostat yields the correct volume distribution after the operation of $e^{\mathcal{L}_{\text{Bar}} \Delta t}$ at least in the limit $\Delta t \to 0$. Any reasonable barostats and thermostats can be employed in the unified "middle" scheme to produce simple, robust, accurate and efficient algorithms for MD or PIMD simulations of general molecular systems under conditions of constant pressure and temperature. In the limit case that the barostat is effectively withdrew, the unified "middle" scheme for the constant-NPT ensemble is reduced to that for the constant-NVT ensemble proposed in refs [36, 44, 45, 50]. For demonstration in this work, we employ the MTTK barostat and the SCR barostat, with the Langevin thermostat, for studying isotropic systems. It is straightforward to implement the recommended "middle" scheme when other barostats (including the Parrinello-Rahman (PR) barostat[37]) or thermostats (Andersen thermostat[14], Nosé-Hoover thermostat and Nosé-Hoover Chains[79-82], and so on) are used, as well as when anisotropic systems are investigated. For instance, in Section S6 of the Supporting Information, we present the "middle" scheme with the MTTK barostat for simulating anisotropic systems under conditions of constant pressure and temperature.

As demonstrated in Section S2 of the Supporting Information, the recommended "middle" scheme leads to a more accurate distribution of the volume and coordinate (configuration), when the same finite time interval $\Delta t$ is used. The conclusion applies to general molecular systems where the Van der Waals or/and electrostatic intermolecular interactions are involved. MD or PIMD simulations of the isobaric-isothermal ensemble are performed for the LJ liquid, liquid *para*-Hydrogen, liquid water with several typical force fields (SPC/E rigid water model, flexible q-SPC/fw model, and flexible and polarizable MB-pol model). We test a few typical thermodynamic properties of the isobaric-isothermal ensemble, which include the density, enthalpy, isobaric heat capacity, isothermal compressibility, and thermal expansion coefficient. In comparison to the two conventional algorithms of the "side" scheme (as implemented in GROMACS), the "middle" scheme demonstrates superior performance and increases the time interval by a factor of 5~10 to achieve the same accuracy of converged results. Because the "middle" scheme does not cost any additional numerical effort beyond what is required by the conventional NPT MD algorithms, it is capable of improving the sampling efficiency by a factor of 5~10 for the isobaric-isothermal ensemble.

We have integrated the recommended "middle" scheme for the isobaric-isothermal ensemble in prevailing molecular simulation packages, e.g., AMBER, GROMACS, and DL-POLY. For instance, the "middle" scheme with the SCR barostat for sampling the isobaric-isothermal ensemble has been coded in "pmemd" of AMBER for both serial and parallel versions of either of CPU and GPU computing, and in "sander" of AmberTools for CPU-based serial as well as parallel computing. It will be available in the 2025 version and later versions of AMBER and AmberTools. In the previous versions of GROMACS, a bug exists in the evaluation of the instantaneous internal pressure of the implementation of the MTTK barostat, which counts twice the tail corrections from the contribution of the dispersion interactions between the atoms beyond the cutoff. Such a bug leads to the higher converged density in MD simulations. We have reported this issue to GROMACS developers (see https://gitlab.com/gromacs/gromacs/-/issues/5072 ), and have rectified the bug in the 2025 version. More importantly, we have integrated the efficient "middle" scheme with the MTTK barostat in GROMACS, which will be released soon. The recommended "middle" scheme for the isobaric-isothermal ensemble should also be implemented in CHARMM[83], LAPMMS, and other molecular simulation packages.

The "middle" scheme can be combined with other enhanced sampling techniques (e.g., thermodynamic integration[84], umbrella sampling[85], metadynamics[86], integrated tempering/Hamiltonian methods[87-90], reweighting techniques[91]). It is expected that the "middle" scheme for the isobaric-isothermal ensemble will be widely used for simulations for real complex systems, and will also help develop accurate (machine learning) force fields for important systems to understand or predict experimental processes/phenomena[58-60, 66-78].

## ASSOCIATED CONTENT

**Supporting Information**.

Supporting Information is available free of charge via the Internet at the ACS website.

Supporting Information includes six sections: **Evaluation of Thermodynamic Properties in MD and PIMD Simulations; Comparison between the "Middle" Schemes with the MTTK Barostat; Derivation of the "Middle" Scheme with the SCR Barostat; "Middle" Scheme for the Isobaric-Isothermal Ensemble with Holonomic Constraints; Numerical Algorithms for the "Middle" Scheme; "Middle" Scheme in the Isobaric-Isothermal Ensemble for Anisotropic Systems.** (PDF)

## AUTHOR INFORMATION

**Corresponding Author**

*E-mail: jianliupku@pku.edu.cn

**ORCID**

Jian Liu: 0000-0002-2906-5858

**Notes**

The authors declare no competing financial interests.

## ACKNOWLEDGMENT

We thank Xiangsong Cheng for helpful discussions. This work was supported by the National Science Fund for Distinguished Young Scholars Grant No. 22225304 and by the National Natural Science Foundation of China (NSFC) Grant No. 22450002. We acknowledge the High-

■ **SUPPORTING INFORMATION**

**S1. Evaluation of Thermodynamic Properties in MD and PIMD Simulations**

**S1-A: PIMD Simulations of the NPT Ensemble**

Path integral molecular dynamics (PIMD) is an efficient tool for sampling statistical properties that involve nuclear quantum effects, such as tunneling and zero-point energy. These effects are particularly important in systems containing light atoms or those at low temperatures. In PIMD, each atom is mapped onto a classical ring polymer, where the "beads" are connected by harmonic springs. When isotropic systems are studied, a scalar pressure is required to evolve the volume of the system. To derive the internal scalar pressure in PIMD, we begin with the partition function under the path integral formulation of the canonical ensemble

$$\begin{aligned} Z_{\mathrm{NVT}} &= \mathrm{Tr}\left[\exp\left(-\beta \hat{H}\right)\right] = \int_{D(V)} d\mathbf{x} \langle \mathbf{x} | \exp(-\beta \hat{H}) | \mathbf{x} \rangle \\ &\equiv I_{\mathrm{PI}} \lim_{L \to \infty} \int_{D(V)} d\mathbf{x}_1 \int_{D(V)} d\mathbf{x}_2 \cdots \int_{D(V)} d\mathbf{x}_L \exp\left[-\beta U_{\mathrm{eff}}(\mathbf{x}_1, \cdots, \mathbf{x}_L)\right], \end{aligned} \quad (S41)$$

where $U_{\mathrm{eff}}(\mathbf{x}_1, \cdots, \mathbf{x}_L) = \sum_{i=1}^{L} \frac{1}{2} \omega_L^2 \left[ (\mathbf{x}_{i+1} - \mathbf{x}_i)^T \mathbf{M} (\mathbf{x}_{i+1} - \mathbf{x}_i) \right] + \frac{1}{L} \sum_{i=1}^{L} U(\mathbf{x}_i)$ is the effective potential of the path integral beads, $L$ is the number of beads, $I_{\mathrm{PI}} = \left(L/2\pi\beta\hbar^2\right)^{3NL/2} |\mathbf{M}|^{L/2}$ is a normalization factor, and $\omega_L = \frac{\sqrt{L}}{\beta\hbar}$ is the frequency of the springs between beads. Thermodynamic properties are of the general form

$$\langle \hat{B} \rangle_{\text{NVT}}^{\text{QM}} = \frac{1}{Z_{\text{NVT}}} \text{Tr}\left[\exp(-\beta\hat{H})\hat{B}\right] = \frac{1}{Z_{\text{NVT}}} I_{\text{PI}} \lim_{L \to \infty} \prod_{j=1}^{L}\left(\int_{D(V)} d\mathbf{x}_j\right) \exp[-\beta U_{\text{eff}}(\mathbf{x}_1,\cdots,\mathbf{x}_L)]\tilde{B}(\mathbf{x}_1,\cdots,\mathbf{x}_L), \tag{S42}$$

where $\hat{B}$ represents the quantum mechanical operator. The estimator $\tilde{B}(\mathbf{x}_1,\cdots,\mathbf{x}_L)$ for any coordinate dependent operator is given by

$$\tilde{B}(\mathbf{x}_1,\cdots,\mathbf{x}_L) = \frac{1}{L}\sum_{i=1}^{L} B(\mathbf{x}_i). \tag{S43}$$

For kinetic energy operator $\hat{B} = \frac{1}{2}\hat{\mathbf{p}}^T \mathbf{M}^{-1}\hat{\mathbf{p}}$, the primitive form ($K_{\text{prim}}^{\text{QM}}$) is

$$\tilde{B}(\mathbf{x}_1,\cdots,\mathbf{x}_L) = K_{\text{prim}}^{\text{QM}} = \frac{N_f L}{2\beta} - \sum_{i=1}^{L} \frac{1}{2}\omega_L^2 \left[(\mathbf{x}_{i+1}-\mathbf{x}_i)^T \mathbf{M}(\mathbf{x}_{i+1}-\mathbf{x}_i)\right], \tag{S44}$$

and the virial form $K_{\text{vir}}^{\text{QM}}$ is[92]

$$\tilde{B}(\mathbf{x}_1,\cdots,\mathbf{x}_L) = K_{\text{vir}}^{\text{QM}} = \frac{N_f}{2\beta} + \frac{1}{2L}\sum_{i=1}^{L}\left[(\mathbf{x}_i - \mathbf{x}^*)^T \frac{\partial U(\mathbf{x}_i)}{\partial \mathbf{x}_i}\right], \tag{S45}$$

$\mathbf{x}^*$ can be the centroid of the ring polymer $\mathbf{x}^* = \frac{1}{L}\sum_{i=1}^{L}\mathbf{x}_i$, or any individual bead of the ring polymer $\mathbf{x}^* = \mathbf{x}_i$.

Similar to the classical case, the partition function for the isobaric-isothermal ensemble is expressed as[7]

$$Z_{\text{NPT}} = \frac{I_{\text{PI}}}{V_0} \int_0^\infty dV \exp[-\beta PV] Z_{\text{NVT}}. \tag{S46}$$

Hence, thermodynamic properties of the isobaric-isothermal ensemble are given by

$$\begin{aligned}
\langle \hat{B} \rangle_{\text{NPT}}^{\text{QM}} &= \frac{1}{V_0 Z_{\text{NPT}}} \int_0^\infty dV \exp[-\beta PV] \text{Tr}\left[e^{-\beta \hat{H}} \hat{B}\right] \\
&= \frac{1}{V_0 Z_{\text{NPT}}} I_{\text{PI}} \lim_{L \to \infty} \int_0^\infty dV \prod_{j=1}^L \left(\int_{D(V)} d\mathbf{x}_j\right) e^{-\beta[PV + U_{\text{eff}}(\mathbf{x}_1, \cdots, \mathbf{x}_L)]} \tilde{B}(\mathbf{x}_1, \cdots, \mathbf{x}_L, V).
\end{aligned} \tag{S47}$$

The staging transformation is used in this paper, in which the transformation of coordinates and forces is defined as[35, 36]

$$\begin{aligned}
\xi_1 &= \mathbf{x}_1, \\
\xi_i &= \mathbf{x}_i - \frac{(i-1)\mathbf{x}_{i+1} + \mathbf{x}_1}{i} \quad (i = 2, \cdots, L), \\
\frac{\partial \phi}{\partial \xi_1} &= \frac{1}{L} \sum_{i=1}^L U'(\mathbf{x}_i), \\
\frac{\partial \phi}{\partial \xi_i} &= \frac{\partial \phi}{\partial \mathbf{x}_i} + \frac{i-2}{i-1} \frac{\partial \phi}{\partial \xi_{i-1}} \quad (i = 2, \cdots, L),
\end{aligned} \tag{S48}$$

where $\phi$ is defined as $\phi(\xi_1, \cdots, \xi_L) = \frac{1}{L} \sum_{i=1}^L U\left[\mathbf{x}_i(\xi_1, \cdots, \xi_L)\right]$. The effective potential after staging transformation can be expressed as

$$U_{\text{eff}}(\xi_1, \cdots, \xi_L) = \sum_{i=1}^L \frac{1}{2} \omega_L^2 \xi_i^T \overline{\mathbf{M}} \xi_i + \frac{1}{L} \sum_{i=1}^L U\left[\mathbf{x}_i(\xi_1, \cdots, \xi_L)\right]. \tag{S49}$$

The corresponding mass matrix of the staging transformation is given by

$$\begin{aligned}
\overline{\mathbf{M}}_1 &= 0, \\
\overline{\mathbf{M}}_i &= \frac{i}{i-1} \mathbf{M} \quad (i = 2, \cdots, L).
\end{aligned} \tag{S50}$$

To apply MD in the path integral presentation, fictitious momenta $(\mathbf{p}_1, \cdots, \mathbf{p}_L)$ are introduced to the beads of the ring polymer, which produces the PIMD approach. The effective Hamiltonian can be given by

$$H_{\text{eff}}(\mathbf{x}_1, \cdots, \mathbf{x}_L, \mathbf{p}_1, \cdots, \mathbf{p}_L) = \sum_{i=1}^{L} \frac{1}{2} \mathbf{p}_i^T \widetilde{\mathbf{M}}_i^{-1} \mathbf{p}_i + U_{\text{eff}}(\mathbf{x}_1, \cdots, \mathbf{x}_L). \tag{S51}$$

where $\widetilde{\mathbf{M}}$ is the fictitious mass,

$$\begin{aligned}\widetilde{\mathbf{M}}_1 &= \mathbf{M}, \\ \widetilde{\mathbf{M}}_i &= \frac{i}{i-1} \mathbf{M} \quad (i = 2, \cdots, L).\end{aligned} \tag{S52}$$

**S1-B. Estimation of the Internal Pressure, and Difference between the "All-mode Scaled" Method and "Reduced Dynamics" Method for the MTTK Barostat in PIMD**

From eq (S51), the internal pressure in the path integral formation is expressed as an analogy to the classical form by using the effective potential

$$P_{\text{int}}^{\text{PI,all-mode}} = \frac{1}{dV} \sum_{i=1}^{L} \left( \mathbf{p}_i^T \widetilde{\mathbf{M}}_i^{-1} \mathbf{p}_i - \omega_L^2 \boldsymbol{\xi}_i^T \overline{\mathbf{M}}_i \boldsymbol{\xi}_i - \left( \frac{\partial \phi}{\partial \boldsymbol{\xi}_i} \right)^T \boldsymbol{\xi}_i \right). \tag{S53}$$

In comparison, the virial expression of the internal pressure in classical statistical mechanics is

$$P_{\text{int}}^{\text{CM}} = \frac{1}{dV} \left( \mathbf{p}^T \mathbf{M}^{-1} \mathbf{p} - \mathbf{x}^T \frac{\partial U}{\partial \mathbf{x}} \right). \tag{S54}$$

When PIMD uses eq (S53) as the virial expression of the internal pressure, it is often referred to as the "all-mode scaled" method.

In classical MD, the molecular virial is often used instead of the atomic virial to estimate the internal pressure due to the scale separation between intra- and intermolecular interactions. Details of the molecular virial are shown in Section S4. Martyna *et al.* proposed it for the PIMD simulation[7], because the frequency of the harmonic spring between beads is higher than the typical frequencies of molecules. Thus, the isomorphic ring polymer can be treated as a "quasi-molecule", with the harmonic potential between beads as the intramolecular potential. This approach is referred to as the "reduced dynamics" method. In this case, the internal pressure takes the form:

$$P_{\text{int}}^{\text{PI,reduced}} = \frac{1}{dV}\left(\mathbf{p}_1^T \widetilde{\mathbf{M}}_1^{-1} \mathbf{p}_1 - \boldsymbol{\xi}_1^T \frac{\partial \phi}{\partial \boldsymbol{\xi}_1}\right), \tag{S55}$$

where $\boldsymbol{\xi}_1, \mathbf{p}_1, \widetilde{\mathbf{M}}_1$ represents the coordinate, momentum and mass matrix of the first staging mode. The rescaling of the coordinate and momentum in the equations of motion, corresponding to phase space evolution operators $e^{\mathcal{L}_{x_r}\Delta t}$ and $e^{\mathcal{L}_{p_r}\Delta t}$ in the "reduced dynamics" method, is then different from that of the "all-mode scaled" method. In the "all-mode scaled" method, the coordinate and momentum variables of all path integral beads are rescaled. As comparison, in the "reduced dynamics" method, only the coordinate and momentum of the first staging mode are rescaled. The corresponding equations of motion are:

$$\begin{cases} \dot{\boldsymbol{\xi}}_i = \frac{p_\varepsilon}{W}\boldsymbol{\xi}_i \\ \dot{\mathbf{p}}_i = -\left(1 + \frac{d}{N_f L}\right)\frac{p_\varepsilon}{W}\mathbf{p}_i \end{cases} \quad (i=1,\cdots,L), \tag{S56}$$

for the "all-mode scaled" method, and

$$\begin{cases} \dot{\xi}_1 = \dfrac{p_\varepsilon}{W}\xi_1 \\ \dot{\mathbf{p}}_1 = \left(1+\dfrac{d}{N_f}\right)\dfrac{p_\varepsilon}{W}\mathbf{p}_1 \end{cases} \tag{S57}$$

for the "reduced dynamics" method.

The internal pressure expression of the "all-mode scaled" method and that of the "reduced dynamics" method can be demonstrated to be equivalent through the virial theorem

$$\frac{N_f(L-1)}{\beta} - \left\langle \sum_{i=1}^{L} \omega_L^2 \boldsymbol{\xi}_i^T \overline{\mathbf{M}}_i \boldsymbol{\xi}_i \right\rangle = \frac{1}{L}\left\langle \left(\frac{\partial U(\mathbf{x}_i)}{\partial \mathbf{x}_i}\right)^T (\mathbf{x}_i - \mathbf{x}_1) \right\rangle, \tag{S58}$$

By substituting eq (S58) and the staging transformation formula eq (11) into eq (S53), we obtain:

$$P_{\text{int}}^{\text{PI,reduced}} = \frac{1}{dV}\left[\mathbf{p}_1^T \widetilde{\mathbf{M}}_1^{-1}\mathbf{p}_1 + \frac{1}{L}\sum_{i=1}^{L}\left(\frac{\partial U(\mathbf{x}_i)}{\partial \mathbf{x}_i}\right)^T(\mathbf{x}_i-\mathbf{x}_1) - \frac{1}{L}\sum_{i=1}^{L}\left(\frac{\partial U(\mathbf{x}_i)}{\partial \mathbf{x}_i}\right)^T\mathbf{x}_i\right]. \tag{S59}$$

Eq (S59) can be obtained directly by subtracting the latter two terms without periodic boundary conditions. When periodic boundary conditions are employed, the corresponding conditions become more complex, but eq (S59) remains valid. In this paper, we use eq (S59) to estimate the internal pressure of the "reduced dynamics".

In MD/PIMD simulations, we often apply the cutoff for short-range interactions. It is essential to incorporate corrections to account for the contributions of particles beyond the cutoff to both the potential energy and pressure. For the pairwise interaction $u(r)$, the total potential energy per particle can be expressed as

$$\frac{U}{N} = 2\pi \rho_{\text{ins}} \int \mathrm{d}r\, u(r) g(r) r^2, \tag{S60}$$

where $g(r)$ is the radial distribution function (RDF). Similarly, the pressure in classical statistical mechanics can be expressed as

$$P_{\text{int}}^{\text{CM}} = \rho k_B T - \frac{2\pi\rho^2}{3}\int dr\, u'(r)g(r)\,r^3\,. \tag{S61}$$

In eq (S60) and eq (S61) the instantaneous density $\rho_{\text{ins}} = N/V$ is used.

Employ the Lennard-Jones (L-J) potential $u(r) = 4\varepsilon\left[\left(\frac{\sigma}{r}\right)^{12} - \left(\frac{\sigma}{r}\right)^{6}\right]$ as an example[7]. When a soft cutoff is applied to the L-J potential, it is defined by

$$\begin{aligned}
\tilde{u}(r) &= u(r)S(r), \\
S(r) &= \begin{cases} 1 & r < r_c - \lambda, \\ 1 + \tilde{r}^2(2\tilde{r} - 3) & r_c - \lambda \leq r \leq r_c, \\ 0 & r > r_c, \end{cases} \\
\tilde{r} &= \frac{r - r_c}{\lambda} + 1.
\end{aligned} \tag{S62}$$

where $r_c$ is the cutoff distance, and $\lambda$ is the healing length for the soft cutoff. We assume that $g(r) \approx 1$ for $r > r_c - \lambda$. Substituting eq (S62) into eq (S60) as well as into eq (S61), we obtain the correction terms

$$\begin{aligned}
\frac{\Delta U_{\text{corr}}^{\text{L-J,soft}}}{N} &= 2\pi\rho\int dr\, u(r)\left(1 - S(r)\right)g(r)r^2 \\
&= 2\pi\rho\left[\int_{r_c-\lambda}^{r_c} dr\, u(r)r^2\tilde{r}^2(3 - 2\tilde{r}) + \int_{r_c}^{\infty} dr\, u(r)r^2\right]
\end{aligned} \tag{S63}$$

$$\Delta P_{\text{corr}}^{\text{L-J,soft}} = -\frac{2\pi\rho^2}{3}\int dr\,\frac{d(u(r)S(r))}{dr}g(r)r^3. \tag{S64}$$

By neglecting the $r^{-12}$ term in the Lennard-Jones (L-J) potential and assuming $g(r)=1$ beyond the cutoff distance, the correction to the potential energy per particle and that to the pressure can be derived from eq (S63) and eq (S64)

$$\frac{\Delta U_{corr}^{L-J,soft}}{N} = -\frac{8\pi\varepsilon\sigma^6\rho}{\lambda^3}\left(\frac{1}{s_c}+\frac{1}{s_c-1}+2\ln\frac{s_c-1}{s_c}\right), \tag{S65}$$

$$\Delta P_{corr}^{L-J,soft} = \frac{\Delta U_{corr}^{L-J,soft}}{V}, \tag{S66}$$

where $s_c = r_c/\lambda$.

When a hard cutoff is applied to the L-J potential, it is defined by

$$\begin{aligned}\tilde{u}(r) &= u(r)S(r),\\ S(r) &= \begin{cases} 1 & r<r_c,\\ 0 & r>r_c.\end{cases}\end{aligned} \tag{S67}$$

The correction terms can be calculated as

$$\frac{\Delta U_{corr}^{L-J,hard}}{N} = 2\pi\rho\int_{r_c}^{\infty} dr\, u(r)r^2, \tag{S68}$$

$$\Delta P_{corr}^{L-J,hard} = 2\pi\rho^2\left[\frac{1}{3}u(r_c)r_c^3 + \int_{r_c}^{\infty} dr\, u(r)r^2\right]. \tag{S69}$$

Similarly, the correction to the potential energy and that to the pressure can be derived by neglecting the $r^{-12}$ term in the L-J potential and assuming $g(r)=1$ beyond the cutoff distance

$$\frac{\Delta U_{corr}^{L-J,hard}}{N} = 2\pi\rho \cdot 4\epsilon\left(-\frac{\sigma^6}{3r_c^3}\right) \tag{S70}$$

$$\Delta P_{corr}^{L-J,hard} = 2\pi\rho^2 \cdot 4\epsilon\left(-\frac{2\sigma^6}{3r_c^3}\right). \tag{S71}$$

Following the same procedure, we can obtain the cutoff corrections for other forms of short-range interactions in real molecular systems.

**S1-C: Estimators for Enthalpy, Density, Isobaric Heat Capacity, Isothermal Compressibility, and Thermal Expansion Coefficient.**

1) **Enthalpy**

In classical MD, the enthalpy is estimated by

$$\tilde{H} = \langle \tilde{H}_{ins} \rangle = \left\langle \frac{1}{2}\mathbf{p}^T \mathbf{M}^{-1} \mathbf{p} + U(\mathbf{x}) + P_{ext} V \right\rangle. \tag{S72}$$

In PIMD, the primitive estimator for enthalpy is

$$\tilde{H}^{QM} = \langle \tilde{H}^{QM}_{prim} \rangle = \left\langle K^{QM}_{prim} + \frac{1}{L}\sum_{i=1}^{L} U(\mathbf{x}_i) + P_{ext} V \right\rangle, \tag{S73}$$

and the virial estimator is defined as

$$\tilde{H}^{QM} = \langle \tilde{H}^{QM}_{vir} \rangle = \left\langle K^{QM}_{vir} + \frac{1}{L}\sum_{i=1}^{L} U(\mathbf{x}_i) + P_{ext} V \right\rangle, \tag{S74}$$

which satisfies the relation

$$\tilde{H}^{QM}_{prim} = \tilde{H}^{QM}_{vir} = -\frac{\partial \ln Z_{NPT}}{\partial \beta}. \tag{S75}$$

In the main text, we use the virial estimator for the enthalpy for better convergence.

2) **Density**

For the density, the estimator is

$$\rho = \frac{M_{\text{system}}}{\langle V \rangle}. \tag{S76}$$

for both classical and quantum cases. $M_{\text{system}}$ is the total mass of the system.

### 3) Isobaric Heat Capacity

The isobaric heat capacity is defined as

$$C_p = \left(\frac{\partial \tilde{H}}{\partial T}\right)_P, \tag{S77}$$

In classical mechanics, the enthalpy can be calculated throughout the partition function of eq (2) in the main text

$$\tilde{H} = -\frac{\partial \ln Z_{\text{NPT}}}{\partial \beta}. \tag{S78}$$

The isobaric heat capacity can then be calculated as

$$C_p = \frac{\partial}{\partial T}\left(-\frac{\partial \ln Z_{\text{NPT}}}{\partial \beta}\right) = \frac{1}{k_B T^2}\frac{\partial^2 \ln Z_{\text{NPT}}}{\partial \beta^2} = \frac{1}{k_B T^2}\left(\langle \tilde{H}_{\text{ins}}^2 \rangle - \langle \tilde{H}_{\text{ins}} \rangle^2\right), \tag{S79}$$

where $\langle \cdot \rangle$ denotes the ensemble average. The isobaric heat capacity is explicitly related to the variance of enthalpy $\sigma_{\tilde{H}}^2 = \langle \tilde{H}_{\text{ins}}^2 \rangle - \langle \tilde{H}_{\text{ins}} \rangle^2$. Equation (S79) can be recast into

$$C_p = \frac{\sigma_{\tilde{H}}^2}{k_B T^2} = \frac{\sigma_K^2}{k_B T^2} + \frac{\sigma_{U+P_{\text{ext}}V}^2}{k_B T^2} \tag{S80}$$

in classical statistical mechanics, where $\sigma_A^2 = \langle A^2 \rangle - \langle A \rangle^2$ is the variance of property $A$.

When PIMD or PIMC is used for quantum statistical mechanics, the classical partition function in eq (S79) should be replaced by that in the path integral formulation[7]. For the primitive estimator of the enthalpy[92] in eq (S73), the heat capacity can also be calculated as

$$C_{p,\text{prim}}^{\text{QM}} = \frac{1}{k_B T^2} \frac{\partial^2 \ln Z_{\text{NPT}}}{\partial \beta^2} = \frac{1}{k_B T^2} \left( \langle (\tilde{H}_{\text{prim}}^{\text{QM}})^2 \rangle - \langle \tilde{H}_{\text{prim}}^{\text{QM}} \rangle^2 - \frac{N_f L}{2\beta^2} + \frac{2}{\beta} \langle K_{\text{prim}}^{\text{QM}} \rangle \right). \quad \text{(S81)}$$

The last two terms at the right-hand side of eq (S81) are from the derivatives (of the component of the effective potential contributed by the kinetic energy operator) with respect to the Boltzmann factor $\beta = k_B T$.

In addition to the estimator provided in eq (S81), the virial estimator of the enthalpy, eq (S74) can produce a different expression of the isobaric heat capacity[93].

$$C_{p,\text{vir}}^{\text{QM}} = -\frac{1}{k_B T^2} \frac{\partial \langle \tilde{H}_{\text{vir}}^{\text{QM}} \rangle}{\partial \beta} = \frac{1}{k_B T^2} \left( \langle \tilde{H}_{\text{prim}}^{\text{QM}} \tilde{H}_{\text{vir}}^{\text{QM}} \rangle - \langle \tilde{H}_{\text{prim}}^{\text{QM}} \rangle \langle \tilde{H}_{\text{vir}}^{\text{QM}} \rangle + \frac{N_f}{2\beta^2} \right). \quad \text{(S82)}$$

### 4) Isothermal Compressibility

The isothermal compressibility is defined as

$$\kappa_T = -\frac{1}{V} \left( \frac{\partial V}{\partial P} \right)_T. \quad \text{(S83)}$$

In the isobaric-isothermal ensemble, the volume fluctuates and should be replaced by its ensemble average

$$\kappa_T = -\frac{1}{\langle V \rangle} \left( \frac{\partial \langle V \rangle}{\partial P_{\text{ext}}} \right)_T = \frac{\beta}{\langle V \rangle} \left( \langle V^2 \rangle - \langle V \rangle^2 \right). \quad \text{(S84)}$$

The isothermal compressibility is expressed in terms of the fluctuation of volume.

In the path integral formulation, as the volume distribution remains the same as in the classical case, the estimator of the isothermal compressibility remains unchanged.

5) **Thermal Expansion Coefficient**

The thermal expansion coefficient is defined as

$$\alpha = \frac{1}{V}\left(\frac{\partial V}{\partial T}\right)_P. \tag{S85}$$

Likewise, the volume is replaced by the expectation. The estimator in the NPT ensemble can be expressed as

$$\alpha = \frac{1}{\langle V \rangle}\left(\frac{\partial \langle V \rangle}{\partial T}\right) = \frac{1}{k_B T^2 \langle V \rangle}\left(\langle V\tilde{H}_{\text{ins}}\rangle - \langle V \rangle\langle \tilde{H}_{\text{ins}}\rangle\right). \tag{S86}$$

In the path integral formulation, the estimator reads

$$\alpha^{\text{QM}} = -\frac{1}{k_B T^2 \langle V \rangle}\frac{\partial \langle V \rangle}{\partial \beta} = \frac{1}{k_B T^2 \langle V \rangle}\left(\langle V\tilde{H}_{\text{prim}}^{\text{QM}}\rangle - \langle V \rangle\langle \tilde{H}_{\text{prim}}^{\text{QM}}\rangle\right). \tag{S87}$$

By integration by parts, it can also be expressed in terms of the virial estimator as

$$\alpha^{\text{QM}} = \frac{1}{k_B T^2 \langle V \rangle}\left(\langle V\tilde{H}_{\text{prim}}^{\text{QM}}\rangle - \langle V \rangle\langle \tilde{H}_{\text{prim}}^{\text{QM}}\rangle\right) = \frac{1}{k_B T^2 \langle V \rangle}\left(\langle V\tilde{H}_{\text{vir}}^{\text{QM}}\rangle - \langle V \rangle\langle \tilde{H}_{\text{vir}}^{\text{QM}}\rangle\right). \tag{S88}$$

6) **Explicit Analytical Evaluation of the Component that is a Function of only Momentum in Classical Statistical Mechanics**

In classical statistical mechanics, as shown by the distribution function $\exp\left[-\beta\left(PV + H(\mathbf{x},\mathbf{p})\right)\right]$ of the isobaric-isothermal ensemble of eq (2) of the main text, the momentum distribution is independent of the distribution of the coordinate and volume. The momentum distribution is simply the Maxwell distribution, a Gaussian distribution of momentum $\exp\left[-\beta \mathbf{p}^T \mathbf{M}^{-1} \mathbf{p}/2\right]$ which is always known without calculation in classical statistical mechanics. Estimation of the enthalpy can be divided into two terms,

$$\tilde{H} = \langle \tilde{H}_{ins} \rangle = \left\langle \frac{1}{2}\mathbf{p}^T \mathbf{M}^{-1} \mathbf{p} \right\rangle + \langle U(\mathbf{x}) + P_{ext}V \rangle = \frac{N_f k_B T}{2} + \langle U(\mathbf{x}) + P_{ext}V \rangle. \tag{S89}$$

Estimation of the variance of the enthalpy can be decomposed into two components. The variance of the enthalpy is equivalent to the sum of two parts as

$$\sigma_{\tilde{H}}^2 = \sigma_K^2 + \sigma_{U+P_{ext}V}^2 = \frac{N_f (k_B T)^2}{2} + \sigma_{U+P_{ext}V}^2. \tag{S90}$$

Moreover, in classical statistical mechanics the variance of the kinetic energy $\sigma_K^2$ only depends on the Maxwell momentum distribution, which yields $\sigma_K^2 = N_f (k_B T)^2/2$. The contribution of $\sigma_K^2$ to the isobaric heat capacity, the first term of the right-hand side of eq (S80), is $N_f k_B/2$. Equation (S80) then becomes

$$C_p = \frac{N_f k_B}{2} + \frac{\sigma_{U+P_{ext}V}^2}{k_B T^2} \tag{S91}$$

for the evaluation of the isobaric heat capacity.

Because eq (2) of the main text indicates that the momentum distribution and the volume distribution are independent in the isobaric-isothermal ensemble, the covariance of the kinetic energy and volume is zero. Provided that the Maxwell momentum distribution is known in classical statistical mechanics, eq (S86) for the estimator for the thermal expansion coefficient can be recast into

$$\alpha = \frac{1}{\langle V \rangle}\left(\frac{\partial \langle V \rangle}{\partial T}\right) = \frac{1}{k_B T^2 \langle V \rangle}\left[\langle V(U(\mathbf{x}) + P_{ext}V) \rangle - \langle V \rangle \langle U(\mathbf{x}) + P_{ext}V \rangle\right]. \tag{S92}$$

Such a strategy is not necessary in PIMD or PIMC of quantum statistical mechanics, because all physical properties are functions of only the coordinate of the PI bead and the volume. The momentum of the PI bead is simply fictitious, of which the distribution is not important at all in the evaluation of thermodynamic properties.

**S2. Comparison between the "Middle" Schemes with the MTTK Barostat**

In the isotropic isobaric-isothermal ensemble, the equilibrium distribution is described by

$$\rho(\mathbf{x},\mathbf{p},V) = \frac{1}{Z_{\text{NPT}}} \exp\left[-\beta\left(\frac{1}{2}\mathbf{p}^T \mathbf{M}^{-1}\mathbf{p} + U(\mathbf{x}) + P_{\text{ext}}V\right)\right] \ . \tag{S93}$$

When using the MTTK barostat (coupled with the Langevin thermostat) to sample the isobaric-isothermal ensemble, the momentum of fictitious 'piston' is introduced, and the full equations of motion are:

$$\begin{aligned}
\dot{\mathbf{x}} &= \mathbf{M}^{-1}\mathbf{p} + \frac{p_\epsilon}{W}\mathbf{x} \\
\dot{\mathbf{p}} &= -\frac{\partial U}{\partial \mathbf{x}} - \left(1 + \frac{d}{N_f}\right)\frac{p_\epsilon}{W}\mathbf{p} - \gamma_{\text{Lang}}\mathbf{p} + \sqrt{\frac{2\gamma_{\text{Lang}}}{\beta}}\mathbf{M}^{1/2}\boldsymbol{\zeta}(t) \\
\dot{V} &= dV\frac{p_\varepsilon}{W} \\
\dot{p}_\varepsilon &= dV(P_{\text{int}} - P_{\text{ext}}) + \frac{d}{N_f}\mathbf{p}^T\mathbf{M}^{-1}\mathbf{p} - \gamma_{\text{Lang}}^V p_\varepsilon + \sqrt{\frac{2\gamma_{\text{Lang}}^V}{\beta}}W^{1/2}\zeta_\varepsilon(t)
\end{aligned} \tag{S94}$$

where $\boldsymbol{\zeta}(t)$ and $\zeta_\varepsilon(t)$ represents the Gaussian white noise (vector and scalar variables, respectively). The equations of motion lead to the stationary distribution

$$\rho(\mathbf{x},\mathbf{p},V,p_\varepsilon) = \mathcal{N}_{\text{MTTK}} \exp\left[-\beta\left(\frac{1}{2}\mathbf{p}^T\mathbf{M}^{-1}\mathbf{p}+U(\mathbf{x})+P_{\text{ext}}V+\frac{p_\epsilon^2}{2W}\right)\right], \tag{S95}$$

whose marginal distribution of $\mathbf{x},\mathbf{p},V$ is the equilibrium distribution of the isobaric-isothermal ensemble, eq (S93). In eq (S95), $\mathcal{N}_{\text{MTTK}}$ is a normalization constant.

Below we show that $\rho(\mathbf{x},\mathbf{p},V,p_\varepsilon)$ of eq (S95) is the stationary distribution. The Kolmogorov operators of the decomposition of the MTTK barostat in the main text reads

$$\mathcal{L}_\mathbf{x}\rho = -\mathbf{p}^T\mathbf{M}^{-1}\frac{\partial\rho}{\partial\mathbf{x}} = \rho\left(\beta\mathbf{p}^T\mathbf{M}^{-1}\frac{\partial U}{\partial\mathbf{x}}\right), \tag{S96}$$

$$\mathcal{L}_\mathbf{p}\rho = \left(\frac{\partial U}{\partial\mathbf{x}}\right)^T\frac{\partial\rho}{\partial\mathbf{p}} = \rho\left(-\beta\mathbf{p}^T\mathbf{M}^{-1}\frac{\partial U}{\partial\mathbf{x}}\right), \tag{S97}$$

$$\mathcal{L}_T\rho = \gamma_{\text{Lang}}\left(\frac{\partial}{\partial\mathbf{p}}\right)^T(\mathbf{p}\rho)+\frac{\gamma_{\text{Lang}}}{\beta}\left(\frac{\partial}{\partial\mathbf{p}}\right)^T\mathbf{M}\frac{\partial\rho}{\partial\mathbf{p}} = 0, \tag{S98}$$

$$\mathcal{L}_{\mathbf{x}_r}\rho = -\frac{p_\varepsilon}{W}\left(\frac{\partial}{\partial\mathbf{x}}\right)^T(\mathbf{x}\rho) = \frac{p_\varepsilon}{W}\rho\left(-N_f + \beta\mathbf{x}^T\frac{\partial U}{\partial\mathbf{x}}\right), \tag{S99}$$

$$\mathcal{L}_{\mathbf{p}_r}\rho = \left(1+\frac{d}{N_f}\right)\frac{p_\varepsilon}{W}\left(\frac{\partial}{\partial\mathbf{p}}\right)^T(\mathbf{p}\rho) = \left(1+\frac{d}{N_f}\right)\frac{p_\varepsilon}{W}\rho\left(N_f - \beta\mathbf{p}^T\mathbf{M}^{-1}\mathbf{p}\right), \tag{S100}$$

$$\mathcal{L}_V\rho = -d\frac{p_\varepsilon}{W}\frac{\partial}{\partial V}(V\rho) = d\frac{p_\varepsilon}{W}\rho\left(-1+\beta P_{\text{ext}}V\right), \tag{S101}$$

$$\begin{aligned}\mathcal{L}_{p_\varepsilon}\rho &= -dV\left(P_{\text{int}}-P_{\text{ext}}\right)\frac{\partial\rho}{\partial p_\varepsilon} - \frac{d}{N_f}\mathbf{p}^T\mathbf{M}^{-1}\mathbf{p}\frac{\partial\rho}{\partial p_\varepsilon} \\ &= \frac{p_\varepsilon}{W}\beta\rho\left[\left(1+\frac{d}{N_f}\right)\mathbf{p}^T\mathbf{M}^{-1}\mathbf{p} - \mathbf{x}^T\frac{\partial U}{\partial\mathbf{x}} - dP_{\text{ext}}V\right],\end{aligned} \tag{S102}$$

$$\mathcal{L}_B\rho = \gamma_{\text{Lang}}^V\frac{\partial}{\partial p_\varepsilon}(p_\varepsilon\rho) + \frac{\gamma_{\text{Lang}}^V W}{\beta}\frac{\partial^2\rho}{\partial p_\varepsilon^2} = 0. \tag{S103}$$

The distribution function $\rho(\mathbf{x},\mathbf{p},V,p_\varepsilon)$ of eq (S95) is substituted in the right-hand side of each of eqs (S96)-(S103). Equations (S96)-(S103) lead to

$$\mathcal{L}_{\text{MTTK}}^{(\text{full})}\rho \equiv \left(\mathcal{L}_\mathbf{x} + \mathcal{L}_{\mathbf{x}_r} + \mathcal{L}_\mathbf{p} + \mathcal{L}_{\mathbf{p}_r} + \mathcal{L}_V + \mathcal{L}_{p_\varepsilon} + \mathcal{L}_T + \mathcal{L}_B\right)\rho = 0, \tag{S104}$$

where the left-hand side of eq (S104) is the Liouville equation of the MTTK barostat and the right-hand side of eq (S104) indicates that $\rho(\mathbf{x},\mathbf{p},V,p_\varepsilon)$ of eq (S95) is the stationary distribution. One can also verify that $\rho(\mathbf{x},\mathbf{p},V,p_\varepsilon)$ of eq (S95) is the stationary distribution by employing the conserve quantity and non-compressibility of the MTTK barostat as demonstrated in ref [94]..

In the main text we describe that the efficient integrator for the isobaric-isothermal ensemble should approach the "middle" thermostat scheme for the canonical ensemble when barostat effectively vanishes. The recommended "middle" scheme for the MTTK barostat in the main text is only one possible choice. Here, we investigate several possible alternative decomposition schemes with numerical tests to show that the recommended "middle" scheme (proposed in the main text) is actually the optimal decomposition.

To identity the major aspect of the decomposition order, we introduce

$$\mathcal{L}_{\mathbf{x}_t} = \mathcal{L}_\mathbf{x} + \mathcal{L}_{\mathbf{x}_r} \tag{S105}$$

and

$$\mathcal{L}_{\mathbf{p}_t} = \mathcal{L}_\mathbf{p} + \mathcal{L}_{\mathbf{p}_r}, \tag{S106}$$

which is closely related to the decomposition proposed in ref [23]. That is,

$$\mathcal{L}_{\mathbf{x}_t}\rho = -N_f \frac{p_\varepsilon}{W}\rho - \left(\mathbf{p}^T\mathbf{M}^{-1} + \frac{p_\varepsilon}{W}\mathbf{x}^T\right)\frac{\partial \rho}{\partial \mathbf{x}} \tag{S107}$$

and

$$\mathcal{L}_{\mathbf{p}_t}\rho = \left(\frac{\partial U}{\partial \mathbf{x}} + \left(1 + \frac{d}{N_f}\right)\frac{p_\varepsilon}{W}\mathbf{p}\right)^T \frac{\partial \rho}{\partial \mathbf{p}} + \left(N_f + d\right)\frac{p_\varepsilon}{W}\rho \quad . \tag{S108}$$

The relevant Kolmogorov operators are defined by eqs (S107), (S108), (S98), (S101), (S102), and (S103). Further comparison of this strategy to the decomposition in the main text (with eqs (S96)-(S103)) is presented in Sub-Section **S2-C**.

The commutation relations between different Kolmogorov operators read

$$\begin{aligned}
\left[\mathcal{L}_{\mathbf{x}_t}, \mathcal{L}_V\right] &= 0 \\
\left[\mathcal{L}_{\mathbf{p}_t}, \mathcal{L}_V\right] &= 0 \\
\left[\mathcal{L}_T, \mathcal{L}_V\right] &= 0 \\
\left[\mathcal{L}_T, \mathcal{L}_B\right] &= 0
\end{aligned} \quad . \tag{S109}$$

By applying these rules and considering the fact that the "$p-x-T-x-p$" scheme (based on the velocity-Verlet algorithm) is the optimal decomposition for the canonical ensemble[44, 46, 49, 50], we have 6 alternatives with the MTTK barostat, which approaches the "$p-x-T-x-p$" scheme[46, 49, 50] (based on the velocity-Verlet algorithm) when the barastat is withdrew.

We simplify the notation by considering the symmetry of the evolution. Only the first half of the propagation (in the right-to-left sequence in the whole integrator for a finite time interval) is utilized for notation. For example, the scheme

$$\begin{aligned}
e^{\mathcal{L}_{\mathrm{NPT}}\Delta t} &= e^{\mathcal{L}_{p_\varepsilon}\Delta t/2} e^{\mathcal{L}_V \Delta t/2} e^{\mathcal{L}_{\mathbf{p}_t}\Delta t/2} e^{\mathcal{L}_{\mathbf{x}_t}\Delta t/2} e^{\mathcal{L}_T \Delta t} e^{\mathcal{L}_B \Delta t} e^{\mathcal{L}_{\mathbf{x}_t}\Delta t/2} e^{\mathcal{L}_{\mathbf{p}_t}\Delta t/2} e^{\mathcal{L}_V \Delta t/2} e^{\mathcal{L}_{p_\varepsilon}\Delta t/2} \\
&= e^{\mathcal{L}_{p_\varepsilon}\Delta t/2} e^{\mathcal{L}_V \Delta t/2} e^{\mathcal{L}_{\mathbf{p}_t}\Delta t/2} e^{\mathcal{L}_{\mathbf{x}_t}\Delta t/2} e^{\mathcal{L}_B \Delta t/2} e^{\mathcal{L}_T \Delta t/2} e^{\mathcal{L}_T \Delta t/2} e^{\mathcal{L}_B \Delta t/2} e^{\mathcal{L}_{\mathbf{x}_t}\Delta t/2} e^{\mathcal{L}_{\mathbf{p}_t}\Delta t/2} e^{\mathcal{L}_V \Delta t/2} e^{\mathcal{L}_{p_\varepsilon}\Delta t/2}
\end{aligned} \tag{S110}$$

is denoted as $T-B-x-p-V-p_V$. The six velocity-Verlet-based "middle" schemes are listed in **Table S1**, and the corresponding schemes based on the leap-frog algorithm are listed in **Table S2**.

**Table S1.** Tested "Middle" Schemes Based on the Velocity-Verlet Algorithm with the MTTK Barostat

| # | Decomposition Order | Simplified notation |
|---|---|---|
| #1 | $e^{\mathcal{L}_{p_\varepsilon} \Delta t/2} e^{\mathcal{L}_V \Delta t/2} e^{\mathcal{L}_{\mathbf{p}_t} \Delta t/2} e^{\mathcal{L}_{\mathbf{x}_t} \Delta t/2} e^{\mathcal{L}_T \Delta t} e^{\mathcal{L}_B \Delta t} e^{\mathcal{L}_{\mathbf{x}_t} \Delta t/2} e^{\mathcal{L}_{\mathbf{p}_t} \Delta t/2} e^{\mathcal{L}_V \Delta t/2} e^{\mathcal{L}_{p_\varepsilon} \Delta t/2}$ | $T - B - x - p - V - p_V$ |
| #2 | $e^{\mathcal{L}_V \Delta t/2} e^{\mathcal{L}_{p_\varepsilon} \Delta t/2} e^{\mathcal{L}_{\mathbf{p}_t} \Delta t/2} e^{\mathcal{L}_{\mathbf{x}_t} \Delta t/2} e^{\mathcal{L}_T \Delta t} e^{\mathcal{L}_B \Delta t} e^{\mathcal{L}_{\mathbf{x}_t} \Delta t/2} e^{\mathcal{L}_{\mathbf{p}_t} \Delta t/2} e^{\mathcal{L}_{p_\varepsilon} \Delta t/2} e^{\mathcal{L}_V \Delta t/2}$ | $T - B - x - p - p_V - V$ |
| #3 | $e^{\mathcal{L}_{\mathbf{p}_t} \Delta t/2} e^{\mathcal{L}_{\mathbf{x}_t} \Delta t/2} e^{\mathcal{L}_{p_\varepsilon} \Delta t/2} e^{\mathcal{L}_V \Delta t/2} e^{\mathcal{L}_T \Delta t} e^{\mathcal{L}_B \Delta t} e^{\mathcal{L}_V \Delta t/2} e^{\mathcal{L}_{p_\varepsilon} \Delta t/2} e^{\mathcal{L}_{\mathbf{x}_t} \Delta t/2} e^{\mathcal{L}_{\mathbf{p}_t} \Delta t/2}$ | $T - B - V - p_V - x - p$ |
| #4 | $e^{\mathcal{L}_{\mathbf{p}_t} \Delta t/2} e^{\mathcal{L}_{\mathbf{x}_t} \Delta t/2} e^{\mathcal{L}_V \Delta t/2} e^{\mathcal{L}_{p_\varepsilon} \Delta t/2} e^{\mathcal{L}_T \Delta t} e^{\mathcal{L}_B \Delta t} e^{\mathcal{L}_{p_\varepsilon} \Delta t/2} e^{\mathcal{L}_V \Delta t/2} e^{\mathcal{L}_{\mathbf{x}_t} \Delta t/2} e^{\mathcal{L}_{\mathbf{p}_t} \Delta t/2}$ | $T - B - p_V - V - x - p$ |
| #5 | $e^{\mathcal{L}_{\mathbf{p}_t} \Delta t/2} e^{\mathcal{L}_{p_\varepsilon} \Delta t/2} e^{\mathcal{L}_V \Delta t/2} e^{\mathcal{L}_{\mathbf{x}_t} \Delta t/2} e^{\mathcal{L}_T \Delta t} e^{\mathcal{L}_B \Delta t} e^{\mathcal{L}_{\mathbf{x}_t} \Delta t/2} e^{\mathcal{L}_V \Delta t/2} e^{\mathcal{L}_{p_\varepsilon} \Delta t/2} e^{\mathcal{L}_{\mathbf{p}_t} \Delta t/2}$ | $T - B - x - V - p_V - p$ |
| #6 | $e^{\mathcal{L}_{\mathbf{p}_t} \Delta t/2} e^{\mathcal{L}_V \Delta t/2} e^{\mathcal{L}_{p_\varepsilon} \Delta t/2} e^{\mathcal{L}_{\mathbf{x}_t} \Delta t/2} e^{\mathcal{L}_T \Delta t} e^{\mathcal{L}_B \Delta t} e^{\mathcal{L}_{\mathbf{x}_t} \Delta t/2} e^{\mathcal{L}_{p_\varepsilon} \Delta t/2} e^{\mathcal{L}_V \Delta t/2} e^{\mathcal{L}_{\mathbf{p}_t} \Delta t/2}$ | $T - B - x - p_V - V - p$ |

**Table S2.** Tested "Middle" Schemes Based on the Leap-frog Algorithm

with the MTTK Barostat

| # | Decomposition Order | Simplified notation |
|---|---|---|
| #1 | $e^{\mathcal{L}_{x_t}\Delta t/2} e^{\mathcal{L}_T \Delta t} e^{\mathcal{L}_B \Delta t} e^{\mathcal{L}_{x_t}\Delta t/2} e^{\mathcal{L}_{p_t}\Delta t/2} e^{\mathcal{L}_V \Delta t/2} e^{\mathcal{L}_{p_\varepsilon}\Delta t} e^{\mathcal{L}_V \Delta t/2} e^{\mathcal{L}_{p_t}\Delta t/2}$ | $T-B-x-p-V-p_V$ |
| #2 | $e^{\mathcal{L}_{x_t}\Delta t/2} e^{\mathcal{L}_T \Delta t} e^{\mathcal{L}_B \Delta t} e^{\mathcal{L}_{x_t}\Delta t/2} e^{\mathcal{L}_{p_t}\Delta t/2} e^{\mathcal{L}_{p_\varepsilon}\Delta t/2} e^{\mathcal{L}_V \Delta t} e^{\mathcal{L}_{p_\varepsilon}\Delta t/2} e^{\mathcal{L}_{p_t}\Delta t/2}$ | $T-B-x-p-p_V-V$ |
| #3 | $e^{\mathcal{L}_{x_t}\Delta t/2} e^{\mathcal{L}_{p_\varepsilon}\Delta t/2} e^{\mathcal{L}_V \Delta t/2} e^{\mathcal{L}_T \Delta t} e^{\mathcal{L}_B \Delta t} e^{\mathcal{L}_V \Delta t/2} e^{\mathcal{L}_{p_\varepsilon}\Delta t/2} e^{\mathcal{L}_{x_t}\Delta t/2} e^{\mathcal{L}_{p_t}\Delta t}$ | $T-B-V-p_V-x-p$ |
| #4 | $e^{\mathcal{L}_{x_t}\Delta t/2} e^{\mathcal{L}_V \Delta t/2} e^{\mathcal{L}_{p_\varepsilon}\Delta t/2} e^{\mathcal{L}_T \Delta t} e^{\mathcal{L}_B \Delta t} e^{\mathcal{L}_{p_\varepsilon}\Delta t/2} e^{\mathcal{L}_V \Delta t/2} e^{\mathcal{L}_{x_t}\Delta t/2} e^{\mathcal{L}_{p_t}\Delta t}$ | $T-B-p_V-V-x-p$ |
| #5 | $e^{\mathcal{L}_{p_\varepsilon}\Delta t/2} e^{\mathcal{L}_V \Delta t/2} e^{\mathcal{L}_{x_t}\Delta t/2} e^{\mathcal{L}_T \Delta t} e^{\mathcal{L}_B \Delta t} e^{\mathcal{L}_{x_t}\Delta t/2} e^{\mathcal{L}_V \Delta t/2} e^{\mathcal{L}_{p_\varepsilon}\Delta t/2} e^{\mathcal{L}_{p_t}\Delta t}$ | $T-B-x-V-p_V-p$ |
| #6 | $e^{\mathcal{L}_V \Delta t/2} e^{\mathcal{L}_{p_\varepsilon}\Delta t/2} e^{\mathcal{L}_{x_t}\Delta t/2} e^{\mathcal{L}_T \Delta t} e^{\mathcal{L}_B \Delta t} e^{\mathcal{L}_{x_t}\Delta t/2} e^{\mathcal{L}_{p_\varepsilon}\Delta t/2} e^{\mathcal{L}_V \Delta t/2} e^{\mathcal{L}_{p_t}\Delta t}$ | $T-B-x-p_V-V-p$ |

Scheme **#1** has been proposed in ref [30], although the dynamics in ref [30] is not based on the equations of the MTTK barostat. As will be demonstrated in numerical tests below, its corresponding algorithm in the MTTK barostat analog does NOT considerably outperform the conventional "side" scheme.

In the following sub-sections, we test these six schemes with different benchmark models to identify the best option for MD simulations of the isobaric-isothermal ensemble.

**S2-A: Numerical Tests of Different "Middle" Schemes on the One-dimensional Nanowire Model**

In this sub-section, we test the proposed schemes listed in Table S1 and Table S2. We utilize the one-dimensional nanowire model of ref [7]. It involves a single particle moving within a one-dimensional cosine potential,

$$U(x,V) = \frac{m\omega^2 V^2}{4\pi^2}\left[1-\cos\left(\frac{2\pi x}{V}\right)\right] , \qquad (S111)$$

where $m=1$, $\omega=1$. Classical MD simulations of the isobaric-isothermal ensemble are performed at $k_B T = 0.01$, and $P_{ext} = 0.01$, and the "piston mass" is set to $W = 1000$, where all parameters and variables are in reduced units. Six proposed velocity-Verlet-based "middle" schemes listed in Table S1 and their corresponding leap-frog-based schemes are tested. For each scheme, we run 20 independent trajectories for the model system. After each trajectory is equilibrated for $2\times10^7$ r.u., it is propagated for $8\times10^7$ r.u. to evaluate physical properties. The friction parameter $\gamma_{Lang}$ for the Langevin thermostat and $\gamma^V_{Lang}$ for the barostat are 1.0 r.u. and 0.001 r.u., respectively. The number of degrees of freedom ($N_f$) is 1 for this model system.

MD simulations for the 1-dimensional nanowire model of eq (S111) are performed with different time intervals. In Figures S1-S2, eqs (S76) and (S84) are used to calculate the density and isothermal compressibility, and eqs (S72), (S91), and (S86) are utilized to evaluate the enthalpy, isobaric heat capacity, and thermal expansion coefficient. In Figure S3, both eq (S79) and eq (S91) are employed for the evaluation of the isobaric heat capacity.

As shown in **Figure S10**, the length (volume in the 1-dimensional case), potential energy, enthalpy, isobaric heat capacity, isothermal compressibility and thermal expansion coefficient are plotted as functions of the time interval. The gray horizontal lines represent accurate values obtained by numerical integration. As the time interval decreases, all the schemes converge to the same result, which is consistent with the fact that all integrators are in principle equivalent when the time interval is infinitesimal.

Schemes **#1** and **#2** yield nearly identical numerical results for either velocity-Verlet based schemes or leap-frog based schemes. Schemes **#3** and **#4** also perform nearly the same. Similarly, Schemes **#5** and **#6** produce almost indistinguishable results. This indicates that, although operators $e^{\mathcal{L}_V \Delta t}$ and $e^{\mathcal{L}_{p_\varepsilon} \Delta t}$ do not commute, their relative order in the sequence of the full evolution propagator is not important. Regardless of whether the relative order is '… $p_V - V$ …' or '… $V - p_V$ …', it generates almost the same numerical results within the statistical error bar.

Figure S1 suggest that Schemes **#5** and **#6** based on the velocity-Verlet algorithm produce relatively accurate results for the volume and isothermal compressibility at a relatively large time interval. When $\Delta t = 1.0$ r.u. is used, the absolute deviation of the length for these two schemes is only $\sim 0.001$ r.u., while that of the other four schemes is more than $0.1$ r.u.; the absolute deviation of the isothermal compressibility of these two schemes is only $\sim 0.5$ r.u., and in comparison, those of the other four schemes are more than $7$ r.u.. These imply that Schemes **#5** and **#6** are much better for sampling the volume distribution. Moreover, Schemes **#5** and **#6** also perform better for sampling the coordinate distribution, as indicated by the results for the potential energy as a function of the time interval in Panel (b) of Figure S1.

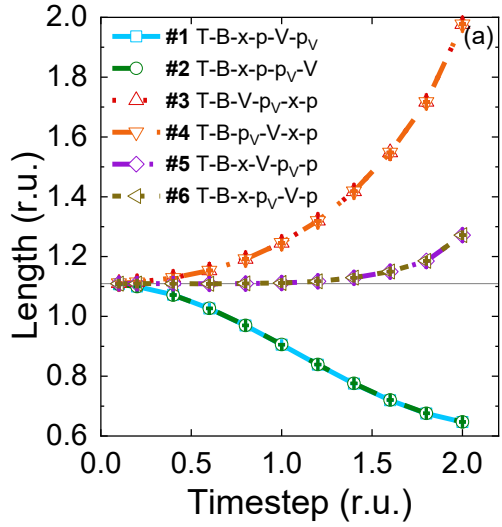
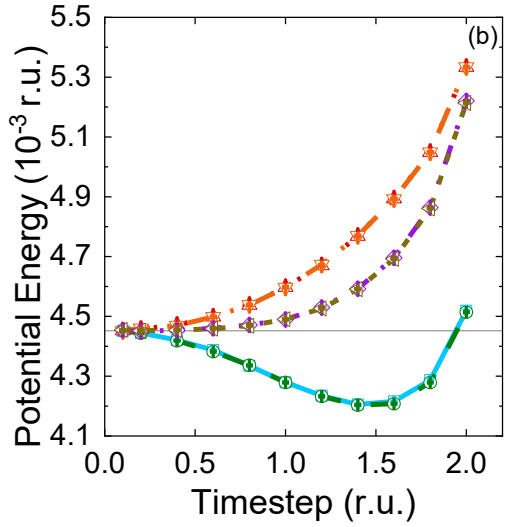
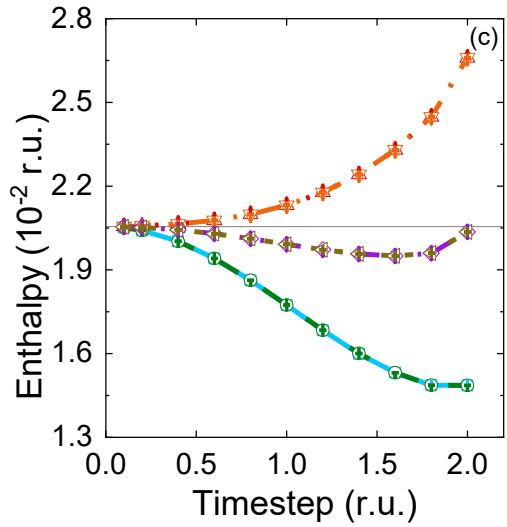
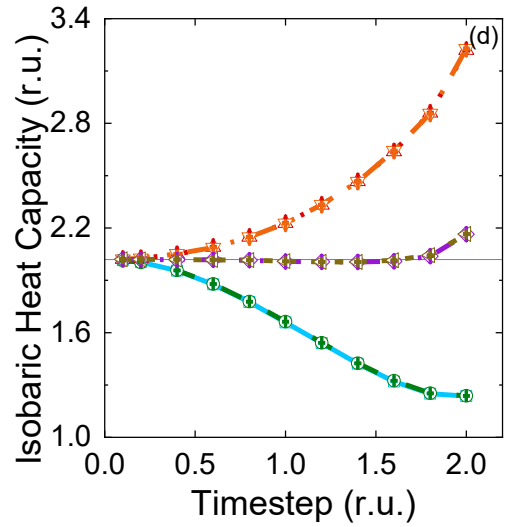
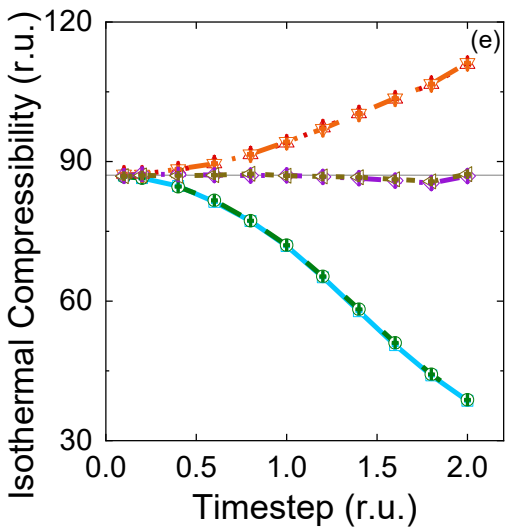
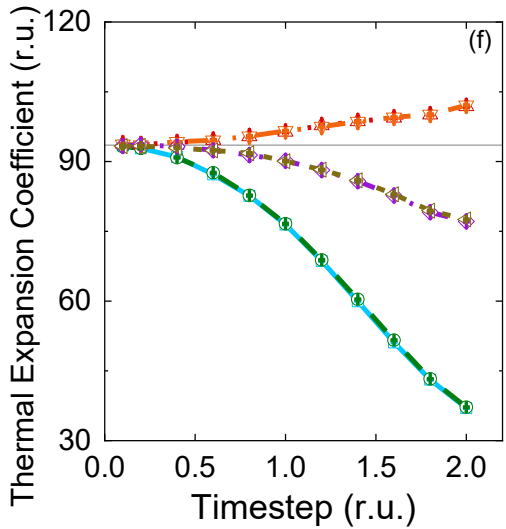

**Figure S10.** Comparison of different velocity-Verlet-based "middle" schemes with the MTTK barostat using the one-dimensional nanowire model at $k_B T = 0.01$ r.u. and $P_{ext} = 0.01$ r.u.. The calculated thermodynamic properties include (a) length, (b) potential energy, (c) enthalpy, (d) isobaric heat capacity, (e) thermal expansion coefficient, and (f) isothermal compressibility. Each line corresponds to one scheme listed in Table S1. Gray horizontal lines: accurate values calculated by numerical integration. 20 trajectories for each scheme are equilibrated for $2 \times 10^7$ r.u. equilibration and then propagated for $8 \times 10^7$ r.u.. All data are divided into 20 groups to calculate the standard error. The friction parameter $\gamma_{Lang}$ for the Langevin thermostat and $\gamma_{Lang}^V$ for the barostat are 1.0 r.u. and 0.001 r.u., respectively. All MD simulations are performed by our independently developed MD/PIMD program.

The conclusion is similar when the corresponding six schemes based on the leap-frog algorithm are applied. Figure S2 shows that leap-frog-based schemes **#5** and **#6** of Table S2 have the best performance among all six schemes. As described in refs [45, 50], in comparison to the velocity-Verlet-based "middle" thermostat scheme, the leap-frog-based "middle" thermostat scheme generates the more accurate momentum marginal distribution of the canonical ensemble. As expected, leap-frog-based schemes **#5** and **#6** of **Table S2** and **Figure S11** produce better results for the kinetic energy, enthalpy, and thermal expansion coefficient than velocity-Verlet-based schemes **#5** and **#6** of

**Table S1** and **Figure S10**. The comparison of **Figure S10**(c) to **Figure S11**(c) demonstrates that, when $\Delta t = 1.0$ r.u. is used, the absolute deviation in the enthalpy generated by velocity-Verlet-based schemes **#5** and **#6** is $\sim 6 \times 10^{-4}$ r.u., while that by leap-frog-based schemes **#5** and **#6** is only $\sim 6\times 10^{-5}$ r.u., about an order of magnitude smaller. **Figure S10**(f) shows that the absolute deviation in the thermal expansion coefficient produced by velocity-Verlet-based schemes **#5** and **#6** is $\sim 3.2$ r.u., but **Figure S11**(f) demonstrates that by leap-frog-based schemes **#5** and **#6** is as small as $\sim 0.46$ r.u..

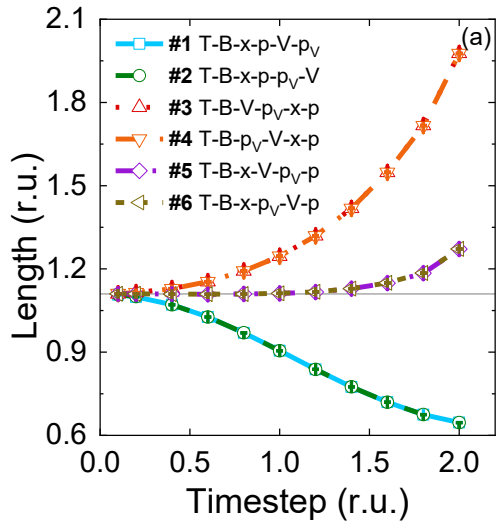
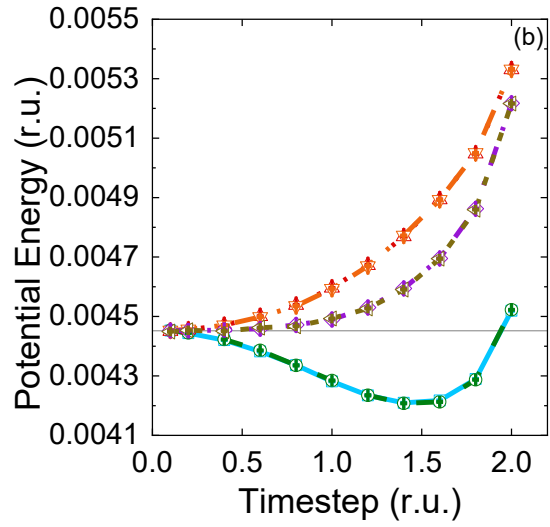
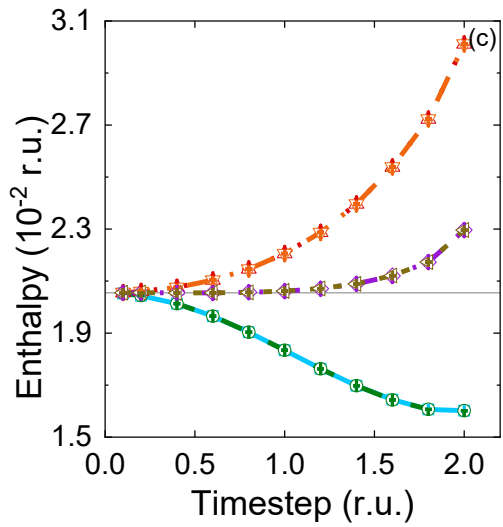
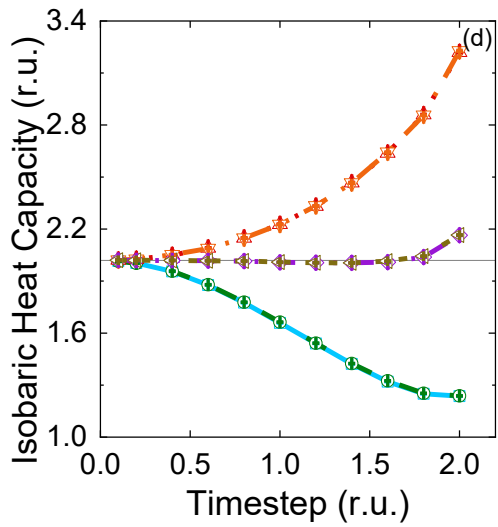
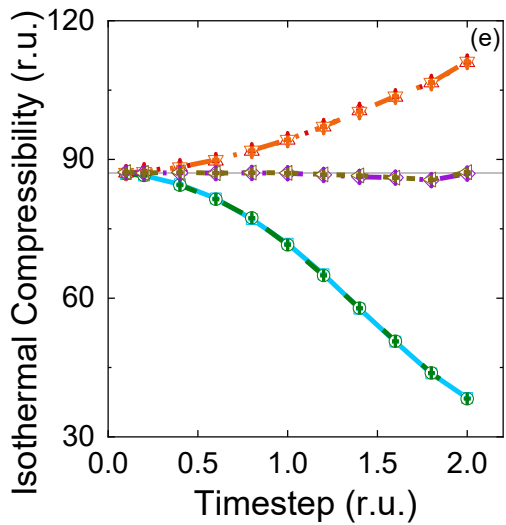
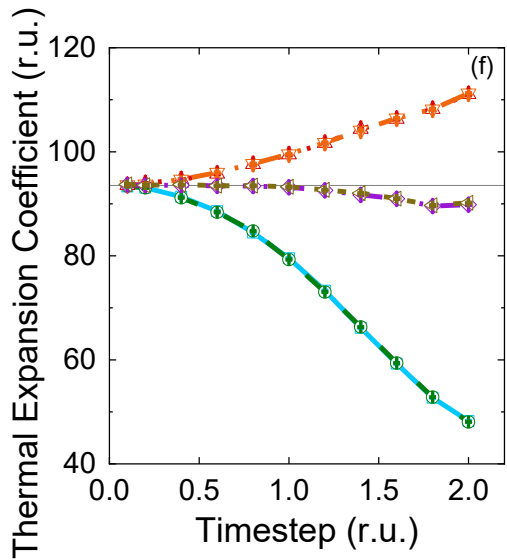

**Figure S11.** Comparison of different leap-frog-based schemes with the MTTK barostat using the one-dimensional nanowire model at $k_B T = 0.01$ r.u. and $P_{\text{ext}} = 0.01$ r.u. . The calculated thermodynamic properties include (a) length, (b) potential energy, (c) enthalpy, (d) isobaric heat Capacity, (e) thermal expansion coefficient and (f) isothermal compressibility. Each line corresponds to one scheme listed in **Table S2**. Gray horizontal lines: accurate values calculated by numerical integration. 20 trajectories for each scheme are equilibrated for $2 \times 10^7$ r.u., then propagated for $8 \times 10^7$ r.u. for estimating physical properties. All data are divided into 20 groups to calculate the standard error. The friction parameter $\gamma_{\text{Lang}}$ for the Langevin thermostat and $\gamma_{\text{Lang}}^V$ for the barostat are $1.0$ r.u. and $0.001$ r.u., respectively. All MD simulations are performed by our independently developed MD/PIMD program.

Both classical MD simulations in **Figure S10** for the velocity-Verlet-based schemes and those in **Figure S102** for leap-frog-based schemes employ eq (S91) for the isobaric heat capacity. When the brute-force estimator of eq (S79) is used to evaluate the isobaric heat capacity, it leads significantly worse results. **Figure S12** shows that, when eq (S79) is used for the evaluation of the isobaric heat capacity, velocity-Verlet-based schemes **#5** and **#6**, as well as leap-frog-based schemes **#5** and **#6**, lead to the absolute deviation no smaller than $0.26$ r.u. at $\Delta t = 1.4$ r.u.. In comparison, when the estimator of eq (S91) is employed, the absolute deviation of the isobaric heat capacity is only $\sim 0.013$ r.u. at $\Delta t = 1.4$ r.u..

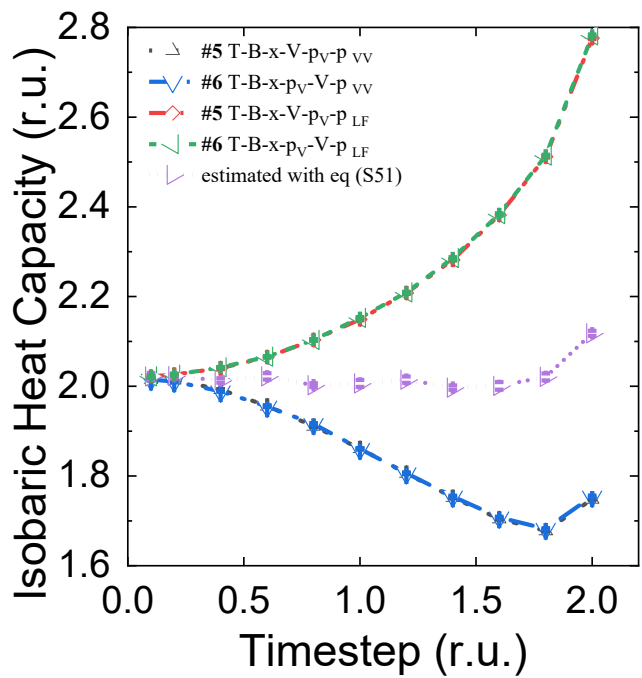

**Figure S12.** The isobaric heat capacity calculated by the estimator of eq (S91) is compared to that computed by eq (S79), for leap-frog-based schemes **#5** and **#6**, as well as velocity-Verlet-based schemes **#5** and **#6**. Simulation details are similar to those of **Figures S1 and S2**.

Figures S1-S3 suggest that, in the evaluation of physical properties in classical statistical mechanics, we should take full advantage of the known Maxwell momentum distribution and replace the components that are functions of only momentum by the exact averaged results, as described in Sub-Section S1C-6. In the main text as well as the rest of the Supporting Information, while we employ eqs (S76) and (S84) for calculating the density and isothermal compressibility, we use eqs (S89), (S91), and (S92) for the evaluation of the enthalpy, isobaric heat capacity, and thermal expansion coefficient.

We test the performance of different "middle" schemes for liquid water at $T = 298.15\text{ K}$ and $P_{\text{ext}} = 1\text{ bar}$ described in the main text is used to compare the different schemes. The flexible q-SPC/fw model is used for the MD simulations. The results show the comparison of six leap-frog-based "middle" schemes with different time intervals. Schemes #5 and #6 perform best in all 6 leap-frog-based schemes, which is consistent with the conclusion based on the results of the nanowire system. Figure S4 of the MD simulations of liquid water also shows that the relative order between the $p_V$-updating step ($e^{\mathcal{L}_{p_\varepsilon}\Delta t}$) and the $V$-updating step ($e^{\mathcal{L}_V \Delta t}$) does not noticeably change the results, i.e., the scheme with '... $p_V - V$ ...' and the counterpart with '... $V - p_V$ ...' generate similar results within the statistical error bar. One can also verify that Schemes #5 and #6 are also the optimal choice among six velocity-Verlet-based "middle" schemes.

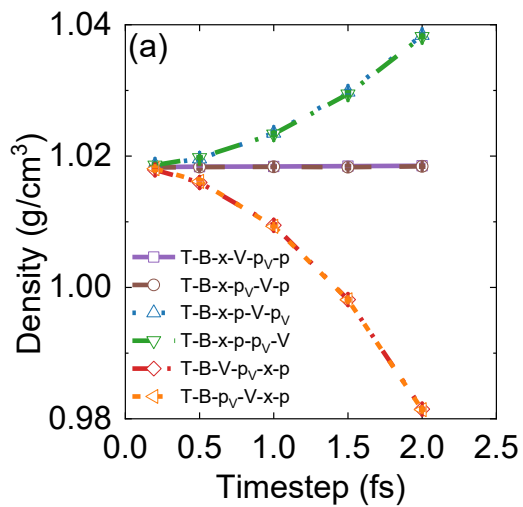
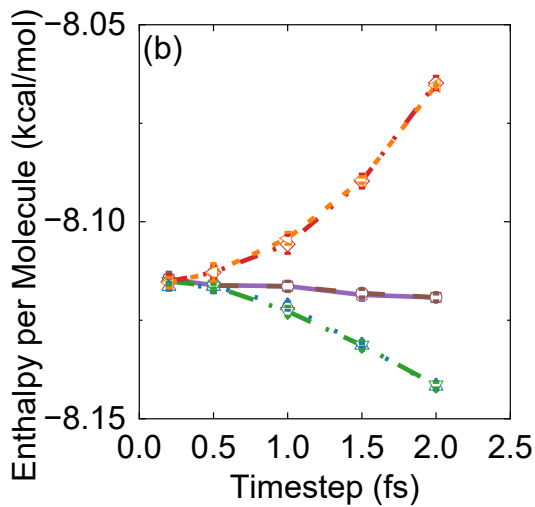
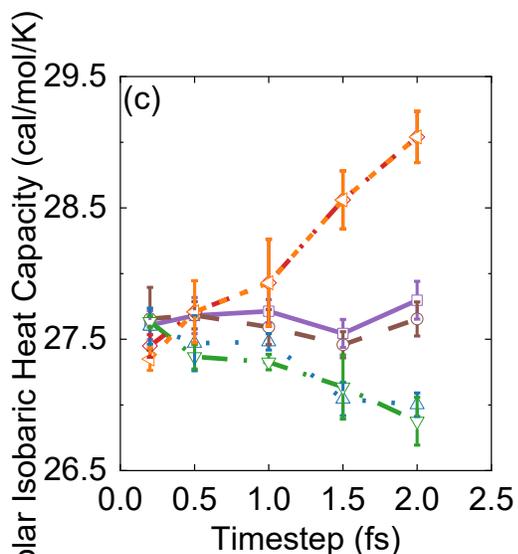
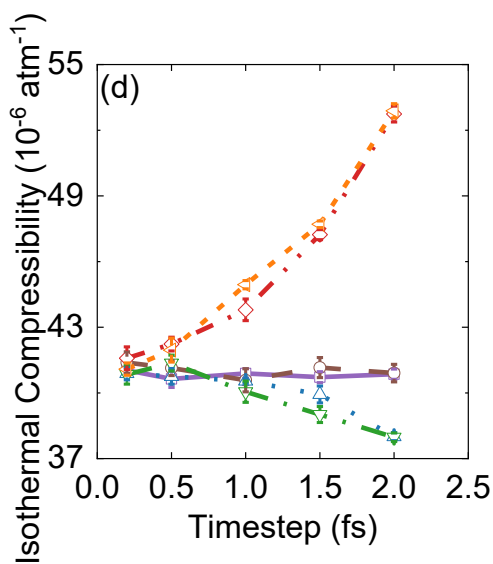
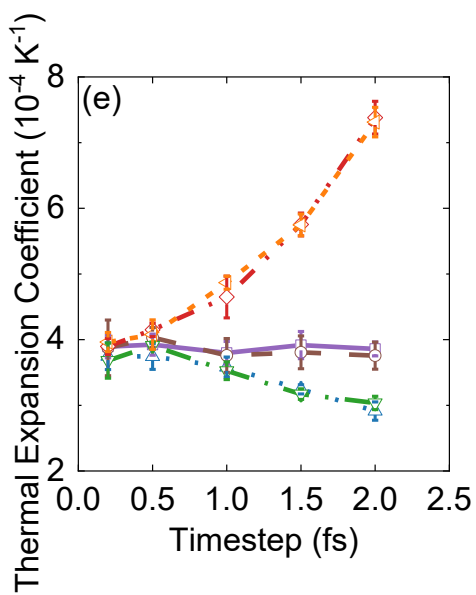
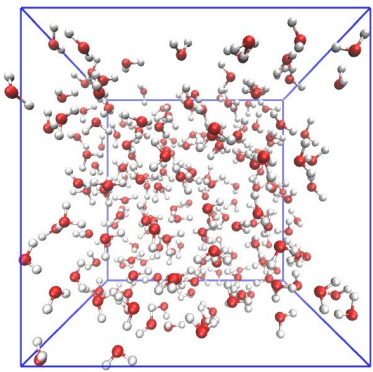

**Figure S13.** Comparison of different leap-frog-based "middle" schemes with the MTTK barostat using liquid water at $T = 298.15$ K and $P_{\text{ext}} = 1$ bar. The q-SPC/fw model is used. The calculated thermodynamic properties include (a) density, (b) enthalpy per molecule (c) molar isobaric heat capacity, (d) isothermal compressibility and (e) thermal expansion coefficient. Each line corresponds to one scheme listed in **Table S2**. Twenty trajectories are equilibrated for 1 ns and propagated for 10 ns, using a cubic box with periodic boundary conditions containing 216 molecules. All data are divided into 20 groups to calculate the standard error. The friction parameter $\gamma_{\text{Lang}}$ for the Langevin thermostat and $\gamma_{\text{Lang}}^V$ for the barostat are all $5.0$ ps$^{-1}$, respectively. The "piston" mass $W$ is set to the recommended value as shown in eq (29) in the main text. The cutoff for the short-range interactions is 9 Å. All MD simulations are performed by our modified DL_POLY_2 package[61].

**S2-C: On Different Decompositions of Kolmogorov Operators of the MTTK Barostat**

In the previous sub-sections of Section S2, our theoretical analysis and numerical tests are based on the integrators for the relevant Kolmogorov operators defined by eqs (S107), (S108), (S98), (S101), (S102), and (S103), which are similar to those used in the original version of the MTTK barostat proposed in ref [24]. We focus on the optimal version of the "middle" scheme, Scheme #5 of Table S1. Because the integration of $\mathcal{L}_{x_t}$ (of eq (S107)) is related to a hyper-sine function[8], we denote it the "middle-sinh" decomposition.

In refs [18, 26], however, it is suggested that in the isobaric-isothermal ensemble, the propagation of extended variables should be separated from the Hamiltonian dynamics of particles. Such a perspective indicates that the operators or integrators related to the barostat can be identified as the barostat terms, which can be combined to be treated as a whole part so that a clean-cut unified form can be developed. This is indeed what is employed in the main text. When the MTTK barostat is used in the main text, the relevant Kolmogorov operators are defined by eqs (S96)-(S103) instead. The algorithm described in Figure 2 of the main text is similar to Scheme #**5** in Table S2 and can be labeled as '$p_r - p_V - V - x_r - B - x - T - x - B - x_r - V - p_V - p_r - p$' in the right-to-left sequence based on the leap-frog algorithm. Because the integration of $\mathcal{L}_{x_r}$ (of eq (S99)) involves an exponential function, we denote it the "middle-exp" decomposition.

In this sub-section, we directly compare the performance of the "middle-sinh" decomposition to that of the "middle-exp" decomposition used in the main text. Since the time scale of the variables of the 'barostat' terms is much slower than that of the motion of the particles/atoms as suggested by ref [26], the two decompositions should lead to almost the same results in numerical simulations of real molecular systems. The two decompositions are rigorously equivalent in the limit $W \to \infty$.

**Figure S14** shows the simulation results of the two decompositions for the Lennard-Jones liquid. The results for the "middle-exp" decomposition and the "middle-sinh" decomposition are nearly the same in the statistical error bar for each property. It is consistent with our analysis. Because the "middle-exp" decomposition leads to a clean and clear form for implementing any extended-system barostatting methods, we recommend it as the optimal scheme in the main text.

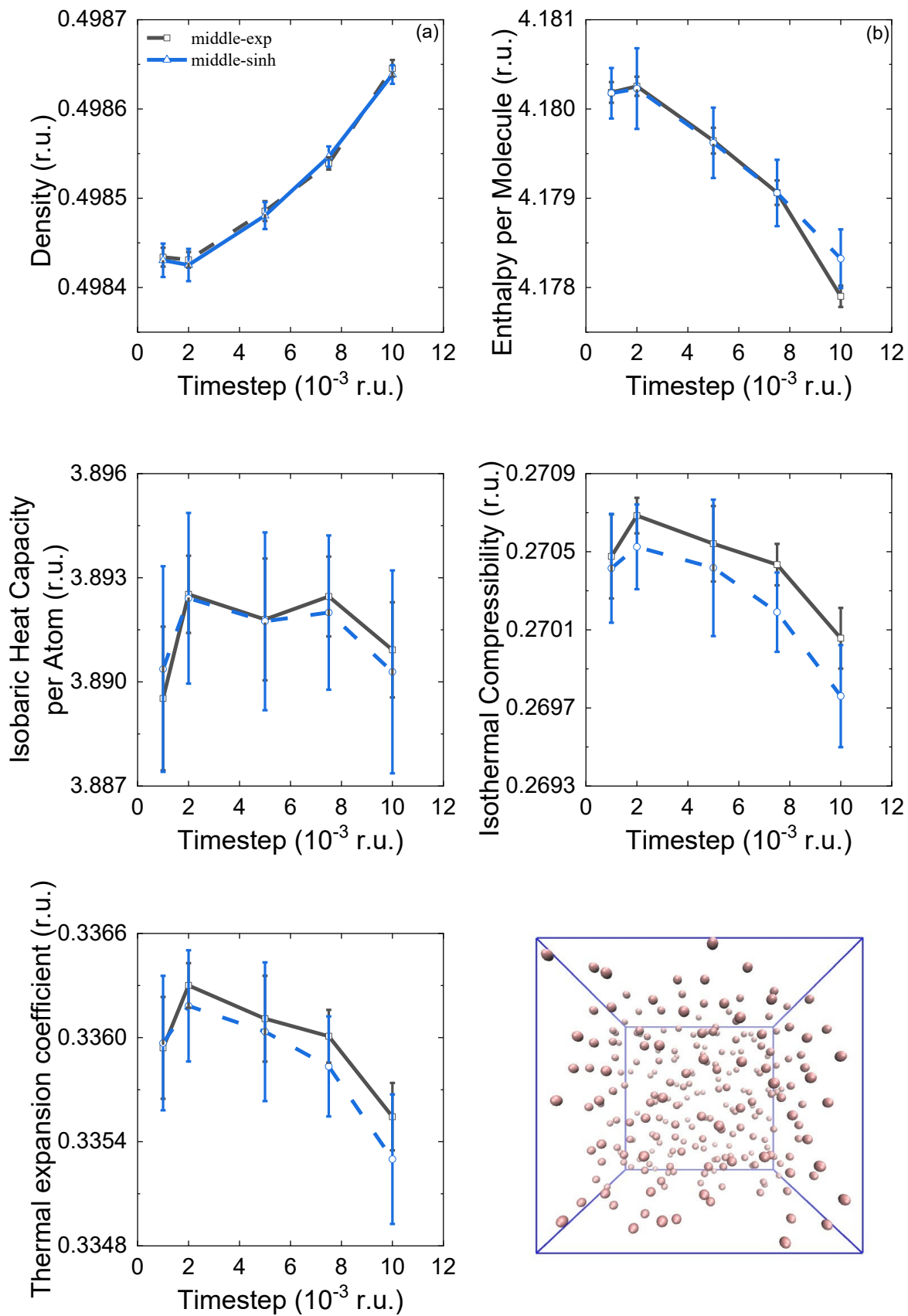

**Figure S14.** Comparison of the "middle-exp" decomposition with the "middle-sinh" decomposition for using the MTTK barostat in the MD simulations of liquid argon. The parameter of L-J potential is $\epsilon = k_B \times 119.8\,\text{K}$, $\sigma = 340.5\,\text{pm}$ and $m = 39.962\,A_r$. The state point is chosen to be $\beta = 0.4\,\varepsilon$ and $P_{ext} = 1.706\,\varepsilon/\sigma^3$. The calculated thermodynamic properties include (a) density, (b) potential energy, (c) enthalpy per molecule, (d) molar isobaric heat Capacity, (e) thermal expansion coefficient and (f) isothermal compressibility. Black line: the "middle-exp" decomposition. Blue line: the "middle-sinh" decomposition (equivalent to the scheme #5 in **Table S1**). Sixty-four trajectories are equilibrated for $2.5\times10^5$ r.u and then propagated for $2.5\times10^5$ r.u, using a cubic box with periodic boundary conditions containing 256 particles. All data are divided into 16 groups to calculate the standard error. The friction parameter ($\gamma_{Lang}$) for the Langevin thermostat and ($\gamma^V_{Lang}$) for the barostat are 5.0 r.u. and 0.5 r.u., respectively. The "piston" mass $W$ is set to 1000 r.u. A smooth cutoff is chosen, i.e., the potential is truncated at $r_c = 3.0\sigma$ and is shifted to zero from $r_s = 2.5\sigma$. For liquid argon, the reduced unit of time is $1.000\times10^{-3}$ r.u. $= 1.000\times10^{-3}\sqrt{m\sigma^2/\varepsilon} = 2.1564\,\text{fs}$. All MD simulations are performed by our independently developed MD/PIMD program. Note that the scale of the y-axis of each panel of Figure S5 is much smaller than that of the corresponding panel of Figure 4.

### S3. Derivation of the "Middle" Scheme with the SCR Barostat

Instead of the category of extended-system barostatting methods such as the MTTK barostat, we can use the category of weak coupling barostat.

One example is the Berendersen barostat[40]. In the Berendersen barostat, the internal pressure of the system and the external pressure satisfy the first-order response relation

$$\frac{dP_{int}}{dt} = \frac{P_{ext} - P_{int}}{\tau_P}, \tag{S112}$$

where $\tau_P$ is the characteristic time of the barostat. However, it has been shown that, in many cases, the Berendsen barostat fails to properly reproduce the volume distribution of the isobaric-isothermal ensemble[42]. The Berendsen barostat should not be recommended.

Another example is the Stochastic Cell-Rescaling (SCR) barostat[41], which is constructed from the Berendsen barostat and correctly describes the volume fluctuation[41]. The equation of motion of the volume variable in the SCR barostat is

$$dV = -\frac{\kappa_T V}{\tau_P}\left(P_{ext} - P_{int} - \frac{k_B T}{V}\right)dt + \sqrt{\frac{2k_B T \kappa_T V}{\tau_P}}dW_t, \tag{S113}$$

where $W_t$ is the Wiener process. After using Ito's lemma and applying the Fokker-Planck equation to eq (S113), we obtain the Kolmogorov operators of the decomposition of the SCR barostat in the main text reads

$$\mathcal{L}_x \rho = -\mathbf{p}^T \mathbf{M}^{-1} \frac{\partial \rho}{\partial \mathbf{x}} = \rho\left(\beta \mathbf{p}^T \mathbf{M}^{-1} \frac{\partial U}{\partial \mathbf{x}}\right), \tag{S114}$$

$$\mathcal{L}_p \rho = \left(\frac{\partial U}{\partial \mathbf{x}}\right)^T \frac{\partial \rho}{\partial \mathbf{p}} = \rho\left(-\beta \mathbf{p}^T \mathbf{M}^{-1} \frac{\partial U}{\partial \mathbf{x}}\right), \tag{S115}$$

$$\mathcal{L}_T \rho = \gamma_{Lang}\left(\frac{\partial}{\partial \mathbf{p}}\right)^T (\mathbf{p}\rho) + \frac{\gamma_{Lang}}{\beta}\left(\frac{\partial}{\partial \mathbf{p}}\right)^T \mathbf{M}\frac{\partial \rho}{\partial \mathbf{p}} = 0, \tag{S116}$$

$$\begin{aligned}
&\mathcal{L}_{Bar}^{MD}\rho(V^{1/d}\mathbf{s}, V^{-1/d}\boldsymbol{\pi}, V) \\
&= \frac{\kappa_T}{\tau_P}\frac{\partial}{\partial V}\left[\left(P_{ext} - P_{int} - \frac{k_B T}{V}\right)V\rho + k_B T\frac{\partial}{\partial V}(V\rho)\right] \\
&= \frac{\kappa_T}{\tau_P}\frac{\partial}{\partial V}\left[-VP_{int}\rho - V\left((V^{-1/d}\boldsymbol{\pi})^T \mathbf{M}^{-1}\left(-\frac{1}{d}\right)(V^{-1/d-1}\boldsymbol{\pi}) + \left(\frac{\partial U}{\partial \mathbf{s}}\right)^T\left(\frac{1}{d}\right)(V^{1/d-1}\mathbf{s})\right)\rho\right] \\
&= \frac{\kappa_T}{\tau_P}\frac{\partial}{\partial V}\left[-VP_{int}\rho - V\left(-\frac{2K}{dV} + \frac{1}{dV}\left(\frac{\partial U}{\partial \mathbf{x}}\right)^T \mathbf{x}\right)\rho\right] = 0
\end{aligned} \quad . \text{(S117)}$$

where $\mathbf{s} = V^{-1/d}\mathbf{x}$ and $\boldsymbol{\pi} = V^{1/d}\mathbf{p}$ is the scaled coordinate and momentum[14, 41], corresponding to the phase space evolution operators eqs (31)-(32) of the main text. The distribution function $\rho(\mathbf{x}, \mathbf{p}, V, p_\varepsilon)$ of eq (S95) is substituted in the right-hand side of each of eqs (S114)-(S117). Equations (S114)-(S117) lead to

$$\mathcal{L}_{SCR}^{(full)}\rho \equiv \left(\mathcal{L}_\mathbf{x} + \mathcal{L}_\mathbf{p} + \mathcal{L}_{Bar}^{MD} + \mathcal{L}_T\right)\rho = 0, \tag{S118}$$

Equation (S113) can be recast into an equivalent equation with $\varepsilon = \ln(V/V_0)$, where $V_0$ is the unit volume,

$$d\varepsilon = -\frac{\kappa_T}{\tau_P}(P_{ext} - P_{int})dt + \sqrt{\frac{2k_B T \kappa_T}{V\tau_P}}dW_t \quad . \tag{S119}$$

In ref [41], it is shown that the equations of motion of the SCR barostat can be viewed as the high-friction limit of those in the following form of the MTTK barostat:

$$\dot{\mathbf{x}} = \mathbf{M}^{-1}\mathbf{p} + \frac{p_\varepsilon}{W}\mathbf{x} , \tag{S120}$$

$$\dot{\mathbf{p}} = -\frac{\partial U}{\partial \mathbf{x}} - \frac{p_\varepsilon}{W}\mathbf{p} - \gamma_{Lang}\mathbf{p} + \sqrt{\frac{2\gamma_{Lang}}{\beta}}\mathbf{M}^{1/2}\boldsymbol{\zeta}(t) , \tag{S121}$$

$$\dot{\varepsilon} = \frac{dp_\varepsilon}{W} , \tag{S122}$$

$$\dot{p}_{\varepsilon} = dV\left(P_{\text{int}} - P_{\text{ext}} + \frac{k_B T}{V}\right) - \gamma_{\text{Lang}}^V p_{\varepsilon} + \sqrt{\frac{2W\gamma_{\text{Lang}}^V}{\beta}}\varsigma_{\varepsilon}(t) \ . \tag{S123}$$

Equations (S120)-(8) are directly obtained from eq (S94) by substituting the term $\frac{d}{N_f}\mathbf{p}^T\mathbf{M}^{-1}\mathbf{p}$ by $dk_B T$ in its last equation. Equations (S122)-(8) can be recast into a second-order differential equation:

$$W\ddot{\varepsilon} = d^2 V(\tilde{P}_{\text{int}} - P_{\text{ext}}) - W\gamma_{\text{Lang}}^V \dot{\varepsilon} + d\sqrt{\frac{2W\gamma_{\text{Lang}}^V}{\beta}}\xi_{\varepsilon}(t) \ . \tag{S124}$$

The high-friction limit of eq (S124) is[41, 95]

$$d\varepsilon = \frac{d^2}{W\gamma_{\text{Lang}}^V}\left(V(\tilde{P}_{\text{int}} - P_{\text{ext}}) - \frac{1}{\beta}\frac{d\ln\gamma_{\text{Lang}}^V}{d\varepsilon}\right)dt + \sqrt{\frac{2d^2}{\beta W\gamma_{\text{Lang}}^V}}dW_t \ . \tag{S125}$$

If we choose

$$W\gamma_{\text{Lang}}^V = \frac{d^2 \tau_P V}{\kappa_T} \ , \tag{S126}$$

the equations of the MTTK barostat then become those of the SCR barostat.

The SCR barostat requires rescaling of the coordinate of the particle and, sometimes, the velocity of the particle after each volume-updating step. In ref [41], three integrators for the SCR barostat are proposed, namely, "Euler," "Reversible," and "Trotter". Since the Euler integrator $e^{\mathcal{L}_{\text{NPT}}^{\text{Euler}}\Delta t} = e^{\mathcal{L}_T \Delta t/2}e^{\mathcal{L}_{\text{Bar}}^{\text{MD}}\Delta t}e^{\mathcal{L}_{\mathbf{p}}\Delta t/2}e^{\mathcal{L}_{\mathbf{x}}\Delta t}e^{\mathcal{L}_{\mathbf{p}}\Delta t/2}e^{\mathcal{L}_T \Delta t/2}$, which minimizes the computation of forces and virial terms in the evolution of a full time step $\Delta t$, is the most efficient among the three integrators of ref [41], we employ it for comparison. It is evident that the Euler integrator of ref [41] falls into the category of the conventional "side" scheme described in the main text. We then use "side SCR" to denote the Euler integrator of the SCR barostat of ref [41].

In the main text, we derive the "middle" scheme with the SCR barostat, e.g., $e^{\mathcal{L}_{\mathrm{NPT}}^{\mathrm{Middle}}\Delta t} = e^{\mathcal{L}_{\mathrm{Bar}}^{\mathrm{MD}}\Delta t}e^{\mathcal{L}_{x}\Delta t/2}e^{\mathcal{L}_{T}\Delta t}e^{\mathcal{L}_{x}\Delta t/2}e^{\mathcal{L}_{\mathbf{p}}\Delta t}$ based on the leap-frog algorithm. **Figure S15** shows the results for liquid water when MD simulations with the SCR barostat are performed. Figure S6 suggests that the "middle" scheme is superior to the conventional "side" scheme for the implementation of the SCR barostat. For instance, the density produced by the "middle" scheme with the SCR barostat exhibits an absolute deviation of only $3\times10^{-4}$ g·cm$^{-3}$ at $\Delta t = 2.0$ fs from the converged values, while the corresponding deviation for the "side" scheme with the SCR barostat is $\sim 3.4\times10^{-2}$ g·cm$^{-3}$. When other properties are tested, the recommended "middle" scheme also significantly outperforms the conventional "side" scheme for the implementation of the SCR barostat. Figure S6 also demonstrates that the performance of the "middle" scheme with the MTTK barostat is similar to that of the "middle" scheme with the SCR barostat. When the "middle" scheme is utilized, the MTTK barostat and the SCR barostat lead to almost the same results for the isobaric-isothermal ensemble.

The choice of parameter $\kappa_T$ is, however, important for efficient simulations with the SCR barostat. The strategy for choosing the value of $\kappa_T$ is discussed in the main text. However, when systems are under extreme conditions, e.g., high-pressure superfluid, where $\kappa_T$ is often unknown, an unreasonable value of $\kappa_T$ introduces errors for the NPT simulation. The relation of eq (S126) can be used to estimate the value of $\kappa_T$ from some primitive calculation.

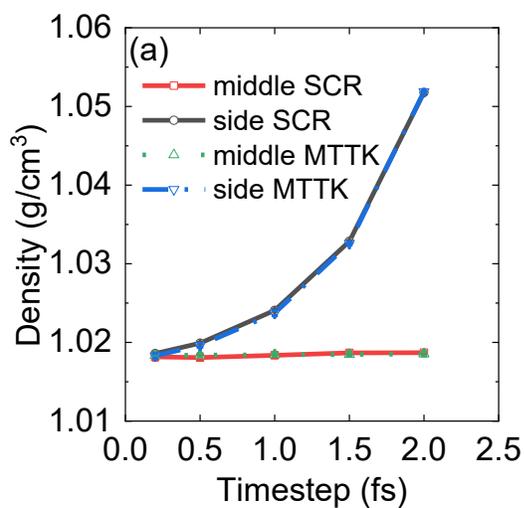
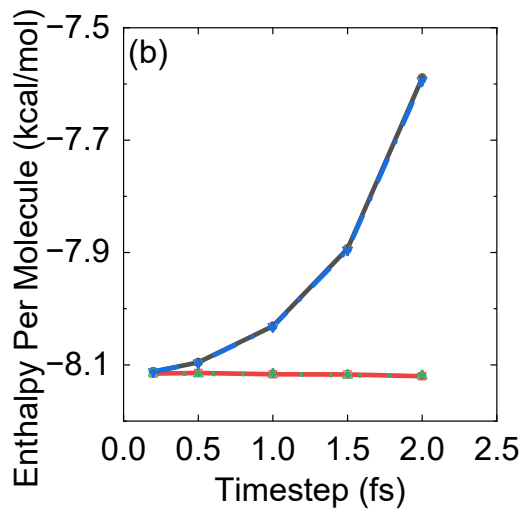
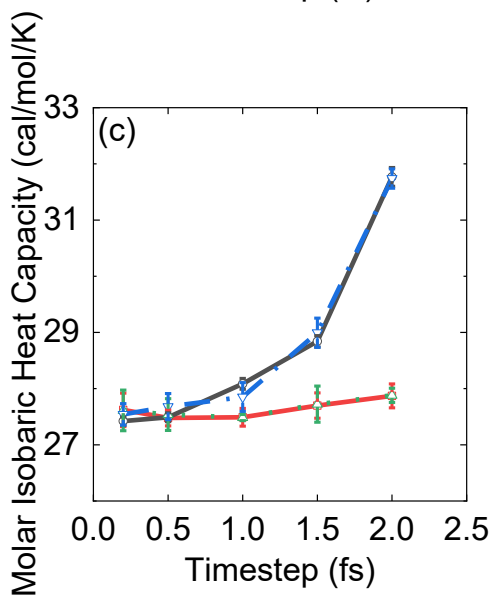
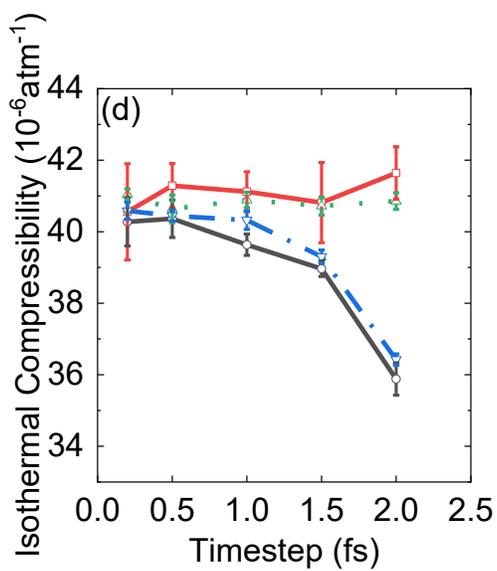
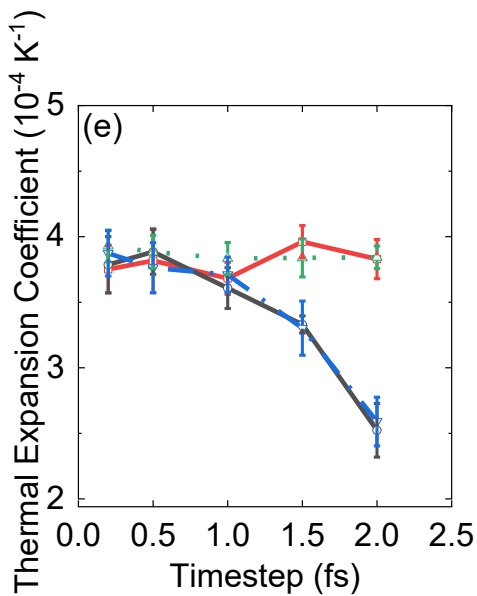
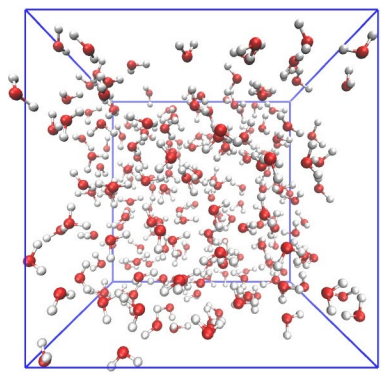

**Figure S15.** Results of NPT MD simulations for liquid water at $T = 298.15$ K and $P_{ext} = 1$ bar using q-SPC/fw model with increasing time intervals. The SCR barostat is used for the simulation. Data of simulations with MTTK barostat is the same as **Figure 6** in the main text. The calculated thermodynamic properties include (a) density, (b) enthalpy per molecule, (d) molar isobaric heat capacity, (e) isothermal compressibility, and (f) thermal expansion coefficient. Red lines: the "middle" scheme with the SCR barostat, Black lines: the conventional "side" scheme with the SCR barostat. Green lines: the "middle" scheme with the MTTK barostat. Blue lines: the conventional "side" scheme with the MTTK barostat. 20 trajectories are equilibrated for 1 ns and propagated for 10 ns using a cubic box with periodic boundary conditions containing 216 molecules. All data are divided into 20 groups to calculate the standard error. The friction parameter ($\gamma_{Lang}$) for the Langevin thermostat is $5.0$ ps$^{-1}$. The characteristic time ($\tau_P$) for the SCR barostat is $2.0$ ps and the isothermal compressibility ($\kappa_T$) is $4.5 \times 10^{-5}$ bar$^{-1}$. The cutoff for the short-range interactions is 9 Å. All MD simulations are performed by our modified DL_POLY_2 package[61].

## S4. "Middle" Scheme for the Isobaric-Isothermal Ensemble with Holonomic Constraints

Holonomic constraints are widely applied in molecular simulations of biological systems. By introducing holonomic constraints to high-frequency oscillatory degrees of freedom, it is possible to accelerate the simulation by significantly increasing the time interval. The "middle" scheme for canonical (constant-NVT) ensembles with holonomic constraints has already been developed in ref [45] and implemented in AMBER[62, 63] and other simulation packages. In this section, we extend our earlier work[45] to the "middle" scheme for the isobaric-isothermal ensembles with holonomic constraints.

## 1) Atomic Pressure and Molecular Pressure

In the main text, the internal pressure estimator of eq (23), often referred to as the atomic pressure, assumes that all degrees of freedom, including high-frequency modes such as O-H stretches, are rescaled after each volume-updating step. This assumption is indicated in the non-linear term in the evolution of coordinates, e.g., in eqs (24)-(28) of the MTTK barostat in the main text.

In contrast, the molecular pressure can be constructed using the center of mass of the molecule[8, 18, 96-98], which is defined as

$$P_{\text{int,mol}} = \frac{1}{3V}\left(\sum_{I=1}^{N_m}\frac{\mathbf{P}_I^2}{M_I} + \sum_{I=1}^{N_m}\mathcal{F}_I \cdot \mathbf{R}_I\right), \qquad (S127)$$

where $\mathbf{R}_I$ and $\mathbf{P}_I$ denote the coordinate and momentum vectors of the center of mass of the $I$-th molecule, respectively, $\mathcal{F}_I$ is the force acting on the center of mass of the $I$-th molecule, $M_I$ represents the total mass of the $I$-th molecule, and $N_m$ is the number of molecules. The equations of motion of the MTTK barostat using the molecular pressure are

$$\dot{\mathbf{x}}_{i,I} = \frac{\mathbf{P}_{i,I}}{m_{i,I}} + \frac{p_\epsilon}{W}\mathbf{R}_I, \qquad (S128)$$

$$\dot{\mathbf{p}}_{i,I} = -\frac{\partial U}{\partial \mathbf{x}_{i,I}} - \left(1+\frac{1}{N_m}\right)\frac{p_\epsilon}{W}\frac{m_{i,I}}{M_I}\mathbf{P}_I, \qquad (S129)$$

$$\dot{V} = \frac{dVp_\epsilon}{W}, \qquad (S130)$$

$$\dot{p}_\epsilon = dV\left(P_{\text{int,mol}} - P_{\text{ext}}\right) + \frac{1}{N_m}\sum_{I=1}^{N_m}\frac{\mathbf{P}_I^2}{M_I}, \qquad (S131)$$

where $\mathbf{x}_{i,I}$, $\mathbf{p}_{i,I}$ and $m_{i,I}$ are the coordinate, momentum, and mass of the $i$-th atom of the $I$-th molecule, respectively. It is straightforward to use the Fokker-Planck equation to verify eqs (S128)-(S131) correctly reproduce the distribution function of the isobaric-isothermal ensemble.

Similarly, the equations of motion of the SCR barostat with the molecular pressure are

$$\dot{\mathbf{x}}_{i,I} = \frac{\mathbf{p}_{i,I}}{m_{i,I}} + \frac{\dot{V}}{dV}\mathbf{R}_i , \quad (S132)$$

$$\dot{\mathbf{p}}_{i,I} = \mathbf{F}_{i,I} - \frac{\dot{V}}{dV}\frac{m_{i,I}}{M_I}\mathbf{P}_I , \quad (S133)$$

$$dV = V\frac{\kappa_T}{\tau_P}(P_{\text{int,mol}} - P_{\text{ext}})dt + \sqrt{\frac{2\kappa_T V k_B T}{\tau_p}}dW_t . \quad (S134)$$

It is easy to prove that the distribution function of the isobaric-isothermal ensemble is stationary for eqs (S132)-(S134).

### 2) "Middle" Scheme for the Isobaric-Isothermal Ensemble with Holonomic Constraints

As shown by Kalibaeva *et al.* in ref [96] and by Martyna *el al* in ref [23], when performing MD simulations of the isothermal-isobaric ensemble for systems with holonomic constraints, the internal pressure can be estimated by using either the molecular virial expression or atomic virial expression. The molecular virial approach can be straightforwardly implemented with the application of the SHAKE[99] or RATTLE[100] algorithm. This is because intramolecular interactions and constraint processes (e.g., bond length/angle constraints) do not contribute to the internal pressure and do not respond to the pressure fluctuation. We implement the recommended "middle" scheme with holonomic constraints in AMBER/AmberTools where the molecular virial expression is used.

The conventional "side" schemes as well as the "middle" scheme using the MTTK barostat with holonomic constraints and those using the SCR barostat with holonomic constraints are listed in **Table S3** and **Table S4**, respectively. As described in ref [45], the coordinate-constraining step $C_1$ and momentum-constraining step $C_2$ refer to the constraining procedures for the coordinate and momentum vectors in the RATTLE algorithm. When the system is subject to $n_c$ holonomic constraints

$$\boldsymbol{\sigma}(\mathbf{x}) = 0 \ , \tag{S135}$$

$$\frac{d}{dt}\boldsymbol{\sigma}(\mathbf{x}) = \left(\frac{\partial \boldsymbol{\sigma}}{\partial \mathbf{x}}\right)^T \mathbf{M}^{-1}\mathbf{p} = 0 \ , \tag{S136}$$

where $\boldsymbol{\sigma}(\mathbf{x})$ is a vector with $n_c$ dimensions, and the time derivative implies the constraint for the momentum vector. The coordinate-constraining step $C_1$ of the Rattle algorithm is

$$C_1 : \begin{cases} \text{solve } \boldsymbol{\lambda} : \boldsymbol{\sigma}\left(\tilde{\mathbf{x}}(\Delta t) + \mathbf{M}^{-1}\dfrac{\partial \boldsymbol{\sigma}}{\partial \mathbf{x}(0)}\boldsymbol{\lambda}\right) = 0 \\ \mathbf{x}(\Delta t) \leftarrow \tilde{\mathbf{x}}(\Delta t) + \mathbf{M}^{-1}\dfrac{\partial \boldsymbol{\sigma}}{\partial \mathbf{x}(0)}\boldsymbol{\lambda} \\ \mathbf{p}(\Delta t/2) \leftarrow \tilde{\mathbf{p}}(\Delta t/2) + \dfrac{1}{\Delta t}\dfrac{\partial \boldsymbol{\sigma}}{\partial \mathbf{x}(0)}\boldsymbol{\lambda} \end{cases} , \tag{S137}$$

where $\tilde{\mathbf{x}}(\Delta t)$ and $\tilde{\mathbf{p}}(\Delta t/2)$ are the coordinate and momentum vectors before the constraining step, and they do not satisfy the holonomic constraint conditions. In eq (S137), $\boldsymbol{\lambda}$ is the $n_c$-demensional vector of Lagrange multipliers. For triatomic molecules, such as the water molecule, holonomic constraints can be analytically solved by the SETTLE algorithm[64]. For general molecular systems, $\boldsymbol{\lambda}$ can be numerically solved in an iterative fashion using Newton's method, of which the iteration step reads

$$\text{solve}\,\bar{\boldsymbol{\lambda}}:\left(\frac{\partial\boldsymbol{\sigma}}{\partial\bar{\mathbf{x}}}\right)^T \mathbf{M}^{-1}\frac{\partial\boldsymbol{\sigma}}{\partial\mathbf{x}(0)}\bar{\boldsymbol{\lambda}}=-\boldsymbol{\sigma}(\bar{\mathbf{x}})$$

$$\bar{\mathbf{x}} \leftarrow \bar{\mathbf{x}} + \mathbf{M}^{-1}\frac{\partial\boldsymbol{\sigma}}{\partial\mathbf{x}(0)}\bar{\boldsymbol{\lambda}} \qquad (S138)$$

$$\boldsymbol{\lambda} \leftarrow \boldsymbol{\lambda} + \bar{\boldsymbol{\lambda}}$$

where the initial values for $\bar{\mathbf{x}}$ and $\boldsymbol{\lambda}$ can be set as $\bar{\mathbf{x}} = \tilde{\mathbf{x}}(\Delta t)$ and $\boldsymbol{\lambda}=0$, respectively. When the number of constraints $n_c$ is small, the vector $\bar{\boldsymbol{\lambda}}$ can be efficiently solved by LU decomposition, this approach is termed the m-SHAKE algorithm[101]. When the system involves a large number of constraints, the iteration can be performed to satisfy each holonomic constraint in succession, which is more numerically favorable[45].

The momentum-constraining step $C_2$ of the Rattle algorithm is

$$C_2:\begin{cases}\text{solve}\,\boldsymbol{\mu}:\left(\frac{\partial\boldsymbol{\sigma}}{\partial\mathbf{x}(\Delta t)}\right)^T \mathbf{M}^{-1}\left(\tilde{\mathbf{p}}(\Delta t)+\frac{\partial\boldsymbol{\sigma}}{\partial\mathbf{x}(\Delta t)}\boldsymbol{\mu}\right)=0 \\ \mathbf{p}(\Delta t) \leftarrow \tilde{\mathbf{p}}(\Delta t) + \frac{\partial\boldsymbol{\sigma}}{\partial\mathbf{x}(\Delta t)}\boldsymbol{\mu}\end{cases}, \qquad (S139)$$

$\boldsymbol{\mu}$ is an $n_c$-demensional vector, which can also be solved by Newton's method

$$\text{solve}\,\boldsymbol{\mu}:\left(\frac{\partial\boldsymbol{\sigma}}{\partial\mathbf{x}(\Delta t)}\right)^T \mathbf{M}^{-1}\frac{\partial\boldsymbol{\sigma}}{\partial\mathbf{x}(\Delta t)}\boldsymbol{\mu}=-\left(\frac{\partial\boldsymbol{\sigma}}{\partial\mathbf{x}(\Delta t)}\right)^T \mathbf{M}^{-1}\tilde{\mathbf{p}}(\Delta t). \qquad (S140)$$

The SETTLE algorithm can be employed to analytically solve the constraints for triatomic molecules. When the system involves a small number of constraints, vector $\boldsymbol{\mu}$ can be straightforwardly solved by the Cholesky decomposition. When the number of constraints is large, it is more reasonable to apply the iteration to each of holonomic constraints in succession as described in ref [45].

**Table S3.** The "Middle" Schemes with the MTTK Barostat and Holonomic Constraints

| Scheme | Decomposition Order |
|---|---|
| Side | $B - T - p_V - C_2 - p_r - p - V - C_2 - C_1 - x_r - x - p - p_r - p_V - T - B$ |
| Middle | $p_V - V - C_2 - C_1 - x_r - B - x - T - x - B - x_r - V - p_V - C_2 - p_r - p$ |

**Table S4.** The "Middle" Schemes with the SCR Barostat and Holonomic Constraints

| Scheme | Decomposition Order |
|---|---|
| Side in GROMACS | $T - C_2 - p - p_r - x_r - V - C_2 - C_1 - x - p - T$ |
| Side in AMBER | $p_r - x_r - V - C_1 - x - p - T$ |
| Middle | $p_r - x_r - V - C_2 - C_1 - x - T - x - C_2 - p$ |

We have implemented the "middle" scheme with the SCR barostat (with holonomic constraints), as described in **Table S4**, into AMBER and AmberTools[62, 63], where the molecular virial and molecular pressure are used. The MD simulations of liquid water with the SPC/E model, where intramolecular motion is fixed, are performed using the modified "pmemd" module of AMBER with CPU-based parallelization and with GPU-accelerated computing. Simulation results are shown in **Figure 8** in the main text. The time interval can also be as large as ~6 fs. When AMBER is used, the "middle" scheme outperforms the conventional "side" scheme in the calculation of thermodynamic properties (density, enthalpy, isobaric heat capacity, isothermal compressibility, and thermal expansion coefficient).

**S5: Numerical Algorithms for the "Middle" Scheme.**

In this section we present the details of the MD and PIMD algorithms of the recommended "middle" scheme. Below we use the velocity-Verlet-based version for demonstration. It is trivial to extend the algorithms to those of the leap-frog-based version.

**1) "Middle" Scheme with the MTTK Barostat for MD**

The velocity-Verlet-based algorithm for the "middle" scheme with the MTTK barostat for MD reads

$$e^{\mathcal{L}_\mathbf{p}\Delta t/2} : \mathbf{p} \leftarrow \mathbf{p} - \frac{\Delta t}{2}\frac{\partial U}{\partial \mathbf{x}},$$

$$e^{\mathcal{L}_{\mathbf{p}_r}\Delta t/2} : \mathbf{p} \leftarrow \mathbf{p} e^{-\left(1+\frac{d}{N_f}\right)\frac{p_\varepsilon}{W}\frac{\Delta t}{2}},$$

(Momentum-constraining step $C_2$ if holonomic constraints are requied)

$$e^{\mathcal{L}_{p_\varepsilon}\Delta t/2} : p_\varepsilon \leftarrow p_\varepsilon + dV(P_{\text{int}} - P_{\text{ext}})\frac{\Delta t}{2} + \frac{d}{N_f}\mathbf{p}^T\mathbf{M}^{-1}\mathbf{p}\frac{\Delta t}{2},$$

$$e^{\mathcal{L}_V\Delta t/2} : V \leftarrow V e^{d\frac{p_\varepsilon}{W}\frac{\Delta t}{2}},$$

$$e^{\mathcal{L}_{\mathbf{x}_r}\Delta t/2} : \mathbf{x} \leftarrow \mathbf{x} e^{\frac{p_\varepsilon}{W}\frac{\Delta t}{2}},$$

$$e^{\mathcal{L}_B\Delta t/2} : p_\varepsilon \leftarrow e^{-\gamma_{\text{Lang}}^V \Delta t/2} p_\varepsilon + \sqrt{1 - e^{-\gamma_{\text{Lang}}^V \Delta t}}\sqrt{\frac{1}{\beta}}W^{1/2}\eta_\varepsilon(t),$$

$$e^{\mathcal{L}_\mathbf{x}\Delta t/2} : \mathbf{x} \leftarrow \mathbf{x} + \frac{\Delta t}{2}\mathbf{M}^{-1}\mathbf{p},$$

$$e^{\mathcal{L}_T\Delta t} : \mathbf{p} \leftarrow e^{-\gamma_{\text{Lang}}\Delta t}\mathbf{p} + \sqrt{1 - e^{-2\gamma_{\text{Lang}}\Delta t}}\sqrt{\frac{1}{\beta}}\mathbf{M}^{1/2}\boldsymbol{\eta}(t),$$

$$e^{\mathcal{L}_\mathbf{x}\Delta t/2} : \mathbf{x} \leftarrow \mathbf{x} + \frac{\Delta t}{2}\mathbf{M}^{-1}\mathbf{p},$$

$$e^{\mathcal{L}_B\Delta t/2} : p_\varepsilon \leftarrow e^{-\gamma_{\text{Lang}}^V \Delta t/2} p_\varepsilon + \sqrt{1 - e^{-\gamma_{\text{Lang}}^V \Delta t}}\sqrt{\frac{1}{\beta}}W^{1/2}\eta_\varepsilon(t + \Delta t/2),$$

$$e^{\mathcal{L}_{\mathbf{x}_r}\Delta t/2} : \mathbf{x} \leftarrow \mathbf{x} e^{\frac{p_\varepsilon}{W}\frac{\Delta t}{2}},$$

(Coordinate-constraining step $C_1$ and momentum-constraining step $C_2$ if holonomic constraints are requied)

$$e^{\mathcal{L}_V\Delta t/2} : V \leftarrow V e^{d\frac{p_\varepsilon}{W}\frac{\Delta t}{2}},$$

(Force calculation)

$$e^{\mathcal{L}_{p_\varepsilon}\Delta t/2} : p_\varepsilon \leftarrow p_\varepsilon + dV(P_{\text{int}} - P_{\text{ext}})\frac{\Delta t}{2} + \frac{d}{N_f}\mathbf{p}^T\mathbf{M}^{-1}\mathbf{p}\frac{\Delta t}{2},$$

$$e^{\mathcal{L}_{\mathbf{p}_r}\Delta t/2} : \mathbf{p} \leftarrow \mathbf{p} e^{-\left(1+\frac{d}{N_f}\right)\frac{p_\varepsilon}{W}\frac{\Delta t}{2}},$$

$$e^{\mathcal{L}_\mathbf{p}\Delta t/2} : \mathbf{p} \leftarrow \mathbf{p} - \frac{\Delta t}{2}\frac{\partial U}{\partial \mathbf{x}}, \tag{S141}$$

where $\eta_\varepsilon(t)$ is a gaussian standard random number at a fixed time $t$ with zero mean $\langle \eta_\varepsilon(t) \rangle = 0$ and unit deviation $\langle (\eta_\varepsilon(t))^2 \rangle = 1$ for updating the "piston" momentum $p_\varepsilon$, $\boldsymbol{\eta}(t)$ is a gaussian standard random number vector at a fixed time $t$ with zero mean $\langle \boldsymbol{\eta}(t) \rangle = \mathbf{0}$ and diagonal deviation matrix $\langle \boldsymbol{\eta}(t)\boldsymbol{\eta}^T(t) \rangle = \mathbf{1}$ for updating the particle momentum $\mathbf{p}$, and $\eta_\varepsilon(t+\Delta t/2)$ is a gaussian standard random number at a fixed time $t+\Delta t/2$ with zero mean $\langle \eta_\varepsilon(t+\Delta t/2) \rangle = 0$ and unit deviation $\langle (\eta_\varepsilon(t+\Delta t/2))^2 \rangle = 1$ for updating the "piston" momentum $p_\varepsilon$. The coordinate-constraining step $C_1$ and momentum-constraining step $C_2$ of the RATTLE algorithm are defined in ref [45] and briefly described in Section S4 of the Supporting Information.

2) **"Middle" Scheme with the MTTK Barostat for PIMD**

The velocity-Verlet-based algorithm for the "middle" scheme with the MTTK barostat for PIMD using the "all-mode scaled" method for the internal pressure (eq (S53)) reads:

$$e^{\mathcal{L}_{\mathbf{p}}\Delta t/2} : \mathbf{p}_i \leftarrow \mathbf{p}_i - \frac{\Delta t}{2}\left(\frac{\partial \phi}{\partial \boldsymbol{\xi}_i} + \omega_P^2 \overline{\mathbf{M}}_i \boldsymbol{\xi}_i\right) \quad (i=1,\cdots,L),$$

$$e^{\mathcal{L}_{\mathbf{p}_r}\Delta t/2} : \mathbf{p}_i \leftarrow \mathbf{p}_i e^{-\left(1+\frac{d}{N_f L}\right)\frac{p_\varepsilon}{W}\frac{\Delta t}{2}} \quad (i=1,\cdots,L),$$

$$e^{\mathcal{L}_{p_\varepsilon}\Delta t/2} : p_\varepsilon \leftarrow p_\varepsilon + dV(P_{\text{int}}^{\text{PI,all-mode}} - P_{\text{ext}})\frac{\Delta t}{2} + \frac{d}{N_f L}\sum_{i=1}^{L}\mathbf{p}_i^T \widetilde{\mathbf{M}}_i^{-1} \mathbf{p}_i \frac{\Delta t}{2},$$

$$e^{\mathcal{L}_V \Delta t/2} : V \leftarrow V e^{d\frac{p_\varepsilon}{W}\frac{\Delta t}{2}},$$

$$e^{\mathcal{L}_{\mathbf{x}_r}\Delta t/2} : \boldsymbol{\xi}_i \leftarrow \boldsymbol{\xi}_i e^{\frac{p_\varepsilon}{W}\frac{\Delta t}{2}} \quad (i=1,\cdots,L),$$

$$e^{\mathcal{L}_B \Delta t/2} : p_\varepsilon \leftarrow e^{-\gamma_{\text{Lang}}^V \Delta t/2} p_\varepsilon + \sqrt{1-e^{-\gamma_{\text{Lang}}^V \Delta t}}\sqrt{\frac{1}{\beta}}W^{1/2}\eta_\varepsilon(t),$$

$$e^{\mathcal{L}_{\mathbf{x}}\Delta t/2} : \boldsymbol{\xi}_i \leftarrow \boldsymbol{\xi}_i + \frac{\Delta t}{2}\widetilde{\mathbf{M}}_i^{-1}\mathbf{p}_i \quad (i=1,\cdots,L),$$

$$e^{\mathcal{L}_T \Delta t} : \mathbf{p}_i \leftarrow e^{-\gamma_{\text{Lang}}^{(i)}\Delta t}\mathbf{p} + \sqrt{1-e^{-2\gamma_{\text{Lang}}^{(i)}\Delta t}}\sqrt{\frac{1}{\beta}}\widetilde{\mathbf{M}}_i^{1/2}\boldsymbol{\eta}(t) \quad (i=1,\cdots,L),$$

$$e^{\mathcal{L}_{\mathbf{x}}\Delta t/2} : \boldsymbol{\xi}_i \leftarrow \boldsymbol{\xi}_i + \frac{\Delta t}{2}\widetilde{\mathbf{M}}_i^{-1}\mathbf{p}_i \quad (i=1,\cdots,L),$$

$$e^{\mathcal{L}_B \Delta t/2} : p_\varepsilon \leftarrow e^{-\gamma_{\text{Lang}}^V \Delta t/2} p_\varepsilon + \sqrt{1-e^{-\gamma_{\text{Lang}}^V \Delta t}}\sqrt{\frac{1}{\beta}}W^{1/2}\eta_\varepsilon(t+\Delta t/2),$$

$$e^{\mathcal{L}_{\mathbf{x}_r}\Delta t/2} : \boldsymbol{\xi}_i \leftarrow \boldsymbol{\xi}_i e^{\frac{p_\varepsilon}{W}\frac{\Delta t}{2}} \quad (i=1,\cdots,L),$$

$$e^{\mathcal{L}_V \Delta t/2} : V \leftarrow V e^{d\frac{p_\varepsilon}{W}\frac{\Delta t}{2}},$$

(Force calculation)

$$e^{\mathcal{L}_{p_\varepsilon}\Delta t/2} : p_\varepsilon \leftarrow p_\varepsilon + dV(P_{\text{int}}^{\text{PI,all-mode}} - P_{\text{ext}})\frac{\Delta t}{2} + \frac{d}{N_f L}\sum_{i=1}^{L}\mathbf{p}_i^T \widetilde{\mathbf{M}}_i^{-1} \mathbf{p}_i \frac{\Delta t}{2},$$

$$e^{\mathcal{L}_{\mathbf{p}_r}\Delta t/2} : \mathbf{p}_i \leftarrow \mathbf{p}_i e^{-\left(1+\frac{d}{N_f L}\right)\frac{p_\varepsilon}{W}\frac{\Delta t}{2}} \quad (i=1,\cdots,L),$$

$$e^{\mathcal{L}_{\mathbf{p}}\Delta t/2} : \mathbf{p}_i \leftarrow \mathbf{p}_i - \frac{\Delta t}{2}\left(\frac{\partial \phi}{\partial \boldsymbol{\xi}_i} + \omega_P^2 \overline{\mathbf{M}}_i \boldsymbol{\xi}_i\right) \quad (i=1,\cdots,L).$$

(S142)

Below we also present the velocity-Verlet-based algorithm for the "middle" scheme with the MTTK barostat for PIMD using the "reduced dynamics" method for the internal pressure (eq (S55)):

$$e^{\mathcal{L}_{\mathbf{p}}\Delta t/2} : \mathbf{p}_i \leftarrow \mathbf{p}_i - \frac{\Delta t}{2}\left(\frac{\partial \phi}{\partial \boldsymbol{\xi}_i} + \omega_P^2 \overline{\mathbf{M}}_i \boldsymbol{\xi}_i\right) \quad (i=1,\cdots,L),$$

$$e^{\mathcal{L}_{\mathbf{p}_r}\Delta t/2} : \mathbf{p}_1 \leftarrow \mathbf{p}_1 e^{-\left(1+\frac{d}{N_f}\right)\frac{p_\varepsilon}{W}\frac{\Delta t}{2}}$$

$$e^{\mathcal{L}_{p_\varepsilon}\Delta t/2} : p_\varepsilon \leftarrow p_\varepsilon + dV(P_{\text{int}}^{\text{PI,reduced}} - P_{\text{ext}})\frac{\Delta t}{2} + \frac{d}{N_f}\mathbf{p}_1^T \widetilde{\mathbf{M}}_1^{-1} \mathbf{p}_1 \frac{\Delta t}{2},$$

$$e^{\mathcal{L}_V \Delta t/2} : V \leftarrow V e^{d\frac{p_\varepsilon}{W}\frac{\Delta t}{2}},$$

$$e^{\mathcal{L}_{\mathbf{x}_r}\Delta t/2} : \boldsymbol{\xi}_1 \leftarrow \boldsymbol{\xi}_1 e^{\frac{p_\varepsilon}{W}\frac{\Delta t}{2}}$$

$$e^{\mathcal{L}_B \Delta t/2} : p_\varepsilon \leftarrow e^{-\gamma_{\text{Lang}}^V \Delta t/2} p_\varepsilon + \sqrt{1 - e^{-\gamma_{\text{Lang}}^V \Delta t}} \sqrt{\frac{1}{\beta}} W^{1/2} \eta_\varepsilon(t),$$

$$e^{\mathcal{L}_{\mathbf{x}}\Delta t/2} : \boldsymbol{\xi}_i \leftarrow \boldsymbol{\xi}_i + \frac{\Delta t}{2} \widetilde{\mathbf{M}}_i^{-1} \mathbf{p}_i \quad (i=1,\cdots,L),$$

$$e^{\mathcal{L}_T \Delta t} : \mathbf{p}_i \leftarrow e^{-\gamma_{\text{Lang}}^{(i)} \Delta t} \mathbf{p} + \sqrt{1 - e^{-2\gamma_{\text{Lang}}^{(i)} \Delta t}} \sqrt{\frac{1}{\beta}} \widetilde{\mathbf{M}}_i^{1/2} \boldsymbol{\eta}(t) \quad (i=1,\cdots,L),$$

$$e^{\mathcal{L}_{\mathbf{x}}\Delta t/2} : \boldsymbol{\xi}_i \leftarrow \boldsymbol{\xi}_i + \frac{\Delta t}{2} \widetilde{\mathbf{M}}_i^{-1} \mathbf{p}_i \quad (i=1,\cdots,L),$$

$$e^{\mathcal{L}_{p_\varepsilon}\Delta t/2} : p_\varepsilon \leftarrow e^{-\gamma_{\text{Lang}}^V \Delta t/2} p_\varepsilon + \sqrt{1 - e^{-\gamma_{\text{Lang}}^V \Delta t}} \sqrt{\frac{1}{\beta}} W^{1/2} \eta_\varepsilon(t + \Delta t/2),$$

$$e^{\mathcal{L}_{\mathbf{x}_r}\Delta t/2} : \boldsymbol{\xi}_1 \leftarrow \boldsymbol{\xi}_1 e^{\frac{p_\varepsilon}{W}\frac{\Delta t}{2}}$$

$$e^{\mathcal{L}_V \Delta t/2} : V \leftarrow V e^{d\frac{p_\varepsilon}{W}\frac{\Delta t}{2}},$$

(Force calculation)

$$e^{\mathcal{L}_{p_\varepsilon}\Delta t/2} : p_\varepsilon \leftarrow p_\varepsilon + dV(P_{\text{int}}^{\text{PI,reduced}} - P_{\text{ext}})\frac{\Delta t}{2} + \frac{d}{N_f}\sum_{i=1}^{L} \mathbf{p}_1^T \widetilde{\mathbf{M}}_1^{-1} \mathbf{p}_1 \frac{\Delta t}{2},$$

$$e^{\mathcal{L}_{\mathbf{p}_r}\Delta t/2} : \mathbf{p}_1 \leftarrow \mathbf{p}_1 e^{-\left(1+\frac{d}{N_f}\right)\frac{p_\varepsilon}{W}\frac{\Delta t}{2}}$$

$$e^{\mathcal{L}_{\mathbf{p}}\Delta t/2} : \mathbf{p}_i \leftarrow \mathbf{p}_i - \frac{\Delta t}{2}\left(\frac{\partial \phi}{\partial \boldsymbol{\xi}_i} + \omega_P^2 \overline{\mathbf{M}}_i \boldsymbol{\xi}_i\right) \quad (i=1,\cdots,L).$$

(S143)

### 3) "Middle" Scheme with the SCR Barostat for MD

The velocity-Verlet-based algorithm for the "middle" scheme with the MTTK barostat for MD is:

$$e^{\mathcal{L}_\mathbf{p}\Delta t/2} : \mathbf{p} \leftarrow \mathbf{p} - \frac{\Delta t}{2}\frac{\partial U}{\partial \mathbf{x}},$$
(Momentum-constraining step $C_2$ if holonomic constraints are requied)

$$e^{\mathcal{L}_\mathbf{x}\Delta t/2} : \mathbf{x} \leftarrow \mathbf{x} + \frac{\Delta t}{2}\mathbf{M}^{-1}\mathbf{p},$$

$$e^{\mathcal{L}_T \Delta t} : \mathbf{p} \leftarrow e^{-\gamma_{\text{Lang}}\Delta t}\mathbf{p} + \sqrt{1 - e^{-2\gamma_{\text{Lang}}\Delta t}}\sqrt{\frac{1}{\beta}}\mathbf{M}^{1/2}\boldsymbol{\eta}(t),$$

$$e^{\mathcal{L}_\mathbf{x}\Delta t/2} : \mathbf{x} \leftarrow \mathbf{x} + \frac{\Delta t}{2}\mathbf{M}^{-1}\mathbf{p},$$
(Coordinate-constraining step $C_1$ and momentum-constraining step $C_2$ if holonomic constraints are requied)
(Force calculation)

$$e^{\mathcal{L}_\varepsilon \Delta t} : \varepsilon \leftarrow \varepsilon + \Delta\varepsilon,$$
$$e^{\mathcal{L}_{\mathbf{x}_r}\Delta t} : \mathbf{x} \leftarrow \mathbf{x}e^{\Delta\varepsilon/d},$$
$$e^{\mathcal{L}_{\mathbf{p}_r}\Delta t} : \mathbf{p} \leftarrow \mathbf{p}e^{-\Delta\varepsilon/d}.$$
$$e^{\mathcal{L}_\mathbf{p}\Delta t/2} : \mathbf{p} \leftarrow \mathbf{p} - \frac{\Delta t}{2}\frac{\partial U}{\partial \mathbf{x}}, \tag{S144}$$

where $\Delta\varepsilon$ is defined by eq (30) of the main text.

Similarly, we can readily construct the PIMD algorithm for the "middle" scheme with the SCR barostat.

### S6. "Middle" Scheme in the Isobaric-Isothermal Ensemble for Anisotropic Systems

In the main text, we focus on the isobaric-isothermal (constant-NPT) ensemble for isotropic systems. When anisotropic systems like crystals are investigated, it is essential to account for changes in the shape of the simulation cell with periodic boundary conditions. In this section, we extend the recommended "middle" scheme for studying anisotropic systems, where the volume fluctuation is anisotropic.

The partition function of the isobaric-isothermal (constant-NPT) ensemble for anisotropic systems is

$$Z_{\text{NPT}} = \frac{I_N}{V_0} \int_0^\infty dV \int d\mathbf{h}_0 e^{-\beta P_{\text{ext}} V} Z_{\text{NVT}}(N,V,\mathbf{h}_0,T)\delta(\det(\mathbf{h}_0)-1)$$
$$= \frac{I_N}{V_0} \int_0^\infty dV \int d\mathbf{h} V^{1-d} e^{-\beta P_{\text{ext}} V} Z_{\text{NVT}}(N,V,\mathbf{h}_0,T)\delta(\det(\mathbf{h})-V), \quad (S145)$$

where $\mathbf{h} = \{\mathbf{a},\mathbf{b},\mathbf{c}\}$ is the simulation cell matrix[8] and $\mathbf{h}_0 = V^{-1/3}\mathbf{h}$ is the unit cell matrix. With loss of generality, we consider the molecular system in the 3-dimensional space. The MTTK barostat for anisotropic systems is defined as [7, 21, 22, 51, 102]

$$\dot{\mathbf{x}}_i = \frac{\mathbf{p}_i}{m_i} + \frac{\mathbf{p}_g}{W_g}\mathbf{x}_i, \quad (S146)$$

$$\dot{\mathbf{p}}_i = -\frac{\partial U}{\partial \mathbf{x}_i} - \frac{\mathbf{p}_g}{W_g}\mathbf{p}_i - \frac{1}{N_f}\frac{\text{Tr}[\mathbf{p}_g]}{W_g}\mathbf{p}_i - \gamma_{\text{Lang}}\mathbf{p}_i + \sqrt{\frac{2\gamma_{\text{Lang}}}{\beta}}\mathbf{M}_i^{1/2}\boldsymbol{\eta}_i, \quad (S147)$$

$$\dot{\mathbf{h}} = \frac{\mathbf{p}_g \mathbf{h}}{W_g}, \quad (S148)$$

$$\dot{\mathbf{p}}_g = \det(\mathbf{h})(\mathbf{P}^{(\text{int})} - \mathbf{I}P_{\text{ext}}) + \frac{1}{N_f}\sum_i \frac{\mathbf{p}_i^2}{m_i}\mathbf{I} - \gamma_{\text{Lang}}^V \mathbf{p}_g + \sqrt{\frac{2\gamma_{\text{Lang}}^V}{\beta}}W_g^{1/2}\boldsymbol{\eta}_g, \quad (S149)$$

where $\mathbf{x}_i, \mathbf{p}_i$ are 3-dimensional coordinate and momentum vectors of the $i$-th particle or atom, $\mathbf{p}_g$ is a 3x3 matrix, and the internal pressure tensor $\mathbf{P}^{(\text{int})}$ is defined as

$$P_{\alpha\beta}^{(\text{int})} = \frac{1}{\det(\mathbf{h})} \sum_{i=1}^{N} \left[ \frac{(\mathbf{p}_i^T \hat{\mathbf{e}}_\alpha)(\mathbf{p}_i^T \hat{\mathbf{e}}_\beta)}{m_i} + (\mathbf{F}_i^T \hat{\mathbf{e}}_\alpha)(\mathbf{x}_i^T \hat{\mathbf{e}}_\beta) \right]. \tag{S150}$$

Equations (S146)-(S149) lead to the stationary distribution of the isobaric-isothermal ensemble

$$\rho(\mathbf{x}, \mathbf{p}, \mathbf{h}, \mathbf{p}_g) \propto \exp\left[ -\beta\left( \frac{1}{2} \mathbf{p}^T \mathbf{M}^{-1} \mathbf{p} + U(\mathbf{x}) + P_{\text{ext}} \det[\mathbf{h}] + \frac{\text{Tr}[\mathbf{p}_g \mathbf{p}_g^T]}{2 W_g} \right) \right] \det[\mathbf{h}]^{1-d}. \tag{S151}$$

The corresponding Kolmogorov operators of eqs (S146)-(S149) and their operations on the density distribution function of eq (S151) are

$$\mathcal{L}_{\mathbf{x}} \rho = \sum_i \left[ -\left( \frac{\mathbf{p}_i}{m_i} \right)^T \frac{\partial \rho}{\partial \mathbf{x}_i} \right] = \beta \rho \left( \sum_{i=1}^{N} \left( \frac{\mathbf{p}_i}{m_i} \right)^T \frac{\partial U}{\partial \mathbf{x}_i} \right), \tag{S152}$$

$$\mathcal{L}_{\mathbf{p}} \rho = \sum_i \left[ \left( \frac{\partial U}{\partial \mathbf{x}_i} \right)^T \frac{\partial \rho}{\partial \mathbf{p}_i} \right] = -\beta \rho \left( \sum_{i=1}^{N} \left( \frac{\mathbf{p}_i}{m_i} \right)^T \frac{\partial U}{\partial \mathbf{x}_i} \right), \tag{S153}$$

$$\mathcal{L}_{\mathbf{x}_r} \rho = \sum_i \left[ -\left( \frac{\partial}{\partial \mathbf{x}_i} \right)^T \left( \frac{\mathbf{p}_g}{W_g} \mathbf{x}_i \rho \right) \right] = -\frac{\rho}{W_g} \sum_{i=1}^{N} \left[ \text{Tr}[p_g] - \beta \left( \frac{\partial U}{\partial \mathbf{x}_i} \right)^T \mathbf{p}_g \mathbf{x}_i \right], \tag{S154}$$

$$\mathcal{L}_{\mathbf{p}_r} \rho = \sum_i \left[ \left( \frac{\partial}{\partial \mathbf{p}_i} \right)^T \left( \frac{\mathbf{p}_g}{W_g} \mathbf{p}_i \rho \right) + \frac{1}{N_f} \frac{\text{Tr}[\mathbf{p}_g]}{W_g} \left( \frac{\partial}{\partial \mathbf{p}_i} \right)^T (\mathbf{p}_i \rho) \right]$$
$$= \frac{\rho}{W_g} \left[ \sum_{i=1}^{N} \left( \text{Tr}[\mathbf{p}_g] - \frac{\beta}{m_i} \mathbf{p}_i^T \mathbf{p}_g \mathbf{p}_i \right) + \text{Tr}[\mathbf{p}_g] - \frac{\beta \text{Tr}[\mathbf{p}_g]}{N_f} \sum_{i=1}^{N} \frac{\mathbf{p}_i^2}{m_i} \right], \tag{S155}$$

$$\mathcal{L}_{\mathbf{h}} \rho = -\sum_{\alpha,\beta} \frac{\partial}{\partial h_{\alpha\beta}} \left( \frac{\mathbf{p}_g}{W_g} \mathbf{h} \rho \right)_{\alpha\beta} = \frac{\rho \text{Tr}[\mathbf{p}_g]}{W_g} \left( \beta P_{\text{ext}} \det[\mathbf{h}] - 1 \right), \tag{S156}$$

$$\mathcal{L}_{\mathbf{p}_g} \rho = \sum_{\alpha,\beta} -\left( \det[\mathbf{h}] \left( P_{\alpha\beta}^{(\text{int})} - P_{\text{ext}} \delta_{\alpha\beta} \right) + \frac{1}{N_f} \sum_i \frac{\mathbf{p}_i^2}{m_i} \delta_{\alpha\beta} \right) \frac{\partial \rho}{\partial p_{g\alpha\beta}}$$
$$= \frac{\beta \rho}{W_g} \left[ \sum_{i=1}^{N} \left( \frac{1}{m_i} \mathbf{p}_i^T \mathbf{p}_g \mathbf{p}_i - \left( \frac{\partial U}{\partial \mathbf{x}_i} \right)^T \mathbf{p}_g \mathbf{x}_i \right) - P_{\text{ext}} \det[\mathbf{h}] \text{Tr}[\mathbf{p}_g] + \frac{\text{Tr}[\mathbf{p}_g]}{N_f} \sum_{i=1}^{N} \frac{\mathbf{p}_i^2}{m_i} \right], \tag{S157}$$

$$\mathcal{L}_T \rho = \sum_i \left[ \gamma_{\text{Lang}} \left( \frac{\partial}{\partial \mathbf{p}_i} \right)^T (\mathbf{p}_i \rho) + \frac{\gamma_{\text{Lang}}}{\beta} m_i \frac{\partial^2 \rho}{\partial \mathbf{p}_i^2} \right] = 0, \tag{S158}$$

$$\mathcal{L}_B \rho = \sum_{\alpha,\beta} \gamma_{\text{Lang}}^V \frac{\partial}{\partial p_{g\alpha\beta}} (p_{g\alpha\beta} \rho) + \frac{\gamma_{\text{Lang}}^V W_g}{\beta} \frac{\partial^2 \rho}{\partial p_{g\alpha\beta}^2} = 0. \tag{S159}$$

In the derivation of $\mathcal{L}_\mathbf{h} \rho$ of eq (S156), we use the identity:

$$\sum_{\beta=1}^{d} h_{\gamma\beta} \frac{\partial \det[\mathbf{h}]}{\partial h_{\alpha\beta}} = \delta_{\alpha\gamma} \det[\mathbf{h}] \tag{S160}$$

Equations (S152)-(S159) lead to

$$\mathcal{L}_{\text{MTTK}}^{(\text{full,aniso})} \rho \equiv \left( \mathcal{L}_\mathbf{x} + \mathcal{L}_{\mathbf{x}_r} + \mathcal{L}_\mathbf{p} + \mathcal{L}_{\mathbf{p}_r} + \mathcal{L}_\mathbf{h} + \mathcal{L}_{\mathbf{p}_g} + \mathcal{L}_T + \mathcal{L}_B \right) \rho = 0 \;, \tag{S161}$$

i.e., $\rho(\mathbf{x},\mathbf{p},\mathbf{h},\mathbf{p}_g)$ of eq (S151) is the stationary distribution of the anisotropic MTTK equation defined by eqs (S146)-(S149).

In the anisotropic case, the scalar internal pressure is replaced by the corresponding tensor form. Except this change, the performance of molecular dynamics is similar. The velocity-Verlet-based "middle" scheme

$$e^{\mathcal{L}_\mathbf{p} \Delta t/2} e^{\mathcal{L}_{\text{Bar}(2)}^{\text{MD}} \Delta t/2} e^{\mathcal{L}_\mathbf{x} \Delta t/2} e^{\mathcal{L}_T \Delta t} e^{\mathcal{L}_\mathbf{x} \Delta t/2} e^{\mathcal{L}_{\text{Bar}(1)}^{\text{MD}} \Delta t/2} e^{\mathcal{L}_\mathbf{p} \Delta t/2} \;, \tag{S162}$$

as well as the leap-frog-based version

$$e^{\mathcal{L}_{\text{Bar}(2)}^{\text{MD}} \Delta t/2} e^{\mathcal{L}_\mathbf{x} \Delta t/2} e^{\mathcal{L}_T \Delta t} e^{\mathcal{L}_\mathbf{x} \Delta t/2} e^{\mathcal{L}_{\text{Bar}(1)}^{\text{MD}} \Delta t/2} e^{\mathcal{L}_\mathbf{p} \Delta t} \tag{S163}$$

is also recommended for the MTTK barostat for MD/PIMD simulations for the isobaric-isothermal ensemble for isotropic systems. In eq (S162) or eq (S163), $e^{\mathcal{L}_{\text{Bar}(1)}^{\text{MD}} \Delta t/2}$ and $e^{\mathcal{L}_{\text{Bar}(2)}^{\text{MD}} \Delta t/2}$ are defined by

$$e^{\mathcal{L}_{\text{Bar}(1)}^{\text{MD}} \Delta t} = e^{\mathcal{L}_B \Delta t} e^{\mathcal{L}_{\mathbf{x}_r} \Delta t} e^{\mathcal{L}_\mathbf{h} \Delta t} e^{\mathcal{L}_{\mathbf{p}_g} \Delta t} e^{\mathcal{L}_{\mathbf{p}_r} \Delta t} \tag{S164}$$

$$e^{\mathcal{L}_{\text{Bar}(2)}^{\text{MD}} \Delta t} = e^{\mathcal{L}_{\mathbf{p}_r} \Delta t} e^{\mathcal{L}_{p_\varepsilon} \Delta t} e^{\mathcal{L}_{\mathbf{p}_g} \Delta t} e^{\mathcal{L}_\mathbf{h} \Delta t} e^{\mathcal{L}_B \Delta t} \tag{S165}$$

The MD algorithm based on the velocity-Verlet algorithm for the "middle" scheme with anisotropic MTTK barostat is

$$e^{\mathcal{L}_{\mathbf{p}}\Delta t/2} : \mathbf{p}_i \leftarrow \mathbf{p}_i - \frac{\Delta t}{2}\frac{\partial U}{\partial \mathbf{x}_i} \quad (i=1,\cdots,N_{\text{atom}}),$$

$$e^{\mathcal{L}_{\mathbf{p}_r}\Delta t/2} : \mathbf{p}_i \leftarrow \exp\left(-\left(\frac{\mathbf{p}_g}{W_g} + \frac{1}{N}\frac{\text{Tr}[\mathbf{p}_g]}{3W_g}\mathbf{I}\right)\frac{\Delta t}{2}\right)\mathbf{p}_i \quad (i=1,\cdots,N_{\text{atom}}),$$

$$e^{\mathcal{L}_{\mathbf{p}_g}\Delta t/2} : \mathbf{p}_g \leftarrow \mathbf{p}_g + \det(\mathbf{h})(\mathbf{P}_{\text{int}} - \mathbf{I}P_{\text{ext}})\frac{\Delta t}{2} + \frac{1}{3N}\sum_i^N \frac{\mathbf{p}_i^2}{m_i}\frac{\Delta t}{2},$$

$$e^{\mathcal{L}_{\mathbf{h}}\Delta t/2} : \mathbf{h} \leftarrow \exp\left(\frac{\mathbf{p}_g}{W_g}\frac{\Delta t}{2}\right)\mathbf{h},$$

$$e^{\mathcal{L}_{\mathbf{x}_r}\Delta t/2} : \mathbf{x}_i \leftarrow \exp\left(\frac{\mathbf{p}_g}{W_g}\frac{\Delta t}{2}\right)\mathbf{x}_i \quad (i=1,\cdots,N_{\text{atom}}),$$

$$e^{\mathcal{L}_{B}\Delta t/2} : \mathbf{p}_g \leftarrow e^{-\gamma_{\text{Lang}}^V \Delta t/2}\mathbf{p}_g + \sqrt{1-e^{-\gamma_{\text{Lang}}^V \Delta t}}\sqrt{\frac{1}{\beta}}W_g^{1/2}\boldsymbol{\eta}^{(g)}(t),$$

$$e^{\mathcal{L}_{\mathbf{x}}\Delta t/2} : \mathbf{x}_i \leftarrow \mathbf{x}_i + \frac{\Delta t}{2}\frac{1}{m_i}\mathbf{p}_i \quad (i=1,\cdots,N_{\text{atom}}),$$

$$e^{\mathcal{L}_{T}\Delta t} : \mathbf{p}_i \leftarrow e^{-\gamma_{\text{Lang}}\Delta t}\mathbf{p}_i + \sqrt{1-e^{-2\gamma_{\text{Lang}}\Delta t}}\sqrt{\frac{m_i}{\beta}}\boldsymbol{\eta}_i(t) \quad (i=1,\cdots,N_{\text{atom}}),$$

$$e^{\mathcal{L}_{\mathbf{x}}\Delta t/2} : \mathbf{x}_i \leftarrow \mathbf{x}_i + \frac{\Delta t}{2}\frac{1}{m_i}\mathbf{p}_i \quad (i=1,\cdots,N_{\text{atom}}),$$

$$e^{\mathcal{L}_{B}\Delta t/2} : \mathbf{p}_g \leftarrow e^{-\gamma_{\text{Lang}}^V \Delta t/2}\mathbf{p}_g + \sqrt{1-e^{-\gamma_{\text{Lang}}^V \Delta t}}\sqrt{\frac{1}{\beta}}W_g^{1/2}\boldsymbol{\eta}^{(g)}(t+\Delta t/2),$$

$$e^{\mathcal{L}_{\mathbf{x}_r}\Delta t/2} : \mathbf{x}_i \leftarrow \exp\left(\frac{\mathbf{p}_g}{W_g}\frac{\Delta t}{2}\right)\mathbf{x}_i \quad (i=1,\cdots,N_{\text{atom}}),$$

$$e^{\mathcal{L}_{\mathbf{h}}\Delta t/2} : \mathbf{h} \leftarrow \exp\left(\frac{\mathbf{p}_g}{W_g}\frac{\Delta t}{2}\right)\mathbf{h},$$

(Force calculation)

$$e^{\mathcal{L}_{\mathbf{p}_g}\Delta t/2} : \mathbf{p}_g \leftarrow \mathbf{p}_g + \det(\mathbf{h})(\mathbf{P}_{\text{int}} - \mathbf{I}P_{\text{ext}})\frac{\Delta t}{2} + \frac{1}{3N}\sum_i^N \frac{\mathbf{p}_i^2}{m_i}\frac{\Delta t}{2},$$

$$e^{\mathcal{L}_{\mathbf{p}_r}\Delta t/2} : \mathbf{p}_i \leftarrow \exp\left(-\left(\frac{\mathbf{p}_g}{W_g} + \frac{1}{N}\frac{\text{Tr}[\mathbf{p}_g]}{3W_g}\mathbf{I}\right)\frac{\Delta t}{2}\right)\mathbf{p}_i \quad (i=1,\cdots,N_{\text{atom}}), \quad \text{(S166)}$$

$$e^{\mathcal{L}_{\mathbf{p}}\Delta t/2} : \mathbf{p}_i \leftarrow \mathbf{p}_i - \frac{\Delta t}{2}\frac{\partial U}{\partial \mathbf{x}_i} \quad (i=1,\cdots,N_{\text{atom}}),$$

where $\exp(\cdot)$ is the matrix exponential, $\mathbf{\eta}_i(t)$ is a three-dimensional independent standard-Gaussian-random-number vector at a fixed time $t$ with zero mean $\langle \mathbf{\eta}_i(t) \rangle = \mathbf{0}$ and diagonal deviation matrix $\langle \mathbf{\eta}_i(t)\mathbf{\eta}_i^T(t) \rangle = \mathbf{1}$ for updating the momentum vector of the $i$-th atom, and each element of the $3 \times 3$ matrix $\mathbf{\eta}^{(g)}(t)$ is defined as an independent standard-Gaussian-random-number at a fixed time $t$ with zero mean $\langle \eta_{\alpha\beta}^{(g)}(t) \rangle = 0$ and $\langle \eta_{\alpha\beta}^{(g)}(t)\eta_{\alpha'\beta'}^{(g)}(t) \rangle = \delta_{\alpha\alpha'}\delta_{\beta\beta'}$.

■ **References:**

(1) Drummond, N. D.; Monserrat, B.; Lloyd-Williams, J. H.; Ríos, P. L.; Pickard, C. J.; Needs, R. J., Quantum Monte Carlo Study of the Phase Diagram of Solid Molecular Hydrogen at Extreme Pressures. *Nature Communications* **2015**, *6*, 7794. http://dx.doi.org/10.1038/ncomms8794
(2) Bi, W.; Culverhouse, T.; Nix, Z.; Xie, W.; Tien, H.-J.; Chang, T.-R.; Dutta, U.; Zhao, J.; Lavina, B.; Alp, E. E.; Zhang, D.; Xu, J.; Xiao, Y.; Vohra, Y. K., Drastic Enhancement of Magnetic Critical Temperature and Amorphization in Topological Magnet Eusn2p2 under Pressure. *npj Quantum Materials* **2022**, *7*, 43. http://dx.doi.org/10.1038/s41535-022-00451-9
(3) Eremets, M. I.; Minkov, V. S.; Drozdov, A. P.; Kong, P. P., The Characterization of Superconductivity under High Pressure. *Nature Materials* **2024**, *23*, 26-27. http://dx.doi.org/10.1038/s41563-023-01769-w
(4) Tagawa, S.; Sakamoto, N.; Hirose, K.; Yokoo, S.; Hernlund, J.; Ohishi, Y.; Yurimoto, H., Experimental Evidence for Hydrogen Incorporation into Earth's Core. *Nature Communications* **2021**, *12*, 2588. http://dx.doi.org/10.1038/s41467-021-22035-0
(5) Gleason, A. E.; Rittman, D. R.; Bolme, C. A.; Galtier, E.; Lee, H. J.; Granados, E.; Ali, S.; Lazicki, A.; Swift, D.; Celliers, P.; Militzer, B.; Stanley, S.; Mao, W. L., Dynamic Compression of Water to Conditions in Ice Giant Interiors. *Scientific Reports* **2022**, *12*, 715. http://dx.doi.org/10.1038/s41598-021-04687-6
(6) Frenkel, D.; Smit, B., *Understanding Molecular Simulation: From Algorithms to Applications*. Third edition ed.; Academic Press, an imprint of Elsevier: London San Diego,CA Cambridge, MA Kidlington, 2023; p 728.
(7) Martyna, G. J.; Hughes, A.; Tuckerman, M. E., Molecular Dynamics Algorithms for Path Integrals at Constant Pressure. *The Journal of Chemical Physics* **1999**, *110*, 3275-3290. http://dx.doi.org/10.1063/1.478193
(8) Tuckerman, M. E., *Statistical Mechanics: Theory and Molecular Simulation*. Oxford University Press: Oxford ; New York, 2010.
(9) McDonald, I. R., Npt-Ensemble Monte Carlo Calculations for Binary Liquid Mixtures. *Molecular Physics* **1972**, *23*, 41-58. http://dx.doi.org/10.1080/00268977200100031

(10) Wood, W. W., Monte Carlo Calculations for Hard Disks in the Isothermal-Isobaric Ensemble. *The Journal of Chemical Physics* **1968**, *48*, 415-434. http://dx.doi.org/10.1063/1.1667938

(11) Chow, K.-H.; Ferguson, D. M., Isothermal-Isobaric Molecular Dynamics Simulations with Monte Carlo Volume Sampling. *Computer Physics Communications* **1995**, *91*, 283-289. http://dx.doi.org/10.1016/0010-4655(95)00059-O

(12) Faller, R.; De Pablo, J. J., Constant Pressure Hybrid Molecular Dynamics–Monte Carlo Simulations. *The Journal of Chemical Physics* **2002**, *116*, 55-59. http://dx.doi.org/10.1063/1.1420460

(13) Åqvist, J.; Wennerström, P.; Nervall, M.; Bjelic, S.; Brandsdal, B. O., Molecular Dynamics Simulations of Water and Biomolecules with a Monte Carlo Constant Pressure Algorithm. *Chemical Physics Letters* **2004**, *384*, 288-294. http://dx.doi.org/10.1016/j.cplett.2003.12.039

(14) Andersen, H. C., Molecular Dynamics Simulations at Constant Pressure and/or Temperature. *The Journal of Chemical Physics* **1980**, *72*, 2384-2393. http://dx.doi.org/10.1063/1.439486

(15) Parrinello, M.; Rahman, A., Crystal Structure and Pair Potentials: A Molecular-Dynamics Study. *Physical Review Letters* **1980**, *45*, 1196-1199. http://dx.doi.org/10.1103/PhysRevLett.45.1196

(16) Parrinello, M.; Rahman, A., Polymorphic Transitions in Single Crystals: A New Molecular Dynamics Method. *Journal of Applied Physics* **1981**, *52*, 7182-7190. http://dx.doi.org/10.1063/1.328693

(17) Nosé, S.; Klein, M. L., Constant Pressure Molecular Dynamics for Molecular Systems. *Molecular Physics* **1983**, *50*, 1055-1076. http://dx.doi.org/10.1080/00268978300102851

(18) Di Pierro, M.; Elber, R.; Leimkuhler, B., A Stochastic Algorithm for the Isobaric–Isothermal Ensemble with Ewald Summations for All Long Range Forces. *The Journal of Chemical Theory and Computation* **2015**, *11*, 5624-5637. http://dx.doi.org/10.1021/acs.jctc.5b00648

(19) Ke, Q.; Gong, X.; Liao, S.; Duan, C.; Li, L., Effects of Thermostats/Barostats on Physical Properties of Liquids by Molecular Dynamics Simulations. *Journal of Molecular Liquids* **2022**, *365*, 120116. http://dx.doi.org/10.1016/j.molliq.2022.120116

(20) Hoover, W. G., Constant-Pressure Equations of Motion. *Physical Review A* **1986**, *34*, 2499-2500. http://dx.doi.org/10.1103/PhysRevA.34.2499

(21) Melchionna, S.; Ciccotti, G.; Lee Holian, B., Hoover Npt Dynamics for Systems Varying in Shape and Size. *Molecular Physics* **1993**, *78*, 533-544. http://dx.doi.org/10.1080/00268979300100371

(22) Martyna, G. J.; Tobias, D. J.; Klein, M. L., Constant Pressure Molecular Dynamics Algorithms. *The Journal of Chemical Physics* **1994**, *101*, 4177-4189. http://dx.doi.org/10.1063/1.467468

(23) Martyna, G. J.; Tuckerman, M. E.; Tobias, D. J.; Klein, M. L., Explicit Reversible Integrators for Extended Systems Dynamics. *Molecular Physics* **1996**, *87*, 1117-1157. http://dx.doi.org/10.1080/00268979600100761

(24) Tuckerman, M. E.; Alejandre, J.; López-Rendón, R.; Jochim, A. L.; Martyna, G. J., A Liouville-Operator Derived Measure-Preserving Integrator for Molecular Dynamics Simulations in the Isothermal–Isobaric Ensemble. *Journal of Physics A: Mathematical and General* **2006**, *39*, 5629-5651. http://dx.doi.org/10.1088/0305-4470/39/19/S18

(25) Yu, T.-Q.; Alejandre, J.; López-Rendón, R.; Martyna, G. J.; Tuckerman, M. E., Measure-Preserving Integrators for Molecular Dynamics in the Isothermal–Isobaric Ensemble Derived from the Liouville Operator. *Chemical Physics* **2010**, *370*, 294-305. http://dx.doi.org/10.1016/j.chemphys.2010.02.014
(26) Lippert, R. A.; Predescu, C.; Ierardi, D. J.; Mackenzie, K. M.; Eastwood, M. P.; Dror, R. O.; Shaw, D. E., Accurate and Efficient Integration for Molecular Dynamics Simulations at Constant Temperature and Pressure. *The Journal of Chemical Physics* **2013**, *139*, 164106. http://dx.doi.org/10.1063/1.4825247
(27) Feller, S. E.; Zhang, Y.; Pastor, R. W.; Brooks, B. R., Constant Pressure Molecular Dynamics Simulation: The Langevin Piston Method. *The Journal of Chemical Physics* **1995**, *103*, 4613-4621. http://dx.doi.org/10.1063/1.470648
(28) Kolb, A.; Dünweg, B., Optimized Constant Pressure Stochastic Dynamics. *The Journal of Chemical Physics* **1999**, *111*, 4453-4459. http://dx.doi.org/10.1063/1.479208
(29) Grønbech-Jensen, N.; Farago, O., Constant Pressure and Temperature Discrete-Time Langevin Molecular Dynamics. *The Journal of Chemical Physics* **2014**, *141*, 194108. http://dx.doi.org/10.1063/1.4901303
(30) Gao, X.; Fang, J.; Wang, H., Sampling the Isothermal-Isobaric Ensemble by Langevin Dynamics. *The Journal of Chemical Physics* **2016**, *144*, 124113. http://dx.doi.org/10.1063/1.4944909
(31) Feynman, R. P., Atomic Theory of the Lambda-Transition in Helium. *Physical Review* **1953**, *91*, 1291-1301. http://dx.doi.org/10.1103/PhysRev.91.1291
(32) Feynman, R. P.; Hibbs, A. R., *Quantum Mechanics and Path Integrals*. McGraw-Hill: New York, 1965.
(33) Chandler, D.; Wolynes, P. G., Exploiting the Isomorphism between Quantum Theory and Classical Statistical Mechanics of Polyatomic Fluids. *The Journal of Chemical Physics* **1981**, *74*, 4078-4095. http://dx.doi.org/10.1063/1.441588
(34) Ceperley, D. M., Path Integrals in the Theory of Condensed Helium. *Reviews of Modern Physics* **1995**, *67*, 279. http://dx.doi.org/10.1103/RevModPhys.67.279
(35) Tuckerman, M. E.; Berne, B. J.; Martyna, G. J.; Klein, M. L., Efficient Molecular Dynamics and Hybrid Monte Carlo Algorithms for Path Integrals. *The Journal of Chemical Physics* **1993**, *99*, 2796-2808. http://dx.doi.org/10.1063/1.465188
(36) Liu, J.; Li, D.; Liu, X., A Simple and Accurate Algorithm for Path Integral Molecular Dynamics with the Langevin Thermostat. *The Journal of Chemical Physics* **2016**, *145*, 024103. http://dx.doi.org/10.1063/1.4954990
(37) Parrinello, M.; Rahman, A., Study of an *F* Center in Molten Kcl. *The Journal of Chemical Physics* **1984**, *80*, 860-867. http://dx.doi.org/10.1063/1.446740
(38) Berne, B. J.; Thirumalai, D., On the Simulation of Quantum Systems: Path Integral Methods. *Annual Review of Physical Chemistry* **1986**, *37*, 401-424. http://dx.doi.org/10.1146/annurev.pc.37.100186.002153
(39) Tuckerman, M. E.; Marx, D.; Klein, M. L.; Parrinello, M., Efficient and General Algorithms for Path Integral Car–Parrinello Molecular Dynamics. *The Journal of Chemical Physics* **1996**, *104*, 5579-5588. http://dx.doi.org/10.1063/1.471771
(40) Berendsen, H. J. C.; Postma, J. P. M.; Van Gunsteren, W. F.; DiNola, A.; Haak, J. R., Molecular Dynamics with Coupling to an External Bath. *The Journal of Chemical Physics* **1984**, *81*, 3684-3690. http://dx.doi.org/10.1063/1.448118

(41)	Bernetti, M.; Bussi, G., Pressure Control Using Stochastic Cell Rescaling. *The Journal of Chemical Physics* **2020**, *153*, 114107. http://dx.doi.org/10.1063/5.0020514
(42)	Rogge, S. M. J.; Vanduyfhuys, L.; Ghysels, A.; Waroquier, M.; Verstraelen, T.; Maurin, G.; Van Speybroeck, V., A Comparison of Barostats for the Mechanical Characterization of Metal–Organic Frameworks. *The Journal of Chemical Theory and Computation* **2015**, *11*, 5583-5597. http://dx.doi.org/10.1021/acs.jctc.5b00748
(43)	Del Tatto, V.; Raiteri, P.; Bernetti, M.; Bussi, G., Molecular Dynamics of Solids at Constant Pressure and Stress Using Anisotropic Stochastic Cell Rescaling. *Applied Sciences* **2022**, *12*, 1139. http://dx.doi.org/10.3390/app12031139
(44)	Zhang, Z.; Liu, X.; Chen, Z.; Zheng, H.; Yan, K.; Liu, J., A Unified Thermostat Scheme for Efficient Configurational Sampling for Classical/Quantum Canonical Ensembles Via Molecular Dynamics. *The Journal of Chemical Physics* **2017**, *147*, 034109. http://dx.doi.org/10.1063/1.4991621
(45)	Zhang, Z.; Liu, X.; Yan, K.; Tuckerman, M. E.; Liu, J., Unified Efficient Thermostat Scheme for the Canonical Ensemble with Holonomic or Isokinetic Constraints Via Molecular Dynamics. *The Journal of Physical Chemistry A* **2019**, *123*, 6056-6079. http://dx.doi.org/10.1021/acs.jpca.9b02771
(46)	Li, D.; Han, X.; Chai, Y.; Wang, C.; Zhang, Z.; Chen, Z.; Liu, J.; Shao, J., Stationary State Distribution and Efficiency Analysis of the Langevin Equation Via Real or Virtual Dynamics. *The Journal of Chemical Physics* **2017**, *147*, 184104. http://dx.doi.org/10.1063/1.4996204
(47)	Leimkuhler, B.; Matthews, C., Rational Construction of Stochastic Numerical Methods for Molecular Sampling. *Applied Mathematics Research eXpress* **2013**, *2013*, 34-56. http://dx.doi.org/10.1093/amrx/abs010
(48)	Grønbech-Jensen, N.; Farago, O., A Simple and Effective Verlet-Type Algorithm for Simulating Langevin Dynamics. *Molecular Physics* **2013**, *111*, 983-991. http://dx.doi.org/10.1080/00268976.2012.760055
(49)	Li, D.; Chen, Z.; Zhang, Z.; Liu, J., Understanding Molecular Dynamics with Stochastic Processes Via Real or Virtual Dynamics. *Chinese Journal of Chemical Physics* **2017**, *30*, 735-760. http://dx.doi.org/10.1063/1674-0068/30/cjcp1711223
(50)	Zhang, Z.; Yan, K.; Liu, X.; Liu, J., A Leap-Frog Algorithm-Based Efficient Unified Thermostat Scheme for Molecular Dynamics. *Chinese Science Bulletin* **2018**, *63*, 3467-3483. http://dx.doi.org/10.1360/N972018-00908
(51)	Quigley, D.; Probert, M. I. J., Langevin Dynamics in Constant Pressure Extended Systems. *The Journal of Chemical Physics* **2004**, *120*, 11432-11441. http://dx.doi.org/10.1063/1.1755657
(52)	Wang, C.; Liu, J.; Shao, J., **Unpublished**.
(53)	Wang, W.; She, X.; Liu, J., **Unpublished**.
(54)	Abraham, M. J.; Murtola, T.; Schulz, R.; Páll, S.; Smith, J. C.; Hess, B.; Lindahl, E., Gromacs: High Performance Molecular Simulations through Multi-Level Parallelism from Laptops to Supercomputers. *SoftwareX* **2015**, *1-2*, 19-25. http://dx.doi.org/10.1016/j.softx.2015.06.001
(55)	Silvera, I. F.; Goldman, V. V., The Isotropic Intermolecular Potential for $H_2$ and $D_2$ in the Solid and Gas Phases. *The Journal of Chemical Physics* **1978**, *69*, 4209–4213. http://dx.doi.org/10.1063/1.437103
(56)	Paesani, F.; Zhang, W.; Case, D. A.; Cheatham, T. E.; Voth, G. A., An Accurate and Simple Quantum Model for Liquid Water. *The Journal of Chemical Physics* **2006**, *125*, 184507. http://dx.doi.org/10.1063/1.2386157

(57) Berendsen, H. J.; Grigera, J. R.; Straatsma, T. P., The Missing Term in Effective Pair Potentials. *The Journal of Physical Chemistry* **1987**, *91*, 6269-6271. http://dx.doi.org/10.1021/j100308a038

(58) Medders, G. R.; Babin, V.; Paesani, F., Development of a "First-Principles" Water Potential with Flexible Monomers. Iii. Liquid Phase Properties. *The Journal of Chemical Theory and Computation* **2014**, *10*, 2906-2910. http://dx.doi.org/10.1021/ct5004115

(59) Babin, V.; Medders, G. R.; Paesani, F., Development of a "First Principles" Water Potential with Flexible Monomers. Ii: Trimer Potential Energy Surface, Third Virial Coefficient, and Small Clusters. *The Journal of Chemical Theory and Computation* **2014**, *10*, 1599-1607. http://dx.doi.org/10.1021/ct500079y

(60) Babin, V.; Leforestier, C.; Paesani, F., Development of a "First Principles" Water Potential with Flexible Monomers: Dimer Potential Energy Surface, Vrt Spectrum, and Second Virial Coefficient. *The Journal of Chemical Theory and Computation* **2013**, *9*, 5395-5403. http://dx.doi.org/10.1021/ct400863t

(61) Smith, W.; Forester, T., Dl_Poly_2. 0: A General-Purpose Parallel Molecular Dynamics Simulation Package. *Journal of Molecular Graphics* **1996**, *14*, 136-141. http://dx.doi.org/10.1016/S0263-7855(96)00043-4

(62) Case, D. A.; Aktulga, H. M.; Belfon, K.; Cerutti, D. S.; Cisneros, G. A.; Cruzeiro, V. W. D.; Forouzesh, N.; Giese, T.; Goetz, A. W.; Gohlke, H.; Izadi, S.; Kasavajhala, K.; Kaymak, M. C.; King, E.; Kurtzman, T.; Lee, T.; Li, P.; Liu, J.; Luchko, T.; Luo, R.; Manathunga, M.; Machado, M. R.; Nguyen, H.; Hearn, K. A. O.; Onufriev, A.; Pan, F.; Pantano, S.; Qi, R.; Rahnamoun, A.; Risheh, A.; Schott-Verdugo, S.; Shajan, A.; Swails, J.; Wang, J.; Wei, H.; Wu, X.; Wu, Y.; Zhang, S.; Zhao, S.; Zhu, Q.; Cheatham III, T. E.; Roe, D. R.; Roitberg, A.; Simmerling, C.; York, D. M.; Nagan, M. C.; Merz, K. M., Ambertools. *Journal of Chemical Information and Modeling* **2023**, *63*, 6183-6191. http://dx.doi.org/10.1021/acs.jcim.3c01153

(63) Case, D. A.; Aktulga, H. M.; Belfon, K.; Ben-Shalom, I. Y.; Berryman, J. T.; Brozell, S. R.; Cerutti, D. S.; Cheatham III, T. E.; Cisneros, G. A.; Cruzeiro, V. W. D.; Darden, T. A.; Forouzesh, N.; Giambasu, G.; Giese, T.; Gilson, M. K.; Gohlke, H.; Goetz, A. W.; Harris, J.; Izadi, S.; Izmailov, S. A.; Kasavajhala, K.; Kaymak, M. C.; King, E.; Kovalenko, A.; Kurtzman, T.; Lee, T.; Li, P.; Lin, C.; Liu, J.; Luchko, T.; Luo, R.; Machado, M.; Man, V.; Manathunga, M.; Merz, K. M.; Miao, Y.; Mikhailovskii, O.; Monard, G.; Nguyen, H.; Hearn, K. A. O.; Onufriev, A.; Pan, F.; Pantano, S.; Qi, R.; Rahnamoun, A.; Roe, D. R.; Roitberg, A.; Sagui, C.; Schott-Verdugo, S.; Shajan, A.; Shen, J.; Simmerling, C. L.; Skrynnikov, N. R.; Smith, J.; Swails, J.; Walker, R. C.; Wang, J.; Wang, J.; Wei, H.; Wu, X.; Xiong, Y.; Xue, Y.; York, D. M.; Zhao, S.; Zhu, Q.; Kollman, P. A., *Amber 2023*. University of California, San Francisco: 2023.

(64) Miyamoto, S.; Kollman, P. A., Settle: An Analytical Version of the Shake and Rattle Algorithm for Rigid Water Models. *Journal of Computational Chemistry* **1992**, *13*, 952-962. http://dx.doi.org/10.1002/jcc.540130805

(65) Reddy, S. K.; Straight, S. C.; Bajaj, P.; Huy Pham, C.; Riera, M.; Moberg, D. R.; Morales, M. A.; Knight, C.; Götz, A. W.; Paesani, F., On the Accuracy of the Mb-Pol Many-Body Potential for Water: Interaction Energies, Vibrational Frequencies, and Classical Thermodynamic and Dynamical Properties from Clusters to Liquid Water and Ice. *The Journal of Chemical Physics* **2016**, *145*, 194504. http://dx.doi.org/10.1063/1.4967719

(66) Qu, C.; Yu, Q.; Houston, P. L.; Conte, R.; Nandi, A.; Bowman, J. M., Interfacing Q-Aqua with a Polarizable Force Field: The Best of Both Worlds. *Journal of Chemical Theory and Computation* **2023**, *19*, 3446-3459. http://dx.doi.org/10.1021/acs.jctc.3c00334
(67) Hoffman, M. P.; Xantheas, S. S., The Many-Body Expansion for Aqueous Systems Revisited: Iv. Stabilization of Halide–Anion Pairs in Small Water Clusters. *The Journal of Physical Chemistry A* **2024**, *128*, 9876-9892. http://dx.doi.org/10.1021/acs.jpca.4c05427
(68) Heindel, J. P.; Herman, K. M.; Xantheas, S. S., Many-Body Effects in Aqueous Systems: Synergies between Interaction Analysis Techniques and Force Field Development. *Annual Review of Physical Chemistry* **2023**, *74*, 337-360. http://dx.doi.org/10.1146/annurev-physchem-062422-023532
(69) Heindel, J. P.; Sami, S.; Head-Gordon, T., Completely Multipolar Model as a General Framework for Many-Body Interactions as Illustrated for Water. *The Journal of Chemical Theory and Computation* **2024**, *20*, 8594-8608. http://dx.doi.org/10.1021/acs.jctc.4c00812
(70) Weldon, R.; Wang, F., Water Potential from Adaptive Force Matching for Ice and Liquid with Revised Dispersion Predicts Supercooled Liquid Anomalies in Good Agreement with Two Independent Experimental Fits. *The Journal of Physical Chemistry B* **2024**, *128*, 3398-3407. http://dx.doi.org/10.1021/acs.jpcb.3c06495
(71) Feng, Y.; Cong, Y.; Zhao, Y.; Zhang, C.; Song, H.; Fang, B.; Yang, F.; Zhang, H.; Zhang, J. Z. H.; Zhang, L., "Blade of Polarized Water Molecule" Is the Key to Hydrolase Catalysis Regulation. *Journal of Chemical Information and Modeling* **2024**, *64*, 7987-7997. http://dx.doi.org/10.1021/acs.jcim.4c01123
(72) Fu, B.; Zhang, D. H., Accurate Fundamental Invariant-Neural Network Representation of *Ab Initio* Potential Energy Surfaces. *National Science Review* **2023**, *10*. http://dx.doi.org/10.1093/nsr/nwad321
(73) Zhang, L.; Han, J.; Wang, H.; Car, R.; E, W., Deep Potential Molecular Dynamics: A Scalable Model with the Accuracy of Quantum Mechanics. *Physical review letters* **2018**, *120*, 143001. http://dx.doi.org/10.1103/PhysRevLett.120.143001
(74) Wang, H.; Zhang, L.; Han, J.; E, W., Deepmd-Kit: A Deep Learning Package for Many-Body Potential Energy Representation and Molecular Dynamics. *Computer Physics Communications* **2018**, *228*, 178-184. http://dx.doi.org/10.1016/j.cpc.2018.03.016
(75) Jiang, F.; Zhou, C.-Y.; Wu, Y.-D., Residue-Specific Force Field Based on the Protein Coil Library. Rsff1: Modification of Opls-Aa/L. *The Journal of Physical Chemistry B* **2014**, *118*, 6983-6998. http://dx.doi.org/10.1021/jp5017449
(76) Zhou, C.-Y.; Jiang, F.; Wu, Y.-D., Residue-Specific Force Field Based on Protein Coil Library. Rsff2: Modification of Amber Ff99sb. *The Journal of Physical Chemistry B* **2015**, *119*, 1035-1047. http://dx.doi.org/10.1021/jp5064676
(77) Ji, C.; Mei, Y.; Zhang, J. Z. H., Developing Polarized Protein-Specific Charges for Protein Dynamics: Md Free Energy Calculation of Pka Shifts for Asp26/Asp20 in Thioredoxin. *Biophysical Journal* **2008**, *95*, 1080-1088. http://dx.doi.org/10.1529/biophysj.108.131110
(78) Chen, K.-W.; Sun, T.-Y.; Wu, Y.-D., New Insights into the Cooperativity and Dynamics of Dimeric Enzymes. *Chemical Reviews* **2023**, *123*, 9940-9981. http://dx.doi.org/10.1021/acs.chemrev.3c00042
(79) Nosé, S., A Molecular Dynamics Method for Simulations in the Canonical Ensemble. *Molecular Physics* **1984**, *52*, 255-268. http://dx.doi.org/10.1080/00268978400101201
(80) Nosé, S., A Unified Formulation of the Constant Temperature Molecular Dynamics Methods. *The Journal of Chemical Physics* **1984**, *81*, 511-519. http://dx.doi.org/10.1063/1.447334

(81) Hoover, W. G., Canonical Dynamics: Equilibrium Phase-Space Distributions. *Physical Review A* **1985**, *31*, 1695-1697. http://dx.doi.org/10.1103/PhysRevA.31.1695
(82) Martyna, G. J.; Klein, M. L.; Tuckerman, M., Nosé–Hoover Chains: The Canonical Ensemble Via Continuous Dynamics. *The Journal of Chemical Physics* **1992**, *97*, 2635-2643. http://dx.doi.org/10.1063/1.463940
(83) Brooks, B. R.; Brooks, C. L.; Mackerell, A. D.; Nilsson, L.; Petrella, R. J.; Roux, B.; Won, Y.; Archontis, G.; Bartels, C.; Boresch, S.; Caflisch, A.; Caves, L.; Cui, Q.; Dinner, A. R.; Feig, M.; Fischer, S.; Gao, J.; Hodoscek, M.; Im, W.; Kuczera, K.; Lazaridis, T.; Ma, J.; Ovchinnikov, V.; Paci, E.; Pastor, R. W.; Post, C. B.; Pu, J. Z.; Schaefer, M.; Tidor, B.; Venable, R. M.; Woodcock, H. L.; Wu, X.; Yang, W.; York, D. M.; Karplus, M., Charmm: The Biomolecular Simulation Program. *Journal of Computational Chemistry* **2009**, *30*, 1545-1614. http://dx.doi.org/10.1002/jcc.21287
(84) Kirkwood, J. G., Statistical Mechanics of Fluid Mixtures. *The Journal of Chemical Physics* **1935**, *3*, 300-313. http://dx.doi.org/10.1063/1.1749657
(85) Kästner, J., Umbrella Sampling. *Wiley Interdisciplinary Reviews: Computational Molecular Science* **2011**, *1*, 932-942. http://dx.doi.org/10.1002/wcms.66
(86) Laio, A.; Parrinello, M., Escaping Free-Energy Minima. *Proceedings of the National Academy of Sciences* **2002**, *99*, 12562-12566. http://dx.doi.org/10.1073/pnas.202427399
(87) Yang, L.; Qin Gao, Y., A Selective Integrated Tempering Method. *The Journal of Chemical Physics* **2009**, *131*, 214109. http://dx.doi.org/10.1063/1.3266563
(88) Han, X.; Lei, Y.-K.; Li, M.; Gao, Y. Q., A Brief Review of Integrated Tempering Sampling Molecular Simulation. *Chemical Physics Reviews* **2024**, *5*. http://dx.doi.org/10.1063/5.0175983
(89) Mori, T.; Hamers, R. J.; Pedersen, J. A.; Cui, Q., Integrated Hamiltonian Sampling: A Simple and Versatile Method for Free Energy Simulations and Conformational Sampling. *The Journal of Physical Chemistry B* **2014**, *118*, 8210-8220. http://dx.doi.org/10.1021/jp501339t
(90) Yang, Y. I.; Niu, H.; Parrinello, M., Combining Metadynamics and Integrated Tempering Sampling. *The Journal of Physical Chemistry Letters* **2018**, *9*, 6426-6430. http://dx.doi.org/10.1021/acs.jpclett.8b03005
(91) Chen, J.-N.; Dai, B.; Wu, Y.-D., Probability Density Reweighting of High-Temperature Molecular Dynamics. *Journal of Chemical Theory and Computation* **2024**, *20*, 4977-4985. http://dx.doi.org/10.1021/acs.jctc.3c01423
(92) Herman, M. F.; Bruskin, E. J.; Berne, B. J., On Path Integral Monte Carlo Simulations. *The Journal of Chemical Physics* **1982**, *76*, 5150-5155. http://dx.doi.org/10.1063/1.442815
(93) Shiga, M.; Shinoda, W., Calculation of Heat Capacities of Light and Heavy Water by Path-Integral Molecular Dynamics. *The Journal of Chemical Physics* **2005**, *123*. http://dx.doi.org/10.1063/1.2035078
(94) Tuckerman, M. E.; Liu, Y.; Ciccotti, G.; Martyna, G. J., Non-Hamiltonian Molecular Dynamics: Generalizing Hamiltonian Phase Space Principles to Non-Hamiltonian Systems. *The Journal of Chemical Physics* **2001**, *115*, 1678-1702. http://dx.doi.org/10.1063/1.1378321
(95) Hottovy, S.; McDaniel, A.; Volpe, G.; Wehr, J., The Smoluchowski-Kramers Limit of Stochastic Differential Equations with Arbitrary State-Dependent Friction. *Communications in Mathematical Physics* **2015**, *336*, 1259-1283. http://dx.doi.org/10.1007/s00220-014-2233-4
(96) Kalibaeva, G.; Ferrario, M.; Ciccotti, G., Constant Pressure-Constant Temperature Molecular Dynamics: A Correct Constrained *Npt* Ensemble Using the Molecular Virial. *Molecular Physics* **2003**, *101*, 765-778. http://dx.doi.org/10.1080/0026897021000044025

(97)     Hünenberger, P. H., Calculation of the Group-Based Pressure in Molecular Simulations. I. A General Formulation Including Ewald and Particle-Particle–Particle-Mesh Electrostatics. *The Journal of Chemical Physics* **2002**, *116*, 6880-6897. http://dx.doi.org/10.1063/1.1463057

(98)     Marry, V.; Ciccotti, G., Trotter Derived Algorithms for Molecular Dynamics with Constraints: Velocity Verlet Revisited. *The Journal of Computational Physics* **2007**, *222*, 428-440. http://dx.doi.org/10.1016/j.jcp.2006.07.033

(99)     Ryckaert, J.-P.; Ciccotti, G.; Berendsen, H. J., Numerical Integration of the Cartesian Equations of Motion of a System with Constraints: Molecular Dynamics of N-Alkanes. *The Journal of Computational Physics* **1977**, *23*, 327-341. http://dx.doi.org/10.1016/0021-9991(77)90098-5

(100)   Andersen, H. C., Rattle: A "Velocity" Version of the Shake Algorithm for Molecular Dynamics Calculations. *Journal of Computational Physics* **1983**, *52*, 24-34. http://dx.doi.org/10.1016/0021-9991(83)90014-1

(101)   Kräutler, V.; Van Gunsteren, W. F.; Hünenberger, P. H., A Fast Shake Algorithm to Solve Distance Constraint Equations for Small Molecules in Molecular Dynamics Simulations. *Journal of computational chemistry* **2001**, *22*, 501-508. http://dx.doi.org/10.1002/1096-987X(20010415)22:5<501::AID-JCC1021>3.0.CO;2-V

(102)   Shinoda, W.; Shiga, M.; Mikami, M., Rapid Estimation of Elastic Constants by Molecular Dynamics Simulation under Constant Stress. *Physical Review B* **2004**, *69*, 134103. http://dx.doi.org/10.1103/PhysRevB.69.134103